\documentclass[letterpaper,english,aps,prx,floatfix,showpacs,twocolumn,amsfonts,amssymb,superscriptaddress,longbibliography]{revtex4-2}

\usepackage{balance}
\usepackage[T1]{fontenc}
\usepackage[utf8]{inputenc}
\usepackage[english]{babel}
\usepackage{graphics}
\usepackage{amssymb}
\usepackage{amsmath, esint}
\usepackage{overpic}
\usepackage[toc]{appendix}
\usepackage{bbold}
\usepackage{comment}

\usepackage{natbib}
\usepackage{xcolor}
\usepackage{soul}
\usepackage{commath}
\usepackage{tikz}
\usepackage{physics}

\usepackage{hyperref}

\newcommand{\be}{\begin{equation}}
\newcommand{\ee}{\end{equation}}
\newcommand{\bspl}{\begin{split}}
\newcommand{\espl}{\end{split}}
\newcommand{\bea}{\begin{eqnarray}}
\newcommand{\eea}{\end{eqnarray}}

\newcommand{\angstrom}{\mbox{\normalfont\AA}}

\def\G{\Gamma}

\definecolor{darkred}{rgb}{0.8, 0.0, 0.0}
\definecolor{darkpowderblue}{rgb}{0.0, 0.2, 0.6}

\begin{document}
\title{Light-induced Faraday effect from dynamical breakdown of Kleinman symmetry
}
\author{Niccolò Sellati}
\email{niccolo.sellati@uniroma1.it}
\affiliation{Department of Physics, ``Sapienza'' University of Rome, P.le A.\ Moro 5, 00185 Rome, Italy}
\author{Jacopo Fiore}
\affiliation{Department of Physics, ``Sapienza'' University of Rome, P.le A.\ Moro 5, 00185 Rome, Italy}
\affiliation{Institute for Theory of Statistical Physics, RWTH Aachen University, Aachen, Germany}
\author{Lara Benfatto}
\email{lara.benfatto@roma1.infn.it}
\affiliation{Department of Physics, ``Sapienza'' University of Rome, P.le A.\ Moro 5, 00185 Rome, Italy}

\begin{abstract}
The observation of anomalously large polarization rotations in pump-probe experiments with circularly polarized light has recently challenged the conventional understanding of the inverse Faraday effect. The striking magnitude of these responses implies the generation of effective magnetic fields orders of magnitude larger than theoretical expectations, raising fundamental questions about the nature of light-induced time-reversal symmetry breaking. 
In this work we demonstrate that a static polarization rotation can originate entirely from the antisymmetric component of the third-order optical susceptibility, without generating a macroscopic magnetization of the material.
We show that this light-induced Faraday effect is inherently dynamical, emerging when Kleinman symmetry breaks down. Using a minimal $sp$ tight-binding model on a square lattice, we demonstrate that the light-induced Faraday response can be sizable even far from dissipative resonances. While the effect emerges at a purely electronic level, we show that resonant coupling with phonons can significantly enhance the pump-probe response. 
Finally, we demonstrate that the light-induced Faraday contribution can strongly exceed the magneto-optical response mediated by the inverse Faraday effect in nonmagnetic materials.
\end{abstract}
\date{\today}

\maketitle

\section{Introduction}
The Faraday effect is a paradigmatic manifestation of time-reversal symmetry breaking, observed as the polarization rotation of light propagating through a medium in the presence of either an external static magnetic field $\text{B}^z$ or spontaneous magnetic order \cite{pershan_pr63,pershan_prl65,pershan_pr66,kimel_nature05}. For a monochromatic electric field with polarization vector $\boldsymbol{\mathcal{E}}$, this effect is captured by the free-energy contribution \cite{pershan_pr63}
{
\begin{align}\label{eqF}
    F=-i\chi^{(\text{F})} \text{B}^z (\boldsymbol{\mathcal{E}}\times \boldsymbol{\mathcal{E}}^*)_z=\chi^{(\text{F})} \text{B}^z(|\mathcal{E}_L|^2-|\mathcal{E}_R|^2),
\end{align}
where $\chi^{(\text{F})}$ is the Faraday susceptibility. In transparent media, $\chi^{(\text{F})}$ is a real quantity governing the splitting of the refractive index $n$ for left- ($\mathcal{E}_L$) and right-circularly ($\mathcal{E}_R$) polarized light, $n_{L/R}=n\pm 2\pi\chi^{(\text{F})} \text{B}^z/n$.
Large Faraday rotations are typically driven by spontaneous magnetization or strong spin-orbit coupling in magnetic systems \cite{freiser_moreview}. However, nonmagnetic materials can also exhibit a small but finite $\chi^{(\text{F})}$, arising from non-resonant electronic excitations \cite{fowles_optics}.

The advent of high-intensity light sources has sparked considerable interest in optomagnetic effects \cite{rasing_rmp10}, where magnetic properties are manipulated via optical pulses. While strong effects are expected primarily in magnetic media \cite{kirilyuk_nature24}, recent experiments in nonmagnetic materials driven by circularly polarized pulses \cite{kruglyak_prb12,luo_science23,basini_nature24,davies_cm26,kamba_cm26,nelson_cm26} have challenged this picture. 
The interpretation of these studies was based on the inverse Faraday effect (IFE), described by the same free-energy of Eq.\ \eqref{eqF}, to induce a static magnetization $\text{M}^z=-\partial F/\partial \text{B}^z\propto|\mathcal{E}_L|^2-|\mathcal{E}_R|^2$ in the material. The resulting magnetization is typically detected through the polarization rotation of a delayed probe pulse \cite{kruglyak_prb12,luo_science23,basini_nature24,davies_cm26,kamba_cm26,nelson_cm26}.
Despite the weak paramagnetic or diamagnetic nature of the driven systems, the observed response is remarkably large, raising fundamental questions about its microscopic origin. 
For pumps in the visible range, the conclusion of Ref.\ \cite{kruglyak_prb12} is that the IFE is not governed by the same magneto-optical constant $\chi^{(\text{F})}$ that controls the direct Faraday effect, even though the reciprocity between the two processes was established long ago using a magnetic-field-sensitive probe, namely, by measuring the current induced in a coil wrapped around the sample \cite{pershan_prl65}.
In the terahertz (THz) regime, experiments resonantly driving soft infrared-active phonons \cite{luo_science23,basini_nature24,kamba_cm26} have explored whether the large Faraday rotation originates from ionic magnetic moments.
Dynamical multiferroicity \cite{juraschek_prm17,juraschek_prm19,juraschek_prr20,juraschek_prr22} predicts that circular lattice motion generates an ionic magnetization, which can in turn produce a local magnetic field and rotate the probe through the direct Faraday effect.
However, the magnetization inferred from the experimental rotation angle exceeds \textit{ab initio} estimates by several orders of magnitude \cite{basini_nature24,kamba_cm26}, indicating that the dynamical-multiferroic mechanism alone cannot account for the observed signal. This discrepancy has sparked intense theoretical efforts aimed at investigating a phononic analogue of the inverse Faraday effect \cite{luo_science23, geilhufe_prl24,wehling_prl25,juraschek_review24,urazhdin_prb25,juraschek_prb25,xiao_cm26}. Although the signals observed in these experiments have so far been ascribed to a magneto-optical origin, actual quantitative estimates of the IFE contribution \cite{basini_nature24,nelson_cm26} likewise fail to account for the anomalously large probe rotation.
An alternative explanation has been proposed by Merlin in Refs.\ \cite{merlin_prb24,merlin_pnas25}, suggesting that the rotation of the probe electric field is quantitatively dominated by a third-order nonlinear optical effect, analogous to the optical Faraday effect discussed for visible pump pulses \cite{wismer_prb17}. In this picture, the pump induces a transient magnetic-like field confined within the material, that has been referred to as non-Maxwellian as it is not accessible to magnetic probes. However, Refs.\ \cite{basini_nature24,kamba_cm26} report a quasi-static dichroic signal in addition to the conventional third-order Kerr response, raising the question of whether the effect discussed in Refs.\ \cite{wismer_prb17,merlin_prb24,merlin_pnas25} is already accounted for in the experimental analysis.
The relevance of the phononic excitations themselves has been questioned by recent experimental findings \cite{davies_cm26,nelson_cm26}, where a Faraday-like rotation of the probe is observed under circular driving in spectral regions where infrared-active phonons are absent.

In this manuscript, we resolve the apparent paradox posed by the observation of Faraday-like rotations in nonmagnetic systems driven by circularly polarized light, by demonstrating that they can be quantitatively understood in terms of a light-induced Faraday effect, sketched in Fig.\ \ref{Fig1}(a).
This phenomenon emerges directly from the third-order optical susceptibility $\chi_{ij;kl}$ and, crucially, does not induce a real macroscopic magnetization of the sample. 
While related optical analogues of the Faraday effect have been discussed previously \cite{wismer_prb17,merlin_prb24}, here we establish their microscopic origin, clarify their relevance to recent experiments \cite{luo_science23,basini_nature24,kamba_cm26,davies_cm26,nelson_cm26}, and quantify their magnitude in nonmagnetic materials.
By exploiting the intimate connection between the frequency dependence and the spatial symmetries of $\chi_{ij;kl}$ \cite{boyd_chap1,pershan_pr63,wismer_prb17}, we decompose the total response into symmetric and antisymmetric components $\chi^{\mathcal{S},\mathcal{A}}_{xy}$ with respect to the frequencies of the pump fields. While the former generates the optical Kerr effect \cite{basini_prb24}, observable also with linearly polarized light, the light-induced Faraday effect is governed by the antisymmetric component $\chi_{xy}^\mathcal{A}$, and only emerges under circularly polarized driving.
The corresponding free-energy contribution leading to a static rotation of the probe polarization can be expressed as
\begin{align}\label{eqK}
    F=\chi^{\mathcal{A}}_{xy} (\boldsymbol{\mathcal{E}}_\text{pu}\times \boldsymbol{\mathcal{E}}_\text{pu}^*)_z (\boldsymbol{\mathcal{E}}_\text{pr}\times \boldsymbol{\mathcal{E}}_\text{pr}^*)_z,
\end{align}
where $\boldsymbol{\mathcal{E}}_\text{pu}$ and $\boldsymbol{\mathcal{E}}_\text{pr}$ denote the pump and probe polarization vectors, respectively. The link to the magneto-optical Faraday effect described by Eq.\ \eqref{eqF} becomes evident by identifying the circularly polarized pump as an effective time-reversal-breaking field.
A key observation is that under the static approximation Kleinman symmetry holds \cite{boyd_chap1,alu_optica19,he_prb26}, forcing the antisymmetric component to vanish. The light-induced Faraday effect thus represents a purely dynamical response that emerges when Kleinman symmetry is broken at finite frequencies, even far from dissipative resonances. 
This fundamental symmetry constraint explains why theoretical estimates \cite{kruglyak_prb12,basini_nature24,kamba_cm26} based on a static approximation for $\chi_{ij;kl}$ cannot capture the light-induced Faraday response and therefore motivated interpretations based on alternative mechanisms.
\begin{figure}[t]
		\includegraphics[width=0.5\textwidth]{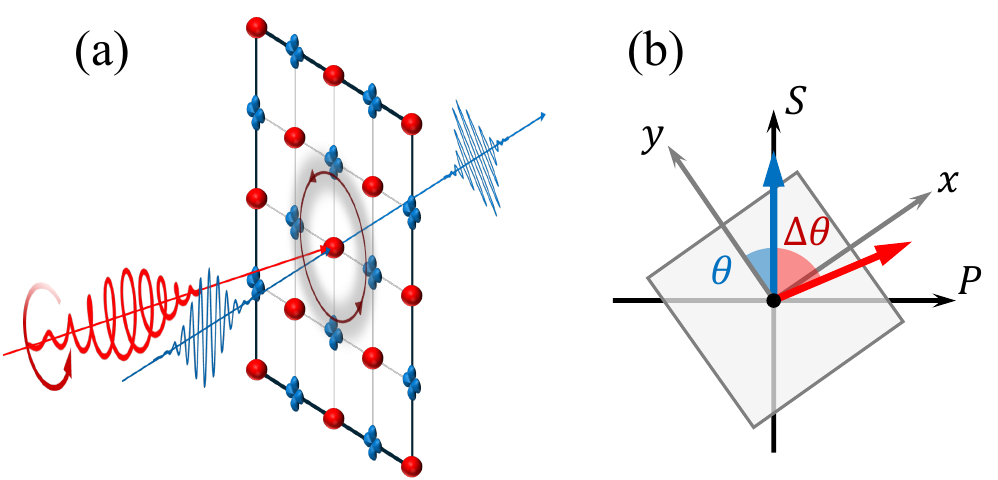}
		\caption{(a) Schematic illustration of the light-induced Faraday effect: a circularly polarized pump induces a collective electronic response, resulting in a rotation of the probe polarization. (b) Polarization geometry. The probe (blue) defines a fixed $PS$ reference frame, while the sample is rotated by an angle $\theta$. The pump (red) is described in a reference frame rotated by a fixed angle $\Delta\theta$ with respect to $PS$. \label{Fig1}}
\end{figure}

To quantitatively estimate the magnitude of the light-induced Faraday effect, we compute the relative strength of $\chi^\mathcal{A}_{xy}$ and $\chi^\mathcal{S}_{xy}$ within a multi-orbital toy model. The resulting response is of the same order of magnitude as that observed in Ref.\ \cite{basini_nature24} for paraelectric SrTiO$_3$. Notably, while the light-induced Faraday effect is intrinsically present at the purely electronic level, the resonant driving of infrared-active phonons can provide a significant enhancement of the response, a mechanism that is not accessible with optical pumps \cite{wismer_prb17}. In SrTiO$_3$, the softening of the phonon mode at low temperatures naturally accounts for the measured temperature dependence of the light-induced Faraday response, as we demonstrate by explicitly computing the lattice contribution to the dynamical antisymmetric response.

The mechanism behind the light-induced Faraday effect is fundamentally distinct from the IFE, as it does not involve the generation of a magnetization or external magnetic fields. Instead, the light-induced effect exploits the explicit time-reversal symmetry breaking of circularly polarized light to drive intrinsically dynamical nonlinear processes, the magnitude of which cannot be captured by static considerations. Our findings suggest that these dynamical effects are a general feature of transparent media, challenging the common assumption that a static description based on Kleinman symmetry provides an adequate picture of the nonlinear response of wide-bandgap insulators.
\section{Antisymmetric response beyond Kleinman approximation}
Connecting the light-induced Faraday effect to experimentally observable signals requires a careful modeling of the pump-probe balanced-detection setup, analogous to that employed in Refs.\ \cite{basini_nature24,basini_prb24}. The incoming probe field defines the $PS$ reference frame, with the $S$ axis aligned with the probe polarization $\text E_\text{pr}(\omega)$ [Fig.\ \ref{Fig1}(b)]. In this geometry, a pump-induced probe rotation generates a nonlinear electric field along the orthogonal $P$ axis, arising from the nonlinear polarization component $\text P^P_\text{NL}$. The resulting signal is recorded using a balanced-detection scheme \cite{miranda_2014} as a function of the pump-probe time delay $t_{pp}$,
and the measured quantity $\Delta\Gamma(t_{pp})$, corresponding to the differential intensity between the two photodiodes, is directly proportional to $\text P^P_\text{NL}$, as shown in Ref.\ \cite{basini_prb24}. For simplicity, we here assume that the $PS$ frame coincides with the crystallographic $xy$ axes of the cubic crystal. The more general case of a finite rotation angle $\theta$ between the two frames is discussed in the Supplemental Material \cite{suppl}.
In this configuration, the incoming probe is polarized along $y$, and the differential signal is proportional to the $x$ component of the nonlinear polarization. In particular, for third-order nonlinearities \cite{boyd_chap1,basini_prb24,fiore_prb26},
\begin{align}\label{dGgen}
    \Delta\Gamma(t_{pp}&)\propto \text{P}^x_\text{NL}=3\sum_{l\neq k}\int d\omega_1d\omega_2d\omega_3 e^{-i(\omega_2+\omega_3)t_{pp}}\nonumber\\
    \times\,&\chi_{xy;kl}(\omega;\omega_1,\omega_2,\omega_3)\text{E}_\text{pr}(\omega_1)\text{E}_\text{pu}^k(\omega_2)\text{E}_\text{pu}^l(\omega_3),
\end{align}
where $k,l=x,y$.
The four-wave-mixing processes generate sidebands around the fundamental probe frequency $\omega_1$, shifted by $\omega_2+\omega_3$, which appear as temporal modulations of $\Delta\Gamma(t_{pp})$. Here, $\text E_\text{pu}^{k,l}$ denote the $x$ or $y$ components of the pump field, polarized at an angle $\Delta\theta$ with respect to the probe at $t_{pp}=0$. For the third-order susceptibility $\chi_{ij;kl}(\omega_i;\omega_j,\omega_k,\omega_l)$ we adopt the standard convention in which $i$ labels the outgoing nonlinear signal, $j$ the incoming probe, and $k$, $l$ the pump fields. Energy conservation fixes $\omega_i=\omega_j+\omega_k+\omega_l$. Eq.\ \eqref{dGgen} assumes that all possible generated sidebands of the probe are collected by the photodetectors.

In this notation, each spatial index $s$ is associated with a corresponding frequency $\omega_s$. 
While intrinsic permutation symmetry ensures that the susceptibility remains unchanged under a simultaneous swap of spatial indices and their associated frequencies \cite{boyd_chap1}, $\chi_{xy;kl}(\omega;\omega_1,\omega_2,\omega_3)=\chi_{xy;lk}(\omega;\omega_1,\omega_3,\omega_2)$, exchanging only the spatial indices while keeping the frequencies fixed generally yields inequivalent tensor components $\chi_{xy;kl}(\omega;\omega_1,\omega_2,\omega_3)\neq\chi_{xy;lk}(\omega;\omega_1,\omega_2,\omega_3)$.
In the static limit, however, this distinction disappears and one recovers $\chi_{xy;kl}(0;0,0,0)=\chi_{xy;lk}(0;0,0,0)$. This property, known as Kleinman symmetry \cite{boyd_chap1}, is often assumed to remain valid in dynamical conditions, provided that all optical frequencies lie below the electronic band gap.
The underlying assumption is that, in the transparent regime, the material response is effectively instantaneous and non-adiabatic effects are negligible. However, the static approximation is fundamentally inadequate for describing the response under circularly polarized driving, and leads to qualitatively wrong results.
To show that this is the case, we take advantage of the permutation symmetry of the susceptibility and recast Eq.\ \eqref{dGgen} as $\Delta\Gamma(t_{pp})\propto\int d\omega_1d\omega_2d\omega_3 e^{-i(\omega_2+\omega_3)t_{pp}}\text{E}_\text{pr}(\omega_1)\Delta\gamma(\omega_1,\omega_2,\omega_3)$, where
\begin{align}\label{dg}
    \Delta&\gamma(\omega_1,\omega_2,\omega_3)= \chi^\mathcal{A}_{xy}(\omega;\omega_1,\omega_2,\omega_3)\mathcal{A}\nonumber\\
    &+\chi^\mathcal{S}_{xy}(\omega;\omega_1,\omega_2,\omega_3)\big(\text{sin}(2\Delta\theta)\mathcal{S}_\parallel
    +\text{cos}(2\Delta\theta)\mathcal{S}_\perp\big).
\end{align}
Here, $\mathcal{S}_{\parallel,\perp}$ and $\mathcal{A}$ denote combinations of the pump-field components that are, respectively, symmetric and antisymmetric under exchange $\omega_2\leftrightarrow \omega_3$,
\begin{align}
    \mathcal{S}_\parallel&=\text{E}^y_\text{pu}(\omega_2)\text{E}^y_\text{pu}(\omega_3)-\text{E}^x_\text{pu}(\omega_2)\text{E}^x_\text{pu}(\omega_3),\nonumber\\
    \mathcal{S}_\perp&=\text{E}^x_\text{pu}(\omega_2)\text{E}^y_\text{pu}(\omega_3)+\text{E}^y_\text{pu}(\omega_2)\text{E}^x_\text{pu}(\omega_3),\nonumber\\    
    \mathcal{A}&=\text{E}^x_\text{pu}(\omega_2)\text{E}^y_\text{pu}(\omega_3)-\text{E}^y_\text{pu}(\omega_2)\text{E}^x_\text{pu}(\omega_3).
\end{align}
The coefficients $\chi_{xy}^\mathcal{S}$ and $\chi_{xy}^\mathcal{A}$ represent, analogously, symmetric and antisymmetric combinations of the susceptibility components,
\begin{align}\label{chiS-AS}
    \chi_{xy}^\mathcal{S,A}(\omega;\omega_1,\omega_2,\omega_3)&=\chi_{xy;xy}(\omega;\omega_1,\omega_2,\omega_3)\nonumber\\
    &\pm\chi_{xy;yx}(\omega;\omega_1,\omega_2,\omega_3).
\end{align}
It is evident that the antisymmetric contribution to the differential signal would vanish under Kleinman symmetry, while the symmetric terms remain finite.

For a circularly polarized pump, the symmetric and antisymmetric components of the susceptibility give rise to oscillations at distinct frequencies in $\Delta\G(t_{pp})$. This separation becomes evident for a monochromatic pump at frequency $\Omega$, where the field components $\text{E}_\text{pu}^{x,y}(\omega_s)=\mathcal{E}_{x,y}\,\delta(\omega_s-\Omega)+\mathcal{E}_{x,y}^*\,\delta(\omega_s+\Omega)$ can be expressed in the circular basis through $\mathcal{E}_{L,R}=(\mathcal{E}_x\pm i\mathcal{E}_y)/\sqrt{2}$. By using this decomposition in Eq.\ \eqref{dg} we see that the symmetric field combinations generate oscillations at $\pm2\Omega$, whereas the antisymmetric component generates a static response. By Fourier transforming the differential signal with respect to the pump-probe delay, $\Delta\Gamma(t_{pp})\to\Delta\bar{\Gamma}(\omega_{pp})$, and taking a monochromatic probe at frequency $\omega$, the second-harmonic response reads
\begin{align}\label{DeltaGS}
    \Delta\bar\Gamma(2\Omega)\sim& \chi_{xy}^\mathcal{S}(\omega+2\Omega;\omega,\Omega,\Omega)\Big[\sin(2\Delta\theta)\big(\mathcal{E}_L^2+\mathcal{E}_R^2\big)\nonumber\\
    &+i\cos(2\Delta\theta)\big(\mathcal{E}_L^2-\mathcal{E}_R^2\big)\Big].
\end{align}
An analogous expression is found for the negative-frequency component, with complex conjugate projectors. These two contributions describe the optical Kerr effect \cite{basini_prb24}.
On the other hand, the static response takes the form
\begin{align}\label{DeltaGA}
    \Delta\bar\Gamma(0)&\sim 
    \chi_{xy}^\mathcal{A}(\omega;\omega,\Omega,-\Omega)(\boldsymbol{\mathcal{E}\times\boldsymbol{\mathcal{E}}^*})_z=\nonumber\\
    &= i\chi_{xy}^\mathcal{A}(\omega;\omega,\Omega,-\Omega)\big(|\mathcal{E}_L|^2-|\mathcal{E}_R|^2\big),
\end{align}
where $\boldsymbol{\mathcal{E}}=(\mathcal{E}_x,\mathcal{E}_y)$, and we used the identity $\mathcal{E}_x\mathcal{E}_y^*-\mathcal{E}_x^*\mathcal{E}_y=i(|\mathcal{E}_L|^2-|\mathcal{E}_R|^2)$. 
Eq.\ \eqref{DeltaGA} provides the first key result of our work, showing that the breakdown of Kleinman symmetry at finite frequency gives rise to an antisymmetric third-order response that generates a quasi-static helicity-dependent signal.
We note that this contribution is finite even for a purely real $\chi_{xy}^\mathcal{A}$, as typically realized when both pump and probe frequencies lie below the electronic band gap.
Because $\boldsymbol{\mathcal{E}}\to \boldsymbol{\mathcal{E}}^*$ under time reversal, the field combination $(\boldsymbol{\mathcal{E}\times\boldsymbol{\mathcal{E}}^*})_z$ explicitly breaks time-reversal symmetry. Consequently, the contribution in Eq.\ \eqref{DeltaGA} can be interpreted as a light-induced Faraday effect, in which the finite-frequency circularly polarized pump plays the same role as the external magnetic field $\text{B}^z$ in the conventional magneto-optical Faraday effect described by Eq.\ \eqref{eqF}. The key difference is that the light-induced mechanism involves an effective fictitious magnetic field \cite{merlin_prb24,merlin_pnas25}, that rotates the probe polarization without generating a macroscopic magnetization. 

To isolate helicity-dependent contributions, the standard experimental procedure \cite{basini_nature24} is to consider the dichroic signal $\Delta\bar\Gamma_{R-L}(\omega_{pp})=\frac{1}{2}\big[\Delta\bar\Gamma_{R}(\omega_{pp})-\Delta\bar\Gamma_{L}(\omega_{pp})\big]$, defined as the difference between the pump-probe response to right ($\Delta\Gamma_R$) and left ($\Delta\Gamma_L$) polarized light.
In addition to the time-reversal breaking static response, the dichroic signal also contains a $2\Omega$ component arising from the helicity-odd term proportional to $\mathcal{E}_L^2-\mathcal{E}_R^2$ in Eq.\ \eqref{DeltaGS}, and a $-2\Omega$ component from its complex conjugate.

To correctly interpret the experimental spectra as the ones of Ref.\ \cite{basini_nature24}, one must also account for a finite frequency mismatch $\Delta\Omega$ between the pump components $\text E_\text{pu}^x(\omega_s)$ and $\text E_\text{pu}^y(\omega_s)$. As discussed in the Supplemental Material \cite{suppl}, such a mismatch shifts the light-induced Faraday response to finite frequency $\Delta\Omega$, while simultaneously activating an additional contribution from the symmetric channel at the same frequency. The latter coincides with the quasi-static signal discussed in Refs.\ \cite{basini_nature24,basini_prb24} under the assumption of Kleinman symmetry. As a consequence, the experimentally observed quasi-static response generally contains contributions from both channels.
We note that recent theoretical works \cite{merlin_prb24,merlin_pnas25,geilhufe_prl24,wehling_prl25} did not explicitly include the contribution of the symmetric channel to the quasi-static peak at $\Delta\Omega$, which may have led to an overestimate of the apparently anomalous signal. Within the present framework, the full spectrum instead emerges naturally from the interplay between Kerr and light-induced Faraday responses.

On the experimental side, the symmetric and antisymmetric contributions can be disentangled through angle-resolved measurements. The antisymmetric response is independent of the probe polarization angle $\theta$ \cite{suppl}, whereas the symmetric contribution exhibits the characteristic fourfold angular dependence discussed in Ref.\ \cite{basini_prb24}. 
Moreover, since the symmetric dichroic signal is proportional to $\cos(2\Delta\theta)$, see Eq.\ \eqref{DeltaGS}, averaging measurements taken at different $\Delta\theta$ can suppress the symmetric contribution and isolate the light-induced Faraday response.

Finally, we note that for a linearly polarized pump both the static and the oscillatory $\pm 2\Omega$ components originate entirely from the symmetric terms of Eq.\ \eqref{dg}, which directly correspond to the classical nonlinear optical response, while the antisymmetric contribution vanishes as $(\boldsymbol{\mathcal E}\times\boldsymbol{\mathcal E}^*)_z=0$ \cite{suppl}. In this case, a theoretical description based on Kleinman symmetry is sufficient to reproduce all qualitative features of the experimental response, as demonstrated in Ref.\ \cite{basini_prb24}.

\section{Microscopic estimate of the dynamical electronic response}
The emergence of the static light-induced Faraday effect under circular driving, encoded in Eq.\ \eqref{DeltaGA}, is a general consequence of the dynamical breakdown of Kleinman symmetry. Evaluating the relevance of this mechanism for the experimental observations, however, requires a quantitative estimate of the relative magnitude of the antisymmetric and symmetric susceptibilities, $\chi^\mathcal{A}_{xy}/\chi^\mathcal{S}_{xy}$. Since the latter uniquely determines the Kerr response at $2\Omega$ as given by Eq.\ \eqref{DeltaGS}, this ratio provides a natural measure of the strength of the light-induced Faraday contribution. Because the susceptibilities depend sensitively on the underlying electronic transitions, a realistic description would require a detailed microscopic modeling of the full band structure. 
Here, rather than addressing a specific material, we employ a minimal toy model designed to identify the microscopic ingredients responsible for a sizable antisymmetric susceptibility in both resonant and off-resonant regimes.

As will become clear in the following, distinct orbital characters of valence and conduction bands give rise to additional microscopic channels contributing to the antisymmetric susceptibility, thereby enhancing the light-induced Faraday effect.
As a representative multiorbital toy model we then consider a square lattice with two atoms per unit cell, A and B, hosting $s$- and $p_{x,y}$-type orbitals, respectively. We include nearest-neighbor hopping with amplitude $t$, and assign on-site energies $\mp\Delta$ to the two sublattices. The real-space structure of the model is shown in Fig.\ \ref{Fig2}(a), while the corresponding band structure is reported in Fig.\ \ref{Fig2}(b). 
The lower band $E_1(\textbf{k})$, shown in red, is predominantly $s$-like, whereas the upper bands $E_2(\textbf{k})$ and $E_3(\textbf{k})$, in blue, are mainly of $p$-like character and degenerate at the $\Gamma$-point. Here, a direct $2\Delta$ gap is present. In the following we set the chemical potential such that $E_1(\textbf{k})$ is the valence band. $E_3(\textbf{k})$ is flat throughout the Brillouin zone, as a consequence of restricting the hopping to nearest neighbors. 

The light–matter interaction is introduced via the Peierls substitution $\textbf k\to\textbf k+e\textbf A/c$ in the tight-binding Hamiltonian $\mathcal{H}_\textbf{k}$, where $-e$ is the electron charge, $c$ the speed of light and $\textbf{A}$ the vector potential describing the optical perturbation. As detailed in the Supplemental Material \cite{suppl}, we derive the effective quantum action for the electronic system in the presence of the electromagnetic field by expanding the Hamiltonian in powers of $\textbf A$ and integrating out the electronic degrees of freedom \cite{benfatto_prb04,udina_prb19}. This procedure yields the microscopic counterpart of the free-energy in Eq.\ \eqref{eqK}, where the third-order nonlinear response is mediated by electronic transitions. The corresponding contributions are shown in Fig.\ \ref{Fig3}(a), each diagram defining a third-order nonlinear kernel. For clarity, we present the expressions for a monochromatic pump, while the generalization to fields with finite bandwidth is straightforward.

We first consider a diamagnetic-like contribution, in which light-matter coupling occurs through two-photon excitation processes mediated by the density-like vertex $\rho_{i}=\partial^2\mathcal{H}_\textbf{k}/\partial\text{k}^2_i$. Denoting with $[\rho_i]_{ab}$ the matrix element of the electronic transitions between bands $a$ and $b$, the resulting antisymmetric kernel scales as \cite{suppl}
\begin{align}\label{diaker}
   \text{K}^\mathcal{A}_\text{dia}&(\omega;\omega,\Omega,-\Omega)\sim \sum_{ab} \big[[\rho_x]_{ab}[\rho_y]_{ba}+[\rho_y]_{ab}[\rho_x]_{ba}\big]\nonumber\\
   \times&\big[d_{ab}(\omega-\Omega)-d_{ab}(\omega+\Omega)\big].
\end{align}
Here, the function $d_{ab}$ collects the frequency-dependent poles associated with electronic transitions, and is independent of the Cartesian indices. The diamagnetic kernel already illustrates why the antisymmetric response becomes finite at finite frequency: the two terms $\text{K}^\text{dia}_{xy;xy}(\omega;\omega,\Omega,-\Omega)$ and $\text{K}^\text{dia}_{xy;yx}(\omega;\omega,\Omega,-\Omega)$ that compose $\text{K}^\mathcal{A}_\text{dia}$ explore different intermediate states after the first transition driven by the $y$-polarized pump and probe photons, as sketched in Fig.\ \ref{Fig3}(b). 
In the former, the $y$-polarized pump carries frequency $-\Omega$, such that the intermediate state evolves with $\omega-\Omega$. In the latter, it instead carries $\Omega$, yielding an intermediate state at $\omega+\Omega$. This dynamical dephasing between the two processes produces a finite difference between the corresponding kernels and therefore a nonvanishing antisymmetric response.
\begin{figure}[t]
		\includegraphics[width=0.47\textwidth]{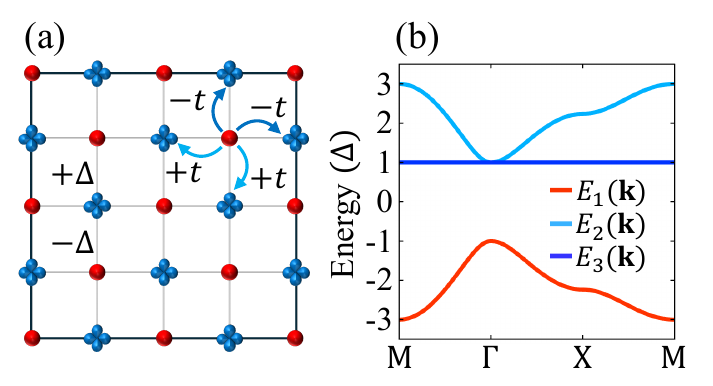}
		\caption{(a) Real-space structure of the square $sp$-model. Each unit cell contains two sites: site A hosting an $s$-type orbital (red) and site B hosting $p_{x,y}$-type orbitals (blue). The two sublattices have on-site energies $-\Delta$ and $+\Delta$ respectively, and are connected by nearest-neighbor hopping with amplitude $t$. (b) Band structure along high-symmetry directions of the Brillouin zone for $t/\Delta=1$. The red band has predominant $s$-like orbital character, the blue bands have $p$-like character.\label{Fig2}}
\end{figure}

The paramagnetic-like contribution, on the other hand, corresponds to processes involving one-photon excitations only, with four velocity vertices $\text{v}_i=\partial \mathcal{H}_\textbf{k}/\partial\text{k}_i$. Introducing the velocity vector $\textbf{v}=(\text{v}_x,\text{v}_y)$, the corresponding antisymmetric kernel can be expressed in a compact form as \cite{suppl}
\begin{align}\label{paraker}
    \text{K}^\mathcal{A}_\text{para}&(\omega;\omega,\Omega,-\Omega)\sim \sum_{abcd}\big[([\textbf{v}]_{ab}\!\times\![\textbf{v}]_{bc})\!\cdot\!([\textbf{v}]_{cd}\!\times\![\textbf{v}]_{da})\big]&\nonumber\\
    \times& \big[p_{abcd}(-\Omega,\Omega,\omega)-p_{abcd}(\Omega,-\Omega,\omega)\big]+\text{perm.},
\end{align}
where the function $p_{abcd}$ plays the same role as $d_{ab}$ for paramagnetic-like transitions, and ``perm.'' accounts for all permutations of the pump-probe interaction sequence. 
\begin{figure}[t]
		\includegraphics[width=0.49\textwidth]{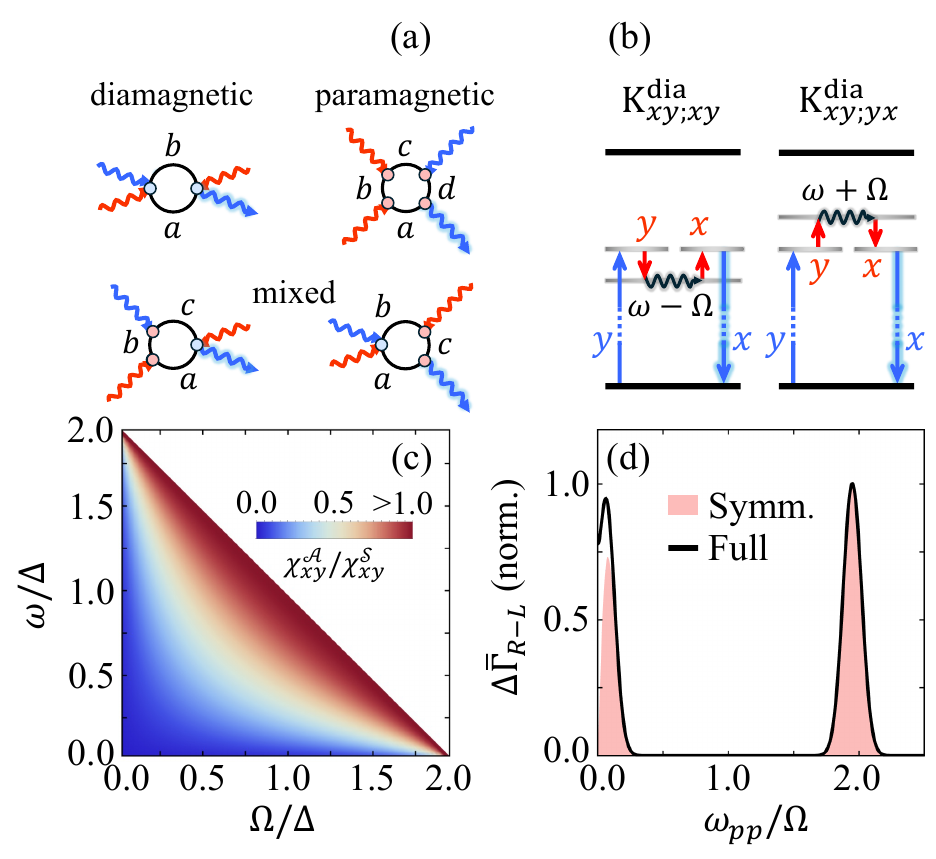}
		\caption{(a) Diagrammatic representation of the electronic processes contributing to the third-order susceptibility. Red (blue) wavy lines denote pump (probe) fields, while black lines represent electronic propagators. Red and blue dots denote velocity and density-like vertices, respectively. The probe photon on which the detection is performed is highlighted by a blue shadow. All possible permutations of photon vertices must also be considered. (b) Schematic representation of the two diamagnetic-like processes $\text{K}^\text{dia}_{xy;xy}$ and $\text{K}^\text{dia}_{xy;yx}$. Black (gray) lines denote real (virtual) electronic states. Red (blue) arrows represent pump (probe) fields. The intermediate state (black wavy line) propagates at frequencies $\omega\mp\Omega$, leading to distinct poles $d_{ab}(\omega\mp\Omega)$ and hence to a finite antisymmetric contribution $\text{K}^{\mathcal{A}}_\text{dia}$, as in Eq.\ \eqref{diaker}.
        (c) Map of the ratio $\chi^\mathcal{A}_{xy}/\chi_{xy}^\mathcal{S}$ in the $sp$-model, for fixed $t/\Delta=0.1$. The color scale is saturated for $\chi^\mathcal{A}_{xy}/\chi_{xy}^\mathcal{S}>1$. (d) Dichroic response $\Delta\bar\Gamma_{R-L}$ for narrowband Gaussian pump pulses ($\Omega\tau=20$) and frequency mismatch between $\text{E}_\text{pu}^{x}$ and $\text{E}_\text{pu}^{y}$ set to $\Delta\Omega/\Omega=0.05$. The response is normalized to the $2\Omega$ peak. The full response (black line) is obtained by fixing $\chi^\mathcal{A}_{xy}/\chi_{xy}^\mathcal{S}=0.5$, while the red shaded area shows the symmetric contribution only.      
        \label{Fig3}}
\end{figure}
Eq.\ \eqref{paraker} also highlights the importance of a multiorbital electronic structure for obtaining a finite antisymmetric paramagnetic response, as also pointed out for the antisymmetric Raman response in Ref.\ \cite{udina_prl26}. In effective two-band models derived from a single orbital channel, the velocity matrix elements are not independent and instead become proportional to each other by symmetry. As a consequence, the corresponding cross products vanish identically, suppressing the paramagnetic contribution.
Finally, the mixed paramagnetic-diamagnetic contribution corresponds to processes involving both one- and two-photon excitations, leading to the kernel $\text K^\mathcal{A}_\text{mix}(\omega;\omega,\Omega,-\Omega)$. Its explicit expression, lengthier but analogous to the previous terms, is reported in the Supplemental Material \cite{suppl}.

By expressing the electric field in terms of the time derivative of the gauge potential, the full antisymmetric susceptibility can be obtained from the corresponding nonlinear kernels as
$\chi^\mathcal{A}_{xy}=[\text{K}^\mathcal{A}_\text{dia}+\text{K}^\mathcal{A}_\text{para}+\text{K}^\mathcal{A}_\text{mix}]/(\omega^2\Omega^2)$, where the frequency dependence has been left implicit for compactness. The resulting susceptibility scales linearly with both pump and probe frequencies in the low-frequency limit, 
\begin{align}\label{eq_linearscale}
\chi^\mathcal{A}_{xy}(\omega;\omega,\Omega,-\Omega)\sim\omega\Omega \quad \text{for } \omega,\Omega\to0.
\end{align}
This demonstrates that the light-induced Faraday effect is intrinsically dynamical and requires finite-frequency pump and probe fields.
Analogous calculations yield the symmetric susceptibility $\chi_{xy}^\mathcal{S}$ which, as expected for a wide-bandgap insulator, remains finite in the static limit $\omega,\Omega\to0$. 

To quantitatively estimate the relative magnitude of the antisymmetric and symmetric susceptibilities within the $sp$-model, we compute the ratio $\chi_{xy}^\mathcal{A}(\omega;\omega,\Omega,-\Omega)/\chi_{xy}^\mathcal{S}(\omega;\omega,\Omega,-\Omega)$ as a function of the pump and probe frequencies $\Omega$ and $\omega$, and show the result in Fig.\ \ref{Fig3}(c). While both susceptibilities increase upon approaching the band gap, the antisymmetric contribution grows comparatively faster, leading to a strong enhancement of the ratio near resonance. Remarkably, a sizable antisymmetric susceptibility persists even far from the resonance condition.
By fixing the ratio between the two susceptibilities to $\chi_{xy}^\mathcal{A}/\chi_{xy}^\mathcal{S}= 0.5$, we simulate \cite{suppl} the dichroic response $\Delta\bar\Gamma_{R-L}(\omega_{pp})$ under circular driving using two narrowband Gaussian fields of duration $\tau$. The component $\text{E}_\text{pu}^y(\omega_s)$ is centered at frequency $\Omega$, while $\text{E}_\text{pu}^x(\omega_s)$ is shifted by a small frequency mismatch $\Delta\Omega$. The resulting spectrum is shown in Fig.\ \ref{Fig3}(d) as a black line. For comparison, we also report as a red shaded area the response obtained by retaining only the symmetric contribution, corresponding to the Kleinman approximation $\chi^\mathcal{A}_{xy}=0$. In this case, the small but finite quasi-static signal associated with the light-induced Faraday effect is absent.

Although the quantitative results presented here are obtained within the $sp$-model, the many-body diagrammatic analysis is general. 
Therefore, while the overall magnitude of the light-induced Faraday response depends on material-specific details, the physical conclusions remain broadly applicable. For pump-probe conditions comparable to those realized in Refs.\ \cite{basini_nature24,kamba_cm26,nelson_cm26}, namely $\omega/\Delta\simeq 1.5$ and $\Omega/\Delta\simeq 0.01$, our toy model with $t/\Delta=0.1$ yields $\chi_{xy}^\mathcal{A}/\chi_{xy}^\mathcal{S}\simeq 0.1$, of the same order of magnitude of the value $\chi_{xy}^\mathcal{A}/\chi_{xy}^\mathcal{S}\simeq 0.5$ used in Fig.\ \ref{Fig3}(d) to better highlight the difference between the static and dynamical calculations. As discussed in Sec.\ \ref{sec:IFE}, the value $\chi_{xy}^\mathcal{A}/\chi_{xy}^\mathcal{S}\simeq 0.1$ provides the correct order of magnitude for the experimental polarization rotation observed in SrTiO$_3$.

\section{Phonon contribution}\label{sec_phonon}
In cubic SrTiO$_3$, the polar TO$_1$ phonon softens from $3.2$ THz at $380$ K to $1.8$ THz at $150$ K \cite{vogt_prb95,yamada_jpsj69}. 
As shown in Ref.\ \cite{basini_prb24}, this softening naturally accounts for the pronounced temperature dependence of the Kerr response measured under linear driving: the ionic contribution is maximized at the temperature $T_\text{res}$ at which the phonon is resonant with the pump frequency, and it is progressively suppressed away from it. Such a strong temperature dependence would be difficult to explain within a purely electronic picture, since a temperature variation of $\Delta T=230$ K is expected to have only a negligible effect on the electronic response given the large direct band gap of SrTiO$_3$ ($2\Delta\simeq3.75$ eV). A similar behavior is observed in Ref.\ \cite{basini_nature24} for the anomalous quasi-static signal measured under circular driving, and analogous results are found in Ref.\ \cite{kamba_cm26} for the similar compound KTaO$_3$. Motivated by these observations, we now discuss the additional contribution of resonantly driven infrared-active phonons to the light-induced Faraday effect.

To describe the phonon-mediated response to circularly polarized light, we model the four-wave-mixing process as a phonon dressing of the antisymmetric electronic susceptibility, in which each pump photon excites an intermediate phonon state that subsequently recombines into off-resonant electronic excitations. 
This mechanism is represented diagrammatically in Fig.\ \ref{Fig4}(a), and details of the derivation are provided in the Supplemental Material \cite{suppl}. The dashed lines denote the phonon propagators
\begin{align}\label{phon_prop}
    D(\Omega)=\frac{2\omega_0(T)}{\Omega^2+2i\gamma(T)\Omega-\omega_0^2(T)},
\end{align}
where $\omega_0(T)$ and $\gamma(T)$ are the temperature-dependent phonon frequency and damping rate respectively.
The vertex $Z$ describes the linear light-phonon coupling, expressed in terms of paramagnetic and diamagnetic light-electron and electron-phonon couplings \cite{cappelluti_prb10,cappelluti_prb12,bistoni_2D19}, and proportional to the Born effective charge $Z_B$ of the mode as $Z=Z_B\sqrt{\omega_0(T)/M}$, with $M$ the ionic mass \cite{resta_rmp94}.
The gray bubble represents the antisymmetric electronic susceptibility $\Pi^{\mathcal{A}}_{xy}$, which includes diamagnetic, paramagnetic and mixed contributions analogous to those in Fig.\ \ref{Fig3}(a), with electron-phonon vertices replacing the light-electron couplings on the pump side.
The resulting antisymmetric optical susceptibility reads \cite{suppl}
\begin{align}\label{chi_phon}
    \chi^{\mathcal{A},\text{ph}}_{xy}(\omega;\omega,\Omega,-\Omega)&=Z^2D(\Omega)D(-\Omega)\nonumber\\
    &\times\Pi^{\mathcal{A}}_{xy}(\omega;\omega,\Omega,-\Omega).
\end{align}
An analogous expression is found for the symmetric response \cite{basini_prb24}. The temperature dependence of this susceptibility is primarily governed by the phonon propagators Eq.\ \eqref{phon_prop}, such that the phonon contribution to the light-induced Faraday effect is maximized at the temperature $T_\text{res}$ satisfying the resonance condition $\Omega=\omega_0(T_\text{res})$, where $D(\Omega)D(-\Omega)\sim 1/\gamma^2(T_\text{res})$, and is strongly suppressed away from it. 

To illustrate the effect quantitatively, we compute the full response including both phonon-mediated and electronic processes. As discussed in the Supplemental Material \cite{suppl}, the relative weight between the two contributions can be expressed as $\chi^{\mathcal{A},\text{ph}}_{xy}/\chi^{\mathcal{A}}_{xy}\sim \alpha(\beta Z_B/M)^2D(\Omega)D(-\Omega)$, where $\beta$ is a dimensionless parametrization of the electron-phonon coupling strength \cite{resta_rmp94,bistoni_2D19}, and $\alpha$ is a numerical constant of order one. The phonon softening is modeled as $\omega_0^2(T)=A\big(\omega_q\coth(\omega_q/T)-T^*\big)$, following the behavior of the TO$_1$ mode in SrTiO$_3$ and KTaO$_3$, with $A$, $\omega_q$, and $T^*$ treated as phenomenological parameters. We also assume a temperature-independent ratio $\gamma(T)/\omega_0(T)$ \cite{vogt_prb95}. The temperature dependence of $\omega_0(T)$ is shown in Fig.\ \ref{Fig4}(b), together with the Gaussian narrowband pump components $\text{E}_\text{pu}^x(\omega_s)$ and $\text{E}_\text{pu}^y(\omega_s)$, shifted by a small frequency mismatch $\Delta\Omega$.
In this parametrization, the phonon frequency crosses the pump spectrum within a finite temperature window, allowing a direct comparison between resonant and off-resonant conditions. In Fig.\ \ref{Fig4}(c) we show the pump-probe response to circular light in both regimes. As expected, the signal is enhanced at resonance. %
\begin{figure}[t]
		\includegraphics[width=0.49\textwidth]{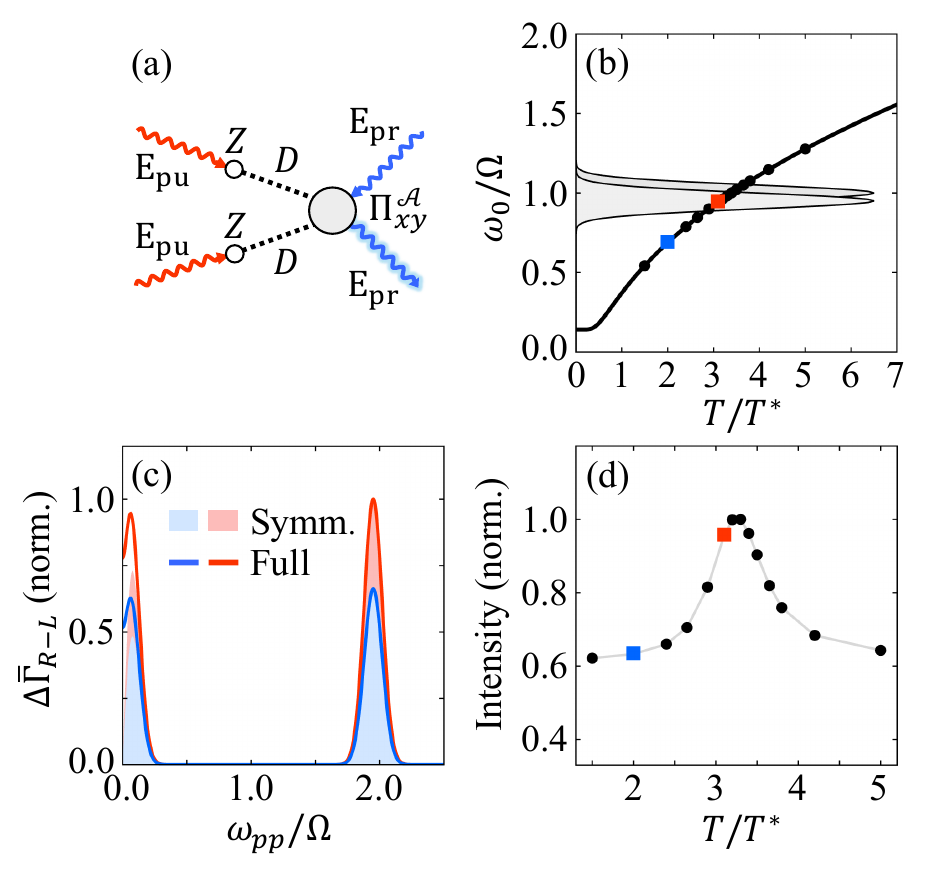}
		\caption{(a) Diagrammatic representation of the phonon-mediated contribution to the pump–probe response. Red (blue) wavy lines denote pump (probe) fields, and dashed black lines the phonon propagators. The linear light-phonon vertices $Z$ are shown as white dots, while the gray bubble represents the electronic susceptibility $\Pi_{xy}^{\mathcal A}$, including diamagnetic, paramagnetic, and mixed contributions. (b) Phenomenological temperature dependence of the phonon frequency $\omega_0(T)$ (black line), with $A/\Omega=0.6$, $\omega_q/\Omega=1.1$, and $T^*/\Omega=1$. The phonon damping rate is $\gamma(T)/\omega_0(T)=0.07$. Gray areas indicate the Gaussian pump field components $\text{E}_\text{pu}^x$ and $\text{E}_\text{pu}^y$, with a relative frequency mismatch $\Delta\Omega/\Omega=0.05$. Black dots and colored squares mark the temperatures analyzed in panels (c) and (d). (c) Dichroic signal including both electronic and phonon-mediated processes, evaluated at phonon resonance (red) and off resonance (blue), for the temperatures indicated in panel (b) with squares. Shaded areas show the symmetric contribution only, while solid lines include both the symmetric and the antisymmetric components. Calculations are performed for $\chi^\mathcal{A}_{xy}/\chi^\mathcal{S}_{xy}=\Pi^\mathcal{A}_{xy}/\Pi^\mathcal{S}_{xy}=0.5$, and the relative weight between phonon-mediated and purely electronic processes is fixed by $\sqrt{\alpha}\beta Z_B/M=0.06\,\Omega$. (d) Temperature dependence of the intensity of the light-induced Faraday response, for the selected temperatures of panel (b). 
        \label{Fig4}}
\end{figure}
Notably, however, the qualitative features of the response remain unchanged: in particular, the relative intensity of the quasi-static peak with respect to the $2\Omega$ component is unaffected, as the phonon contributes equally to the symmetric and antisymmetric channels. 
In Fig.\ \ref{Fig4}(d) we show the intensity of the light-induced Faraday response as a function of temperature, obtained by integrating the difference between the full response and its symmetric contribution over the pump-probe frequency $\omega_{pp}$. 
This signal is maximized in the resonant regime, while remaining finite away from resonance due to the electronic contribution. This behavior is consistent with the experimental observations of Ref.\ \cite{basini_nature24}, where a temperature window approximately corresponding to $2.5 \lesssim T/T^* \lesssim 3.8$ was investigated. The persistence of a finite and substantial response away from resonance is also consistent with the observations of Ref.\ \cite{nelson_cm26}, where the system is driven well outside the phonon resonance.

We note that resonance with other low-energy excitations can likewise enhance the third-order nonlinear response. For instance, the localized $4f$ states of Ce$^{3+}$ in CeF$_3$ \cite{davies_cm26} provide an additional resonant channel, demonstrating that the enhancement of the response needs not be tied exclusively to infrared-active phonons, as previously suggested in Ref.\ \cite{luo_science23}.

\section{Relation to the inverse Faraday effect}
\label{sec:IFE}
%
%
\subsection{Comparison between light-induced and IFE-induced Faraday effects}
In the light-induced Faraday effect, a circularly polarized pump drives a rotation of the collective electronic polarization. Since this dynamics does not correspond to a macroscopic circulating current, it does not generate a macroscopic magnetic field detectable outside the sample. This is fundamentally different from the IFE, in which circularly polarized light can induce a magnetization and, consequently, a real magnetic field that can in turn produce a Faraday rotation. So far, the latter interpretation has been predominant in the discussion of several experiments under circularly polarized light \cite{luo_science23,basini_nature24,davies_cm26,kamba_cm26,nelson_cm26}. Nevertheless, the magnetic-field contribution associated with the IFE is expected to be negligible in nonmagnetic systems, as argued in Ref.\ \cite{merlin_prb24}. 
To understand why, we now expand the discussion around Eq.\ \eqref{eqF} and clarify the assumptions underlying both its derivation and its application to optical experiments in nonmagnetic systems, such as those of Refs.\ \cite{basini_nature24,kamba_cm26,nelson_cm26}. For a quantitative comparison with the experimental results, we formulate the following analysis in MKS units.

On general grounds, the physical electric $\textbf{E}$ and magnetic $\textbf{B}$ fields inside the system couple to the polarization $\textbf{P}$ and magnetization $\textbf{M}$ operators through a modification of the equilibrium Hamiltonian $H_0$, $H(\textbf{E},\textbf{B})=H_0-\textbf{P}\cdot \textbf{E}-\textbf{M}\cdot \textbf{B}$.
The induced macroscopic polarization and magnetization are then defined as
\begin{align}
    \langle \textbf{P}\rangle &=\frac{1}{\mathcal{Z}}\text{Tr} [\textbf{P} e^{-\frac{1}{T}H(\textbf{E},\textbf{B})}],\nonumber\\ \langle \textbf{M}\rangle &=\frac{1}{\mathcal{Z}}\textrm{Tr} [\textbf{M} e^{-\frac{1}{T} H(\textbf{E},\textbf{B})}],
\end{align}
where $\mathcal{Z}=\textrm{Tr}[e^{-\frac{1}{T}H(\textbf{E},\textbf{B})}]$ is the partition function. These definitions can equivalently be expressed in terms of derivatives of the free energy $F(\textbf{E},\textbf{B})$ obtained by integrating out the matter degrees of freedom,
\begin{align}
\label{defpmder}
    \langle \textbf{P}\rangle =-\frac{\partial F(\textbf{E},\textbf{B})}{\partial \textbf{E}}, \quad \langle \textbf{M}\rangle =-\frac{\partial F(\textbf{E},\textbf{B})}{\partial \textbf{B}}.
\end{align}
The free energy can be expanded to arbitrary order in $\textbf{E}$ and $\textbf{B}$, with the expansion coefficients given by the corresponding susceptibilities expressed as correlation functions of the matter operators $\textbf{P}$ and $\textbf{M}$. In the experiments of Refs.\ \cite{luo_science23, basini_nature24,davies_cm26,kamba_cm26}, the observable is the induced nonlinear dipole generated inside the material, as discussed in Ref.\ \cite{basini_prb24} and shown in Eq.\ \eqref{dGgen}. We can therefore retain terms up to fourth order in the electric field, together with the terms involving the magnetic field that are responsible for the Faraday response. For monochromatic fields, and keeping only contributions relevant for the quasi-static signal, 
\begin{align}
\label{eqftot}
    F(\textbf{E},\textbf{B})\simeq -\frac{1}{2}\varepsilon_0\chi^e (|\mathcal{E}_x|^2+ |\mathcal{E}_y|^2)-\frac{1}{2\mu_0}\chi^m |\text{B}^z|^2&\nonumber\\
    -i\varepsilon_0\chi^{(\text{F})} \text{B}^z (\boldsymbol{\mathcal{E}}\times \boldsymbol{\mathcal{E}^*})_z+\varepsilon_0\chi^\mathcal{A}_{xy} |(\boldsymbol{\mathcal{E}}\times \boldsymbol{\mathcal{E}^*})_z|^2&.
\end{align}
Here, $\boldsymbol{\mathcal{E}}$ denotes the polarization of the total optical field, including both the pump and probe components, and we focus on the $z$ component of the magnetic field. The coefficients $\chi^e$ and $\chi^m$ represent the linear electric susceptibility and the magnetic polarizability, respectively, while $\varepsilon_0$ and $\mu_0$ are vacuum permittivity and permeability, respectively.

In the framework described above, the measured optical signal generated in response to a probe field along $y$ is proportional to $\text{P}^x_\text{NL}$, which according to Eq.\ \eqref{defpmder} is obtained from the functional derivative of the free energy with respect to $\mathcal{E}_\text{pr}^x$. Analogously, the magnetization generated along the $z$ direction is obtained from the functional derivative with respect to $\text{B}^z$. Using Eqs.\ \eqref{defpmder}, we thus obtain
\begin{align}\label{eqPtot}
    \langle \text{P}^x \rangle &=-i\varepsilon_0\chi^{(\text{F})} \text{B}^z \mathcal{E}_\text{pr}^y+i\varepsilon_0\chi^\mathcal{A}_{xy} \mathcal{E}_\text{pr}^y(|{\mathcal{E}_L}|^2-|{\mathcal{E}_R}|^2),\\
    \label{eqMtot}
    \langle \text{M}^z \rangle &=\frac{1}{\mu_0}\chi^m \text{B}^z-\varepsilon_0\chi^{(\text{F})} (|{\mathcal{E}_L}|^2-|{\mathcal{E}_R}|^2).
\end{align}
These expressions are written in terms of the physical fields inside the sample. Since no external magnetic field is applied, we can set $\text{B}^z\simeq0$ at leading order,
\begin{align}\label{eqPlead}
    \langle \text{P}^x \rangle &\simeq +i\varepsilon_0\chi^\mathcal{A}_{xy} \mathcal{E}_\text{pr}^y(|{\mathcal{E}_L}|^2-|{\mathcal{E}_R}|^2),\\
\label{eqMlead}
    \langle \text{M}^z \rangle &\simeq -\varepsilon_0\chi^{(\text{F})} (|{\mathcal{E}_L}|^2-|{\mathcal{E}_R}|^2).
\end{align}
Therefore, under a circularly polarized electric field the antisymmetric nonlinear susceptibility $\chi^\mathcal{A}_{xy}$ generates a probe-polarization rotation through the all-optical light-induced Faraday effect, while the magneto-optical susceptibility $\chi^{(\text{F})}$ gives rise to a magnetization through the inverse Faraday effect. 
The latter corresponds to a true magnetic response, and can therefore generate a real magnetic field, whose value inside and outside the material depends in general on the sample geometry.
For example, in a cylinder the internal field $\text{B}^{z}_\text{int}$ can be well approximated with the induced magnetization,
\begin{equation}
\label{bint}
    \text{B}^{z}_\text{int}\simeq \mu_0 \langle \text{M}^z \rangle\simeq-\mu_0\varepsilon_0\chi^{(\text{F})} (|{\mathcal{E}_L}|^2-|{\mathcal{E}_R}|^2). 
\end{equation}
This induced magnetic field can be measured by magnetic probes such as, e.g., variations of the magnetic flux through a coil wrapped around a cylindrically-shaped sample \cite{pershan_prl65}. 
In addition, one can also evaluate the feedback of the induced internal field on the polarization rotation, as described by the first term of Eq.\ \eqref{eqPtot}. Within this picture one obtains a correction to the leading-order estimate Eq.\ \eqref{eqPlead}. Using Eq.\ \eqref{bint} in cylindrical geometry, one obtains
\begin{align}
\label{eqpindtot}
    \langle \text{P}^x \rangle &\simeq -i\mu_0\varepsilon_0
    \chi^{(\text{F})}\langle\text{M}^z \rangle \mathcal{E}_\text{pr}^y+
i\varepsilon_0\chi^{\mathcal{A}}_{xy}\,\mathcal{E}_\text{pr}^y(|{\mathcal{E}_L}|^2-|{\mathcal{E}_R}|^2)\nonumber\\
&\simeq i\varepsilon_0\bigg[\frac{1}{c^2}\big[\chi^{(\text{F})}\big]^2+\chi^{\mathcal{A}}_{xy}\bigg]\,\mathcal{E}_\text{pr}^y(|{\mathcal{E}_L}|^2-|{\mathcal{E}_R}|^2).
\end{align}
%
The crucial point is that nonlinear optical measurements are not directly sensitive to the induced magnetic field, but rather measure the resulting rotation of the probe electric field through the induced polarization $\langle\text{P}^x\rangle$. Consequently, the two contributions to $\langle\text{P}^x\rangle$ in Eq.\ \eqref{eqpindtot}, the direct Faraday response induced by the IFE and the antisymmetric nonlinear optical response, must be considered on equal footing. However $[\chi^{(\text{F})}]^2/c^2$ is largely subleading in nonmagnetic materials, so that the all-optical light-induced Faraday effect is expected to be the main mechanism behind the quasi-static rotation.

An order-of-magnitude estimate of the ratio between the two contributions can be obtained within the many-body framework introduced above. In this approach, the Faraday susceptibility is obtained from the small-momentum expansion of the linear electric susceptibility $\chi^e(\textbf{q})$, evaluated in the presence of the non-uniform vector potential required to describe a magnetic field \cite{fukuyama_rpf69}.
The ratio between $[\chi^{(\text{F})}]^2/c^2$ and $\chi^\mathcal{S}_{xy}$ in the static limit can then be expressed in terms of fundamental constants, the lattice spacing $a$, and the band gap $\Delta$ as \cite{suppl}
\begin{align}\label{eq_ratio}
    \frac{[\chi^{(\text{F})}]^2/c^2}{\chi^\mathcal{S}_{xy}}\sim\frac{4\pi E_\text{mag}}{\Delta},
\end{align}
where $E_\text{mag}=\frac{\mu_0}{4\pi}\frac{\mu_B^2}{a^3}$ is the characteristic magnetic energy associated with the dipole-dipole interaction between two Bohr magnetons $\mu_B=\hbar e/2m_e$ separated by a distance $a$, and $m_e$ is the electron mass. For a typical lattice spacing $a=2.4\,\angstrom$, this gives $E_\text{mag}\simeq3.4\,\mu\text{eV}$. Since the band gap of insulating materials is typically of order eV, and $\chi^\mathcal{A}_{xy}/\chi^\mathcal{S}_{xy}\simeq 0.1$ provides a reasonable estimate for the antisymmetric response in experimental conditions like the ones of Refs.\ \cite{basini_nature24,nelson_cm26}, we obtain 
\begin{align}
\label{estimate}
    \frac{[\chi^{(\text{F})}]^2/c^2}{\chi^\mathcal{A}_{xy}}\sim 10^{-4}.    
\end{align}
Material-specific parameters, such as the electron effective mass and lattice structure, can affect this estimate, but not its overall order of magnitude. Even though Eq.\ \eqref{eq_ratio} expresses the ratio between the two response functions in the static limit, both increase with similar frequency scaling as either the pump or probe frequency approaches the band gap, so that their ratio remains comparable.
Thus, in nonmagnetic systems, the conventional Faraday contribution mediated by the IFE is expected to be strongly subdominant to the light-induced Faraday contribution. 

Our diagrammatic calculations account for the magneto-optical response of valence electrons, while neglecting contributions from localized magnetic moments. Accordingly, the magnetic energy scale entering Eq.\ \eqref{eq_ratio} is set by dipole-dipole interactions, which are generally subleading compared with the exchange interactions responsible for magnetic order. In ordered systems, the Faraday susceptibility is instead controlled by the magnetic order and spin-orbit coupling \cite{cottam-lockwood}, which enables the conversion of electric-field-driven orbital excitations into magnetic excitations.

It is also instructive to recast the magnetic energy as $E_\text{mag}=\big(\frac{\bar\lambda_C}{a}\big)^2E_\text{el}$, where $\bar\lambda_C$ is the reduced Compton wavelength of the electron and $E_\text{el}= e^2/(4\pi\varepsilon_0 a)$ is the characteristic electrostatic energy at the lattice scale, that for $a=2.4\,\angstrom$ gives 
$E_\text{el}\sim 5.8 \text{ eV}$. Eq.\ \eqref{eq_ratio} can then be written as
\begin{align}
    \frac{[\chi^{(\text{F})}]^2/c^2}{\chi^\mathcal{S}_{xy}}\sim\left(\frac{\bar\lambda_C}{a}\right)^2\frac{4\pi E_\text{el}}{\Delta}.
\end{align}
Since $E_\text{el}\sim\Delta$, this relation shows that the magneto-optical response is strongly suppressed relative to the optical one by the small ratio of the Compton wavelength to the lattice spacing. In the particular case $a=a_0$, where $a_0$ is the Bohr radius, our result can be recast as $\frac{1}{c^2}[\chi^{(\text{F})}]^2/{\chi^\mathcal{S}_{xy}}\sim 4\pi \alpha^2\,E_\text{el}/\Delta$, where $\alpha$ is the fine structure constant. This limiting case is consistent with the result obtained in Ref.\ \cite{merlin_prb24}.

For a rough quantitative estimate from experimental values, we consider SrTiO$_3$ at room temperature and use the parameters of Ref.\ \cite{basini_nature24}. The static symmetric third-order susceptibility can be estimated by Kerr measurements \cite{yin_SSC05}, yielding $\chi_{xy}^{\mathcal{S}}=10^{-10}\text{ cm}^2/\text{kV}^2$. Assuming a plausible ratio of $\chi^\mathcal{A}_{xy}/\chi_{xy}^\mathcal{S}=0.1$ for pump and probe frequencies of Ref.\ \cite{basini_nature24}, we obtain $\chi_{xy}^{\mathcal{A}}=10^{-11}\text{ cm}^2/\text{kV}^2$ for the antisymmetric susceptibility. On the other hand, the Faraday susceptibility can be estimated from the Verdet constant $\text{V}=250\text{ rad}/\text{m}\cdot\text{T}$ at probe wavelength $\lambda_\text{pr}=400\text{ nm}$ and corresponding refractive index $n_\text{pr}=2.6$. This gives $\chi^{(\text{F})}=n_\text{pr}\lambda_\text{pr}\text{V}/\pi\simeq 8.3\cdot10^{-5}\text{ T}^{-1}$, and hence $[\chi^{(\text{F})}]^2/c^2\simeq10^{-15}\text{ cm}^2/\text{kV}^2$.
Thus, for SrTiO$_3$, the estimated all-optical contribution is approximately four orders of magnitude larger than the IFE-mediated contribution to the Faraday rotation based on the static value of $\chi^{\mathcal{S}}_{xy}$, in agreement with the estimate Eq.\ \eqref{estimate}.

For simplicity, we have suppressed the frequency dependence in the discussion above. As for the optical susceptibility, the Faraday susceptibility also depends on both the pump and probe frequencies. The magnetization induced via the IFE in Eq.\ \eqref{eqMlead} is controlled by the Faraday susceptibility evaluated at the pump frequency, which, for a THz pump, is expected to be smaller than its value at visible frequencies. Accounting for this frequency dependence would therefore further reduce our estimate of the IFE-mediated contribution for Ref.\ \cite{basini_nature24}.

\subsection{Estimate of the probe polarization rotation}
For a direct comparison with experimentally relevant quantities, we now estimate the probe polarization rotation angle associated with the two effects. In the conventional direct Faraday effect, the rotation angle is proportional to the internal static magnetic field, which, for weakly diamagnetic or paramagnetic systems, is approximately equal to the externally applied field. For a sample of effective length $l_\text{eff}$ over which the magnetic field is present, the rotation angle of the probe polarization is \cite{pershan_prl65}
\begin{align}\label{eqtheta}
    \theta_\text{F}=\xi\, \chi^{(\text{F})}
    \text{B}^z_{\text{int}}=\text{V}l_\text{eff} \text{B}^z_{\text{int}},
 \end{align}
where $\xi= \pi l_\text{eff}/n_\text{pr}\lambda_\text{pr}$.
From Eq.\ \eqref{eqPtot}, the combination $\chi^{(\text F)}\text B^z$ entering the direct Faraday effect can be identified with $\chi^{\mathcal A}_{xy}(|\mathcal E_L|^2-|\mathcal E_R|^2)$ in the light-induced Faraday effect. For the dichroic signal, obtained by comparing left- and right-circularly polarized pumps with equal intensities, this reduces to $2\chi^{\mathcal A}_{xy}|\mathcal E_\text{pu}|^2$. Consequently, the quasi-static probe rotation induced by a monochromatic pump can be written in analogy with Eq.\ \eqref{eqtheta} as
\begin{align}\label{lif_rotation}
\theta_\text{LIF}=2\xi\,\chi^{\mathcal{A}}_{xy}|\mathcal{E}_\text{pu}|^2.
\end{align}
We again consider cubic SrTiO$_3$ with experimental parameters of Ref.\ \cite{basini_nature24}. In particular, $l_\text{eff}$ can be identified with the penetration depth of the THz pump field, which is $l_\text{eff}=2.49\text{ }\mu\text{m}$. With $\lambda_\text{pr}=400\text{ nm}$ and $n_\text{pr}=2.6$, one finds $\xi=7.5\text{ rad}$. Using the value $\chi^\mathcal{A}_{xy}=10^{-11}\text{ cm}^2/\text{kV}^2$ obtained above from the measured $\chi^\mathcal{S}_{xy}$, and a pump field strength $|\mathcal{E}_\text{pu}|=230\text{ kV}/\text{cm}$, Eq.\ \eqref{lif_rotation} yields a rotation angle $\theta_\text{LIF}\simeq8\text{ }\mu\text{rad}$, consistent with the order of magnitude of the experimentally measured values.

On the other hand, Eq.\ \eqref{eqpindtot} gives the IFE-induced Faraday rotation as
\begin{align}\label{ife_rotation}
    \theta_\text{IFE}=2\xi\,\frac{\big[\chi^{(\text{F})}\big]^2}{c^2}|\mathcal{E}_\text{pu}|^2.
\end{align}
Using the estimate $[\chi^{(\text{F})}]^2/c^2=10^{-15}\text{ cm}^2/\text{kV}^2$ obtained above, Eq.\ \eqref{ife_rotation} gives $\theta_\text{IFE}\simeq0.8\text{ nrad}$ for SrTiO$_3$. This angle is four orders of magnitude smaller than the light-induced Faraday rotation, highlighting the strong suppression of the IFE-mediated contribution in nonmagnetic materials.}

Analogous calculations can be performed using the experimental parameters of Ref.\ \cite{nelson_cm26}. In this case, the central pump frequency is $\Omega=0.5\text{ THz}$. Owing to the linearity of the antisymmetric susceptibility, see Eq.\ \eqref{eq_linearscale}, we rescale $\chi_{xy}^\mathcal{A}$ by a factor 6, corresponding to the ratio between the pump frequencies used in Refs.\ \cite{basini_nature24} and \cite{nelson_cm26}. In this case, $l_\text{eff}=100\,\mu\text{m}$ \cite{zhang_sto07}, while $\lambda_\text{pr}=800\text{ nm}$ and $n_\text{pr}=2.3$, giving $\xi=170\text{ rad}$. 
For a pump field strength $|\mathcal{E}_\text{pu}|=300\text{ kV}/\text{cm}$, Eq.\ \eqref{lif_rotation} yields $\theta_\text{LIF}=50\,\mu\text{rad}$, consistent with the measured polarization rotation. By contrast, Eq.\ \eqref{ife_rotation} gives $\theta_\text{IFE}=70\text{ nrad}$, three orders of magnitude smaller than the experimental angle.

\subsection{Relation to dynamical multiferroicity}
Similarly to the phonon-mediated light-induced Faraday effect discussed in Section \ref{sec_phonon}, the inverse Faraday response acquires an additional contribution when the pump field is resonant with an infrared-active phonon. The underlying mechanism, originally proposed in Ref.\ \cite{juraschek_prm17}, is dynamical multiferroicity, whereby a circularly driven infrared phonon carries an ionic magnetic moment
\begin{align}
\label{mph}
   \langle\text{M}^z_{\text{ph}}\rangle \sim \gamma_N (\textbf{Q}_\mathcal{E}\times\dot{\textbf{Q}}_\mathcal{E})_z,
\end{align}
where $\gamma_N=Z/2M$ is the ionic gyromagnetic ratio and $\textbf{Q}_\mathcal{E}\sim \boldsymbol{\mathcal{E}}$ is the field-driven phonon displacement. The resulting magnetization therefore scales with the pump intensity.
In geometries in which the internal field satisfies $\text{B}^{z}_\text{int}\simeq \mu_0\langle\text{M}^z_{\text{ph}}\rangle$, the magnetization produces an additional contribution to the nonlinear polarization,
\begin{align}
\label{Pph}
    \langle \text{P}^x_\text{ph} \rangle=-i \mu_0\varepsilon_0\chi^{(\text{F})}\langle\text{M}^z_{\text{ph}}\rangle\mathcal{E}_\text{pr}^y,
\end{align} 
and hence a phonon-mediated Faraday rotation.

For materials such as SrTiO$_3$ and KTaO$_3$ \cite{basini_nature24,kamba_cm26}, the large effective charge of the infrared phonons enhances $\textbf{Q}_\mathcal{E}$, but the resulting magnetic response remains strongly suppressed. First, the scale of the phonon magnetization Eq.\ \eqref{mph} is set by the nuclear magneton $\mu_N=\hbar\gamma_N$, which is three orders of magnitude smaller than the electronic magneton due to the large ionic-to-electronic mass ratio. Second, the conversion of the phononic magnetization into a probe rotation is controlled by the electronic Faraday susceptibility $\chi^{(\text{F})}$, which is small in diamagnetic materials as discussed above. 
As a result, the phononic magnetization estimated from realistic phonon displacements by \textit{ab initio} calculations is approximately four orders of magnitude smaller than the effective magnetization required to account for the experimental rotations observed in Refs.\ \cite{basini_nature24,kamba_cm26}.
In addition, Ref.\ \cite{nelson_cm26} showed that the dichroic signal in SrTiO$_3$ remains of comparable magnitude when the THz pump is tuned far away from the infrared-phonon resonance, where $\textbf{Q}_\mathcal{E}$ and hence $\langle\text{M}^z_{\text{ph}}\rangle$ is strongly suppressed. By contrast, the light-induced Faraday response can remain large away from phonon resonance, as demonstrated in Fig.\ \ref{Fig4}(d).

We finally stress that the appropriate comparison is between the phonon-induced magnetization $\langle\text{M}^z_{\text{ph}}\rangle$ in Eq.\ \eqref{mph} and the effective internal field inferred from the measured rotation, $\text{B}^z_\text{int}=\mu_0\langle \text{M}_\text{ph}^z\rangle$, which yields $\langle\text{M}^z_\text{ph}\rangle=\text{B}^z_\text{int}/\mu_0=\theta_\text{F}/(\mu_0\xi\,\chi^{(\text{F})})$ by Eq.\ \eqref{eqtheta}. By contrast, it would be misleading to compare the phonon-induced magnetization with the field induced magnetization appearing in the first term of Eq.\ \eqref{eqMtot}, $\langle\text{M}^z_\text{ph}\rangle=\chi^m\text{B}^z_\text{int}/\mu_0=\chi^m\theta_\text{F}/(\mu_0\xi\,\chi^{(\text{F})})$ \cite{luo_science23}. This quantity is much smaller because $\chi^m$ is small in nonmagnetic materials, but it does not represent the magnetization responsible for generating the internal field $\text{B}^z_\text{int}$ required to produce the Faraday rotation according to Eq.\ \eqref{eqtheta}.
\section{Conclusion}
In this work we developed a microscopic theory of the dynamical breaking of time-reversal symmetry in solids driven by circularly polarized light, identifying the resulting nonlinear response as a light-induced Faraday effect.
We showed that a static probe rotation originates from the antisymmetric component of the third-order susceptibility,
and demonstrated that such a contribution emerges only in dynamical conditions as a consequence of the breakdown of Kleinman symmetry. While it is generically expected that the Kleinman approximation fails when the system is driven near electronic resonance \cite{boyd_chap1,alu_optica19,kruglyak_prb12, wismer_prb17,he_prb26}, we found a sizable light-induced Faraday effect even far from dissipative resonances, a prediction that could be further explored experimentally. 
The light-induced Faraday effect can thus be interpreted as the optical analogue of a Faraday rotation \cite{wismer_prb17,merlin_prb24} generated by an effective internal magnetic field associated with the pump-driven electronic excitations, despite the absence of any real macroscopic magnetization.

To establish a quantitative connection with the recent experiments on paraelectric SrTiO$_3$ \cite{basini_nature24}, several additional effects must be taken into account. A crucial aspect is that, under realistic pump conditions, the two components of the circular field may exhibit a small frequency mismatch. In such cases, the symmetric susceptibility also contributes to the quasi-static response through a Kerr-like component, which can often provide the dominant signal. Since this symmetric term remains finite in the static approximation, it could already be captured in previous modeling \cite{basini_nature24,basini_prb24}. 
Neglecting this symmetric contribution, however, leads to an overestimate of the anomalous response in attempts to explain the full signal, an important aspect that was not considered in previous theoretical studies \cite{geilhufe_prl24,wehling_prl25,merlin_prb24,merlin_pnas25}.
At the same time, the residual quasi-static polarization rotation reported in Ref.\ \cite{basini_nature24}, of order $10\,\mu\text{rad}$ and not accounted for by the symmetric susceptibility, is quantitatively consistent with the light-induced Faraday effect, while being underestimated by approximately four orders of magnitude within a magneto-optical IFE picture.

Within a microscopic toy model we estimate both the electronic and phonon-mediated antisymmetric susceptibilities, and find that the resonant driving of an infrared-active phonon is not a necessary condition for the light-induced Faraday response to emerge. A substantial response is also obtained when the pump frequency is detuned from the phonon resonance, consistent with recent experiments on SrTiO$_3$ \cite{nelson_cm26} where rotations of the same order of magnitude as those measured in Ref.\ \cite{basini_nature24} were reported. On the other hand, resonance with a polar phonon can enhance the optical response. As demonstrated in Fig. \ref{Fig4}, the temperature dependence of the soft phonon-mediated light-induced Faraday response is consistent with the results of Ref.\ \cite{basini_nature24}.
Altogether, these results suggest that the quasi-static signal observed in nonmagnetic systems can be naturally interpreted within a light-induced Faraday scenario, without requiring phonon magnetic moments far exceeding theoretical estimates \cite{juraschek_prm17,juraschek_prm19,juraschek_prr20,juraschek_prr22}.

The light-induced Faraday response is also expected to provide a sizable contribution to the pump-probe response of paramagnetic insulators with strong spin-orbit coupling and a pronounced temperature dependence of the magnetic susceptibility, such as CeF$_3$ \cite{luo_science23,davies_cm26}. A quantitative estimate of the antisymmetric response in this case requires accounting for the crystal-field splitting of the electronic spectrum, in light of the recent observation of a strong enhancement of the dichroic signal when the pump frequency is resonant with the localized $4f$ levels of Ce$^{3+}$ \cite{davies_cm26}. The detailed frequency dependence reported in Ref.\ \cite{davies_cm26} further supports the conclusion that resonant driving of a phonon is not necessary to obtain a Faraday rotation: the signal reaches its maximum in a spectral window where phonon absorption is very weak. At the same time, the response exhibits notable differences from that of weakly diamagnetic SrTiO$_3$, including a long-lived signal extending beyond the pump duration \cite{luo_science23,davies_cm26} and a sign change across the electronic resonance \cite{davies_cm26}. Whether these features can be attributed to a purely magneto-optical response, enhanced by localized $4f$ states, or instead arise from a dynamical antisymmetric third-order response remains an open question for future work.

Beyond its immediate relevance for interpreting recent ultrafast experiments, our results open a broader perspective on the possibility of engineering tunable nonequilibrium dichroic optical devices, and potentially realizing magneto-optical, transport, and mechanical Faraday phenomena \cite{milner_prl21,averbukh_prr22} in otherwise nonmagnetic materials. More generally, our findings suggest that magnetic-like responses, often associated with spin degrees of freedom and enhanced by spin-orbit coupling, can also emerge from the dynamics of purely orbital degrees of freedom. This perspective may provide new opportunities for orbitronics \cite{cysne_review_orbitronics,fukami_review_orbitronics,daegeun_review_orbitronics} and for controlling angular-momentum-dependent optical and mechanical responses without relying on spin polarization or magnetic order.

\begin{acknowledgments}
We are grateful to Roberto Merlin, Martina Basini, Mattia Udina, Stefano Bonetti, Davide Bossini, Michele Buzzi, and Michael Fechner for useful discussions and suggestions. This work has been supported by the European Union under the project MORE-TEM ERC-SYN (Grant No.\ 951215); by Sapienza University under the project Ateneo (Grants No.\ RM123188E357C540, and No.\ 
RP124190A63FAA97); and by the Italian Ministry of Education, University and Research under Project PRIN2022-CoInEx (Grant No.\ 2022WS9MS4).
\end{acknowledgments}

\bibliography{bibl.bib}

@article{fukuyama_rpf69,
    author = {Fukuyama, Hidetoshi and Ebisawa, Hiromichi and Wada, Yasushi},
    title = {Theory of Hall Effect. I: Nearly Free Electron},
    journal = {Progress of Theoretical Physics},
    volume = {42},
    number = {3},
    pages = {494-511},
    year = {1969},
    month = {09},
    issn = {0033-068X},
    doi = {10.1143/PTP.42.494},
    url = {https://doi.org/10.1143/PTP.42.494},
    eprint = {https://academic.oup.com/ptp/article-pdf/42/3/494/5471733/42-3-494.pdf},
}

@article{zhang_sto07,
    author = {Han, Jiaguang and Wan, Fan and Zhu, Zhiyuan and Zhang, Weili},
    title = {Dielectric response of soft mode in ferroelectric SrTiO3},
    journal = {Applied Physics Letters},
    volume = {90},
    number = {3},
    pages = {031104},
    year = {2007},
    month = {01},
    issn = {0003-6951},
    doi = {10.1063/1.2431448},
    url = {https://doi.org/10.1063/1.2431448}
}

@article{wismer_prb17,
  title = {Ultrafast optical Faraday effect in transparent solids},
  author = {Wismer, Michael S. and Stockman, Mark I. and Yakovlev, Vladislav S.},
  journal = {Phys. Rev. B},
  volume = {96},
  issue = {22},
  pages = {224301},
  numpages = {7},
  year = {2017},
  month = {Dec},
  publisher = {American Physical Society},
  doi = {10.1103/PhysRevB.96.224301},
  url = {https://link.aps.org/doi/10.1103/PhysRevB.96.224301}
}

@misc{kamba_cm26,
      title={THz-induced phonomagnetism in diamagnetic quantum paraelectric KTaO$_3$}, 
      author={C. Kadlec and F. Kadlec and D. Repček and P. Kužel and M. Basini and J. -C. Deinert and S. Kovalev and T. Tadano and M. Udina and I. Ilyakov and T. V. A. G. de Oliveira and A. Ponomaryov and A. Arshad and A. Maia and S. Bonetti and S. Kamba},
      year={2026},
      eprint={2608.27060},
      archivePrefix={arXiv},
      primaryClass={cond-mat.mtrl-sci},
      url={https://arxiv.org/abs/2608.27060}, 
}

@misc{nelson_cm26,
      title={Light-Driven Ultrafast Control of Time-Reversal Symmetry in SrTiO3}, 
      author={In Hyeok Choi and Sergei Urazhdin and Shivasheesh Varshney and Seung Gyo Jeong and Bharat Jalan and Keith A. Nelson},
      year={2026},
      eprint={2609.09497},
      archivePrefix={arXiv},
      primaryClass={cond-mat.mtrl-sci},
      url={https://arxiv.org/abs/2609.09497}, 
}

@misc{davies_cm26,
      title={A crystal-field route to THz-driven magnetization}, 
      author={T. Zalewski and M. S. Mrudul and Y. Lee and M. Weissenhofer and A. V. Boris and P. M. Oppeneer and A. Kirilyuk and C. S. Davies},
      year={2026},
      eprint={2607.08519},
      archivePrefix={arXiv},
      primaryClass={cond-mat.mtrl-sci},
      url={https://arxiv.org/abs/2607.08519}, 
}

@article{luo_science23,
author = {Jiaming Luo  and Tong Lin  and Junjie Zhang  and Xiaotong Chen  and Elizabeth R. Blackert  and Rui Xu  and Boris I. Yakobson  and Hanyu Zhu },
title = {Large effective magnetic fields from chiral phonons in rare-earth halides},
journal = {Science},
volume = {382},
number = {6671},
pages = {698-702},
year = {2023},
doi = {10.1126/science.adi9601},
URL = {https://www.science.org/doi/abs/10.1126/science.adi9601},
eprint = {https://www.science.org/doi/pdf/10.1126/science.adi9601}}

@article{alu_optica19,
author = {Robert Duggan and Dimitrios Sounas and Andrea Alu},
journal = {Optica},
number = {9},
pages = {1152--1157},
publisher = {Optica Publishing Group},
title = {Optically driven effective Faraday effect in instantaneous nonlinear media},
volume = {6},
month = {Sep},
year = {2019},
url = {https://opg.optica.org/optica/abstract.cfm?URI=optica-6-9-1152},
doi = {10.1364/OPTICA.6.001152},
}

@article{he_prb26,
  title = {Second harmonic Hall response in insulators: Interband quantum geometry and breakdown of Kleinman's conjecture},
  author = {He, Wen-Yu and Law, K. T.},
  journal = {Phys. Rev. B},
  volume = {113},
  issue = {4},
  pages = {L041110},
  numpages = {8},
  year = {2026},
  month = {Jan},
  publisher = {American Physical Society},
  doi = {10.1103/vkml-m2sm},
  url = {https://link.aps.org/doi/10.1103/vkml-m2sm}
}

@article{rasing_rmp10,
  title = {Ultrafast optical manipulation of magnetic order},
  author = {Kirilyuk, Andrei and Kimel, Alexey V. and Rasing, Theo},
  journal = {Rev. Mod. Phys.},
  volume = {82},
  issue = {3},
  pages = {2731--2784},
  numpages = {0},
  year = {2010},
  month = {Sep},
  publisher = {American Physical Society},
  doi = {10.1103/RevModPhys.82.2731},
  url = {https://link.aps.org/doi/10.1103/RevModPhys.82.2731}
}

@article{freiser_moreview,
  title={A survey of magnetooptic effects},
  author={M. J. Freiser},
  journal={IEEE Transactions on Magnetics},
  year={1968},
  volume={4},
  pages={152-161},
  url={https://api.semanticscholar.org/CorpusID:123053631}
}

@article{resta_rmp94,
  title = {Macroscopic polarization in crystalline dielectrics: the geometric phase approach},
  author = {Resta, Raffaele},
  journal = {Rev. Mod. Phys.},
  volume = {66},
  issue = {3},
  pages = {899--915},
  numpages = {0},
  year = {1994},
  month = {Jul},
  publisher = {American Physical Society},
  doi = {10.1103/RevModPhys.66.899},
  url = {https://link.aps.org/doi/10.1103/RevModPhys.66.899}
}

@article{cappelluti_prb12,
  title = {Charged-phonon theory and Fano effect in the optical spectroscopy of bilayer graphene},
  author = {Cappelluti, E. and Benfatto, L. and Manzardo, M. and Kuzmenko, A. B.},
  journal = {Phys. Rev. B},
  volume = {86},
  issue = {11},
  pages = {115439},
  numpages = {15},
  year = {2012},
  month = {Sep},
  publisher = {American Physical Society},
  doi = {10.1103/PhysRevB.86.115439},
  url = {https://link.aps.org/doi/10.1103/PhysRevB.86.115439}
}

@article{yamada_jpsj69,
author = {Yamada ,Yasusada and Shirane ,Gen},
title = {Neutron Scattering and Nature of the Soft Optical Phonon in SrTiO3},
journal = {Journal of the Physical Society of Japan},
volume = {26},
number = {2},
pages = {396-403},
year = {1969},
doi = {10.1143/JPSJ.26.396},

URL = { 
    
        https://doi.org/10.1143/JPSJ.26.396
    
    

},
eprint = { 
    
        https://doi.org/10.1143/JPSJ.26.396
    
    

}
}

@article{vogt_prb95,
  title = {Refined treatment of the model of linearly coupled anharmonic oscillators and its application to the temperature dependence of the zone-center soft-mode frequencies of ${\mathrm{KTaO}}_{3}$ and ${\mathrm{SrTiO}}_{3}$},
  author = {Vogt, H.},
  journal = {Phys. Rev. B},
  volume = {51},
  issue = {13},
  pages = {8046--8059},
  numpages = {0},
  year = {1995},
  month = {Apr},
  publisher = {American Physical Society},
  doi = {10.1103/PhysRevB.51.8046},
  url = {https://link.aps.org/doi/10.1103/PhysRevB.51.8046}
}

@article{fiore_prb26,
  title = {Two-dimensional terahertz spectroscopy in electronic systems: A many-body diagrammatic approach},
  author = {Fiore, Jacopo and Sellati, Niccol\`o and Udina, Mattia and Benfatto, Lara},
  journal = {Phys. Rev. B},
  volume = {113},
  issue = {17},
  pages = {174524},
  numpages = {26},
  year = {2026},
  month = {May},
  publisher = {American Physical Society},
  doi = {10.1103/8ffz-rtyf},
  url = {https://link.aps.org/doi/10.1103/8ffz-rtyf}
}

@article{merlin_prb24,
  title = {Unraveling the effect of circularly polarized light on reciprocal media: Breaking time reversal symmetry with non-Maxwellian magnetic-esque fields},
  author = {Merlin, R.},
  journal = {Phys. Rev. B},
  volume = {110},
  issue = {9},
  pages = {094312},
  numpages = {5},
  year = {2024},
  month = {Sep},
  publisher = {American Physical Society},
  doi = {10.1103/PhysRevB.110.094312},
  url = {https://link.aps.org/doi/10.1103/PhysRevB.110.094312}
}

@article{merlin_pnas25,
    author = {Merlin, R},
    title = {Magnetophononics and the chiral phonon misnomer},
    journal = {PNAS Nexus},
    volume = {4},
    number = {1},
    pages = {pgaf002},
    year = {2025},
    month = {01},
    issn = {2752-6542},
    doi = {10.1093/pnasnexus/pgaf002},
    url = {https://doi.org/10.1093/pnasnexus/pgaf002},
}

@article{geilhufe_prl24,
  title = {Phonon Inverse Faraday Effect from Electron-Phonon Coupling},
  author = {Shabala, Natalia and Geilhufe, R. Matthias},
  journal = {Phys. Rev. Lett.},
  volume = {133},
  issue = {26},
  pages = {266702},
  numpages = {6},
  year = {2024},
  month = {Dec},
  publisher = {American Physical Society},
  doi = {10.1103/PhysRevLett.133.266702},
  url = {https://link.aps.org/doi/10.1103/PhysRevLett.133.266702}
}

@article{wehling_prl25,
  title = {Ultrafast Pseudomagnetic Fields from Electron-Nuclear Quantum Geometry},
  author = {Klebl, Lennart and Schobert, Arne and Eckstein, Martin and Sangiovanni, Giorgio and Balatsky, Alexander V. and Wehling, Tim O.},
  journal = {Phys. Rev. Lett.},
  volume = {134},
  issue = {1},
  pages = {016705},
  numpages = {7},
  year = {2025},
  month = {Jan},
  publisher = {American Physical Society},
  doi = {10.1103/PhysRevLett.134.016705},
  url = {https://link.aps.org/doi/10.1103/PhysRevLett.134.016705}
}

@article{milner_prl21,
  title = {Observation of Mechanical Faraday Effect in Gaseous Media},
  author = {Milner, Alexander A. and Steinitz, Uri and Averbukh, Ilya Sh. and Milner, Valery},
  journal = {Phys. Rev. Lett.},
  volume = {127},
  issue = {7},
  pages = {073901},
  numpages = {5},
  year = {2021},
  month = {Aug},
  publisher = {American Physical Society},
  doi = {10.1103/PhysRevLett.127.073901},
  url = {https://link.aps.org/doi/10.1103/PhysRevLett.127.073901}
}

@article{averbukh_prr22,
  title = {Rotation of the polarization of light as a tool for investigating the collisional transfer of angular momentum from rotating molecules to macroscopic gas flows},
  author = {Tutunnikov, Ilia and Steinitz, Uri and Gershnabel, Erez and Hartmann, Jean-Michel and Milner, Alexander A. and Milner, Valery and Averbukh, Ilya Sh.},
  journal = {Phys. Rev. Res.},
  volume = {4},
  issue = {1},
  pages = {013212},
  numpages = {13},
  year = {2022},
  month = {Mar},
  publisher = {American Physical Society},
  doi = {10.1103/PhysRevResearch.4.013212},
  url = {https://link.aps.org/doi/10.1103/PhysRevResearch.4.013212}
}

@misc{xiao_cm26,
      title={Nonadiabatic Theory of Phonon Magnetic Moments in Insulators and Metals}, 
      author={Haoran Chen and Wenqin Chen and Kaijie Yang and Ting Cao and Di Xiao},
      year={2026},
      eprint={2605.06983},
      archivePrefix={arXiv},
      primaryClass={cond-mat.mes-hall},
      url={https://arxiv.org/abs/2605.06983},
}

@article{urazhdin_prb25,
  title = {Atomic and interatomic orbital magnetization induced in ${\mathrm{SrTiO}}_{3}$ by chiral phonons},
  author = {Urazhdin, Sergei},
  journal = {Phys. Rev. B},
  volume = {111},
  issue = {21},
  pages = {214435},
  numpages = {7},
  year = {2025},
  month = {Jun},
  publisher = {American Physical Society},
  doi = {10.1103/m73s-fxdp},
  url = {https://link.aps.org/doi/10.1103/m73s-fxdp}
}

@article{juraschek_prb25,
  title = {Magneto-opto-phononic inverse Faraday effect},
  author = {Lopez, Daniel A. Bustamante and Hu, Wanzheng and Juraschek, Dominik M.},
  journal = {Phys. Rev. B},
  volume = {112},
  issue = {14},
  pages = {144306},
  numpages = {13},
  year = {2025},
  month = {Oct},
  publisher = {American Physical Society},
  doi = {10.1103/xmxz-zt1m},
  url = {https://link.aps.org/doi/10.1103/xmxz-zt1m}
}

@article{juraschek_review24,
  author    = {Carl P. Romao and Dominik M. Juraschek},
  title     = {Light makes atoms behave like electromagnetic coils},
  journal   = {Nature},
  year      = {2024},
  volume    = {628},
  number    = {8008},
  pages     = {505--506},
  doi       = {10.1038/d41586-024-00889-w},
  url       = {https://www.nature.com/articles/d41586-024-00889-w},
  month     = apr
}

@article{fukami_review_orbitronics,
  author  = {Fukami, Shunsuke and Lee, Kyung Jin and Kl{\"a}ui, Mathias},
  title   = {Challenges and opportunities in orbitronics},
  journal = {Nature Physics},
  year    = {2026},
  volume  = {22},
  number  = {7},
  pages   = {999-1005},
  doi     = {10.1038/s41567-025-03143-w},
  issn    = {1745-2473},
  url     = {https://www.nature.com/articles/s41567-025-03143-w}
}

@article{cysne_review_orbitronics,
  author  = {Cysne, Tarik P. and Canonico, Luis M. and Costa, Marcio and
             Muniz, R. B. and Rappoport, Tatiana G.},
  title   = {Orbitronics in two-dimensional materials},
  journal = {npj Spintronics},
  year    = {2025},
  volume  = {3},
  number  = {39},
  doi     = {10.1038/s44306-025-00103-1},
  url     = {https://doi.org/10.1038/s44306-025-00103-1}
}

@article{daegeun_review_orbitronics,
  author  = {Jo, Daegeun and Go, Dongwook and Choi, Gyung-Min and Lee, Hyun-Woo},
  title   = {Spintronics meets orbitronics: Emergence of orbital angular momentum in solids},
  journal = {npj Spintronics},
  year    = {2024},
  volume  = {2},
  number  = {19},
  doi     = {10.1038/s44306-024-00023-6},
  url     = {https://doi.org/10.1038/s44306-024-00023-6}
}

@article{basini_prb24,
  	title = {Terahertz ionic Kerr effect: Two-phonon contribution to the nonlinear optical response in insulators},
  	author = {Basini, M. and Udina, M. and Pancaldi, M. and Unikandanunni, V. and Bonetti, S. and Benfatto, L.},
  	journal = {Phys. Rev. B},
  	volume = {109},
  	issue = {2},
 	 pages = {024309},
  	numpages = {10},
  	year = {2024},
  	month = {Jan},
  	publisher = {American Physical Society},
  	doi = {10.1103/PhysRevB.109.024309},
  	url = {https://link.aps.org/doi/10.1103/PhysRevB.109.024309}
}

@article{basini_nature24,
  author    = {Basini, M. and Pancaldi, M. and Wehinger, B. and Udina, M. and Unikandanunni, V. and Tadano, T. and Hoffmann, M. C. and Balatsky, A. V. and Bonetti, S.},
  title     = {Terahertz electric-field-driven dynamical multiferroicity in SrTiO3},
  journal   = {Nature},
  year      = {2024},
  volume    = {628},
  number    = {8008},
  pages     = {534--539},
  doi       = {10.1038/s41586-024-07175-9},
  url       = {https://doi.org/10.1038/s41586-024-07175-9}
}

@article{kimel_nature05,
  title={Ultrafast non-thermal control of magnetization by instantaneous photomagnetic pulses},
  author={Kimel, AV and Kirilyuk, A and Usachev, PA and Pisarev, RV and Balbashov, AM and Rasing, Th},
  journal={Nature},
  volume={435},
  number={7042},
  pages={655--657},
  year={2005},
  publisher={Nature Publishing Group UK London}
}

@article{yin_SSC05,
title = {Nonlinear optical properties in SrTiO3 thin films by pulsed laser deposition},
journal = {Solid State Communications},
volume = {135},
number = {4},
pages = {221-225},
year = {2005},
issn = {0038-1098},
doi = {https://doi.org/10.1016/j.ssc.2005.04.031},
url = {https://www.sciencedirect.com/science/article/pii/S0038109805004072},
author = {Y. Deng and Y.L. Du and M.S. Zhang and J.H. Han and Z. Yin}
}

@article{kirilyuk_nature24,
  title={Phononic switching of magnetization by the ultrafast Barnett effect},
  author={Davies, CS and Fennema, FGN and Tsukamoto, A and Razdolski, I and Kimel, AV and Kirilyuk, A},
  journal={Nature},
  volume={628},
  number={8008},
  pages={540--544},
  year={2024},
  publisher={Nature Publishing Group UK London},
  url={https://doi.org/10.1038/s41586-024-07200-x},
  doi={10.1038/s41586-024-07200-x}
}

@article{pershan_pr63,
  title = {Nonlinear Optical Properties of Solids: Energy Considerations},
  author = {Pershan, P. S.},
  journal = {Phys. Rev.},
  volume = {130},
  issue = {3},
  pages = {919--929},
  numpages = {0},
  year = {1963},
  month = {May},
  publisher = {American Physical Society},
  doi = {10.1103/PhysRev.130.919},
  url = {https://link.aps.org/doi/10.1103/PhysRev.130.919}
}

@article{pershan_prl65,
  title = {Optically-Induced Magnetization Resulting from the Inverse Faraday Effect},
  author = {van der Ziel, J. P. and Pershan, P. S. and Malmstrom, L. D.},
  journal = {Phys. Rev. Lett.},
  volume = {15},
  issue = {5},
  pages = {190--193},
  numpages = {0},
  year = {1965},
  month = {Aug},
  publisher = {American Physical Society},
  doi = {10.1103/PhysRevLett.15.190},
  url = {https://link.aps.org/doi/10.1103/PhysRevLett.15.190}
}

@article{pershan_pr66,
  title = {Theoretical Discussion of the Inverse Faraday Effect, Raman Scattering, and Related Phenomena},
  author = {Pershan, P. S. and van der Ziel, J. P. and Malmstrom, L. D.},
  journal = {Phys. Rev.},
  volume = {143},
  issue = {2},
  pages = {574--583},
  numpages = {0},
  year = {1966},
  month = {Mar},
  publisher = {American Physical Society},
  doi = {10.1103/PhysRev.143.574},
  url = {https://link.aps.org/doi/10.1103/PhysRev.143.574}
}

@article{kruglyak_prb12,
  title = {Ultrafast inverse Faraday effect in a paramagnetic terbium gallium garnet crystal},
  author = {Mikhaylovskiy, R. V. and Hendry, E. and Kruglyak, V. V.},
  journal = {Phys. Rev. B},
  volume = {86},
  issue = {10},
  pages = {100405},
  numpages = {5},
  year = {2012},
  month = {Sep},
  publisher = {American Physical Society},
  doi = {10.1103/PhysRevB.86.100405},
  url = {https://link.aps.org/doi/10.1103/PhysRevB.86.100405}
}

@incollection{boyd_chap1,
title = {Chapter 1 - The Nonlinear Optical Susceptibility},
booktitle = {Nonlinear Optics (Fourth Edition)},
publisher = {Academic Press},
edition = {Fourth Edition},
pages = {1-64},
year = {2020},
isbn = {978-0-12-811002-7},
doi = {https://doi.org/10.1016/B978-0-12-811002-7.00010-2},
url = {https://www.sciencedirect.com/science/article/pii/B9780128110027000102},
author = {Robert W. Boyd}
}

@article{miranda_2014,
    author = {Jiménez, E. and Mikuszeit, N. and Cuñado, J. L. F. and Perna, P. and Pedrosa, J. and Maccariello, D. and Rodrigo, C. and Niño, M. A. and Bollero, A. and Camarero, J. and Miranda, R.},
    title = {Vectorial Kerr magnetometer for simultaneous and quantitative measurements of the in-plane magnetization components},
    journal = {Review of Scientific Instruments},
    volume = {85},
    number = {5},
    pages = {053904},
    year = {2014},
    month = {05},
    issn = {0034-6748},
    doi = {10.1063/1.4871098},
    url = {https://doi.org/10.1063/1.4871098},
}

@article{juraschek_prm17,
  title = {Dynamical multiferroicity},
  author = {Juraschek, Dominik M. and Fechner, Michael and Balatsky, Alexander V. and Spaldin, Nicola A.},
  journal = {Phys. Rev. Mater.},
  volume = {1},
  issue = {1},
  pages = {014401},
  numpages = {9},
  year = {2017},
  month = {Jun},
  publisher = {American Physical Society},
  doi = {10.1103/PhysRevMaterials.1.014401},
  url = {https://link.aps.org/doi/10.1103/PhysRevMaterials.1.014401}
}

@article{juraschek_prm19,
  title = {Orbital magnetic moments of phonons},
  author = {Juraschek, Dominik M. and Spaldin, Nicola A.},
  journal = {Phys. Rev. Mater.},
  volume = {3},
  issue = {6},
  pages = {064405},
  numpages = {8},
  year = {2019},
  month = {Jun},
  publisher = {American Physical Society},
  doi = {10.1103/PhysRevMaterials.3.064405},
  url = {https://link.aps.org/doi/10.1103/PhysRevMaterials.3.064405}
}

@article{juraschek_prr20,
  title = {Phono-magnetic analogs to opto-magnetic effects},
  author = {Juraschek, Dominik M. and Narang, Prineha and Spaldin, Nicola A.},
  journal = {Phys. Rev. Res.},
  volume = {2},
  issue = {4},
  pages = {043035},
  numpages = {11},
  year = {2020},
  month = {Oct},
  publisher = {American Physical Society},
  doi = {10.1103/PhysRevResearch.2.043035},
  url = {https://link.aps.org/doi/10.1103/PhysRevResearch.2.043035}
}

@article{juraschek_prr22,
  title = {Giant effective magnetic fields from optically driven chiral phonons in $4f$ paramagnets},
  author = {Juraschek, Dominik M. and Neuman, Tom\'a\ifmmode \check{s}\else \v{s}\fi{} and Narang, Prineha},
  journal = {Phys. Rev. Res.},
  volume = {4},
  issue = {1},
  pages = {013129},
  numpages = {9},
  year = {2022},
  month = {Feb},
  publisher = {American Physical Society},
  doi = {10.1103/PhysRevResearch.4.013129},
  url = {https://link.aps.org/doi/10.1103/PhysRevResearch.4.013129}
}

@article{udina_prl26,
  title = {Antisymmetric Raman Response},
  author = {Udina, Mattia and Paul, Indranil},
  journal = {Phys. Rev. Lett.},
  volume = {136},
  issue = {12},
  pages = {126505},
  numpages = {6},
  year = {2026},
  month = {Mar},
  publisher = {American Physical Society},
  doi = {10.1103/gnpt-5ffq},
  url = {https://link.aps.org/doi/10.1103/gnpt-5ffq}
}

@article{fowles_optics,
    author = {Fowles, Grant R. and Lynch, David W.},
    title = {Introduction to Modern Optics},
    journal = {American Journal of Physics},
    volume = {36},
    number = {8},
    pages = {770-771},
    year = {1968},
    month = {08},
    issn = {0002-9505},
    doi = {10.1119/1.1975142},
    url = {https://doi.org/10.1119/1.1975142}
}

@article{cappelluti_prb10,
	Author = {Cappelluti, E. and Benfatto, L. and Kuzmenko, A. B.},
	Doi = {10.1103/PhysRevB.82.041402},
	Issue = {4},
	Journal = {Phys. Rev. B},
	Month = {Jul},
	Numpages = {4},
	Pages = {041402},
	Publisher = {American Physical Society},
	Title = {Phonon switching and combined Fano-Rice effect in optical spectra of bilayer graphene},
	Url = {https://link.aps.org/doi/10.1103/PhysRevB.82.041402},
	Volume = {82},
	Year = {2010}}

@article{bistoni_2D19,
	Author = {O Bistoni and P Barone and E Cappelluti and L Benfatto and F Mauri},
	Doi = {10.1088/2053-1583/ab2ce0},
	Journal = {2D Materials},
	Month = {jul},
	Number = {4},
	Pages = {045015},
	Publisher = {{IOP} Publishing},
	Title = {Giant effective charges and piezoelectricity in gapped graphene},
	Url = {https://doi.org/10.1088%2F2053-1583%2Fab2ce0},
	Volume = {6},
	Year = 2019}

@article{udina_prb19,
	Author = {Udina, Mattia and Cea, Tommaso and Benfatto, Lara},
	Doi = {10.1103/PhysRevB.100.165131},
	Issue = {16},
	Journal = {Phys. Rev. B},
	Month = {Oct},
	Numpages = {18},
	Pages = {165131},
	Publisher = {American Physical Society},
	Title = {Theory of coherent-oscillations generation in terahertz pump-probe spectroscopy: From phonons to electronic collective modes},
	Url = {https://link.aps.org/doi/10.1103/PhysRevB.100.165131},
	Volume = {100},
	Year = {2019}}

@misc{suppl,
	Howpublished = {See Supplemental Information for the derivation of the differential intensity at generic sample orientation for linearly and circularly polarized monochromatic light; the derivation of the differential intensity for frequency-mismatched monochromatic and Gaussian pumps; details of the calculations of the antisymmetric kernel in the $sp$ model; details of the calculations of the phonon-mediated response; details on the estimate of the ratio between IFE-induced and light-induced Faraday response.}}

@article{benfatto_prb04,
	Author = {Benfatto, L. and Toschi, A. and Caprara, S.},
	Doi = {10.1103/PhysRevB.69.184510},
	Issue = {18},
	Journal = {Phys. Rev. B},
	Month = {May},
	Numpages = {21},
	Pages = {184510},
	Publisher = {American Physical Society},
	Title = {Low-energy phase-only action in a superconductor: A comparison with the $\mathrm{XY}$ model},
	Url = {https://link.aps.org/doi/10.1103/PhysRevB.69.184510},
	Volume = {69},
	Year = {2004}}

@book{cottam-lockwood,
  title     = {Light Scattering in Magnetic Solids},
  author    = {Cottam, Michael G. and Lockwood, David J.},
  publisher = {Wiley},
  year      = {1986},
  address   = {New York},
  isbn      = {0471817015},
  pages     = {272}
}
\clearpage
\onecolumngrid

\begin{center}
{\fontsize{12}{2}\selectfont\bfseries
Light-induced Faraday effect from dynamical breakdown of Kleinman symmetry - Supplemental Material}

\vspace{1em}

Niccolò Sellati$^{1}$, Jacopo Fiore$^{1,2}$, Lara Benfatto$^{1}$

\vspace{0.5em}

{\small
$^{1}$\textit{Department of Physics, ``Sapienza'' University of Rome,
P.le A. Moro 5, 00185 Rome, Italy} \\
$^{2}$\textit{Institute for Theory of Statistical Physics,
RWTH Aachen University, Aachen, Germany}}
\end{center}

\setcounter{equation}{0}
\setcounter{figure}{0}
\renewcommand{\theequation}{S\arabic{equation}}
\renewcommand{\thefigure}{S\arabic{figure}}

	\title{Light-induced Faraday effect from dynamical breakdown of Kleinman symmetry - Supplemental Material}
    \author{Niccolò Sellati}
\email{niccolo.sellati@uniroma1.it}
\affiliation{Department of Physics, ``Sapienza'' University of Rome, P.le A.\ Moro 5, 00185 Rome, Italy}
\author{Jacopo Fiore}
\affiliation{Department of Physics, ``Sapienza'' University of Rome, P.le A.\ Moro 5, 00185 Rome, Italy}
\affiliation{Institute for Theory of Statistical Physics, RWTH Aachen University, Aachen, Germany}
\author{Lara Benfatto}
\email{lara.benfatto@roma1.infn.it}
\affiliation{Department of Physics, ``Sapienza'' University of Rome, P.le A.\ Moro 5, 00185 Rome, Italy}

\maketitle

\section*{Calculation of the differential intensity}
In the balanced detection scheme as the one reported in Refs.\ \cite{basini_nature24,basini_prb24}, the differential intensity $\Delta\Gamma$ is directly proportional to the nonlinear polarization induced in the direction $P$, orthogonal to the polarization of the incoming probe along the direction $S$. Following the geometry described in Fig.\ 1(b) of the main text, reported here in Fig.\ \ref{FigSframe}, $\Delta\Gamma$ can then be connected to the nonlinear polarizations induced along the crystallographic $x$ and $y$ directions as
\begin{align}\label{DGtoP}
    \Delta\Gamma\propto \text{P}_\text{NL}^P=\cos\theta\,\text{P}_\text{NL}^x-\sin\theta\,\text{P}_\text{NL}^y.
\end{align}
The results reported in the main text are obtained from the following calculations by setting $\theta=0$.\\
Along the $i$-th direction ($i=x,y$), we can link the third-order nonlinear polarization $\text P_\text{NL}^i$ to the external electric field through the rank-4 third-order nonlinear susceptibility $\chi_{ijkl}(\omega;\omega_1,\omega_2,\omega_3)$. For the experimental geometry considered, we have explicitly
\begin{align}\label{PchiE3}
    \text{P}^i_\text{NL}(\omega)=\sum_{j,k,l=x,y}\int d\omega_1d\omega_2d\omega_3\,\delta(\omega-\omega_1-\omega_2-\omega_3)\, \chi_{ijkl}(\omega;\omega_1,\omega_2,\omega_3)\text{E}^j(\omega_1)\text{E}^k(\omega_2)\text{E}^l(\omega_3).
\end{align}
We then introduce the time delay $t_{pp}$ between pump $\textbf E_\text{pu}$ and probe $\textbf E_\text{pr}$, such that the total applied electric field is $\textbf E(t,t_{pp})=\textbf E_\text{pu}(t+t_{pp})+\textbf E_\text{pr}(t)$. For a single $s$ component, and by taking the Fourier transform with respect to $t$, this is equivalent to substituting
\begin{align}\label{ppdelay}
    \text{E}^s(\omega_s)\to\text{E}^s(\omega_s,t_{pp})=\text{E}^s_{\text{pr}}(\omega_s)+\text{E}^s_{\text{pu}}(\omega_s)e^{-i\omega_st_{pp}},
\end{align}
and thus rewrite Eq.\ \eqref{PchiE3} as
\begin{align}\label{PchiE3pp}
    \text{P}^i_\text{NL}(\omega,t_{pp})=3\sum_{j,k,l=x,y}\int d\omega_1d\omega_2d\omega_3\,\delta(\omega-\omega_1-\omega_2-\omega_3)\, \chi_{ij;kl}(\omega;\omega_1,\omega_2,\omega_3)\text{E}_\text{pr}^j(\omega_1)\text{E}_\text{pu}^k(\omega_2)\text{E}_\text{pu}^l(\omega_3)e^{-i(\omega_2+\omega_3)t_{pp}}.
\end{align}
Here we retained only terms scaling in $\text{E}_\text{pr}\text{E}_\text{pu}^2$, relevant for the pump-probe experiment. The third order susceptibility $\chi_{ij;kl}$ with the semicolon implies a precise ordering of the Cartesian indices, with $i$ labeling the outgoing nonlinear signal, $j$ the incoming probe, $k$ and $l$ the pump fields.\\
In the following we consider a cubic crystal with four independent components of the susceptibility, 
$\chi_{ii;ii}(\omega;\omega_1,\omega_2,\omega_3)$, $\chi_{ii;jj}(\omega;\omega_1,\omega_2,\omega_3)$, $\chi_{ij;ij}(\omega;\omega_1,\omega_2,\omega_3)$ and $\chi_{ij;ji}(\omega;\omega_1,\omega_3,\omega_2)$ ($i,j=x,y$ and $i\neq j$), and separate $\text{P}^i_\text{NL}(\omega,t_{pp})$ into three contributions:
\begin{align}\label{PchiE3pp2}
    \text{P}^i_\text{NL}(\omega,t_{pp})=3\int d\omega_1d\omega_2d\omega_3\,\delta(\omega-\omega_1-\omega_2-\omega_3)e^{-i(\omega_2+\omega_3)t_{pp}}\bigg[&\chi_{ii;ii}(\omega;\omega_1,\omega_2,\omega_3)\text{E}_\text{pr}^i(\omega_1)\text{E}_\text{pu}^i(\omega_2)\text{E}_\text{pu}^i(\omega_3)\nonumber\\
    &\chi_{ii;jj}(\omega;\omega_1,\omega_2,\omega_3)\text{E}_\text{pr}^i(\omega_1)\text{E}_\text{pu}^j(\omega_2)\text{E}_\text{pu}^j(\omega_3
    )\nonumber\\
    \big(\chi_{ij;ij}(\omega;\omega_1,\omega_2,\omega_3)+&\chi_{ij;ji}(\omega;\omega_1,\omega_3,\omega_2)\big)\text{E}_\text{pr}^j(\omega_1)\text{E}_\text{pu}^i(\omega_2)\text{E}_\text{pu}^j(\omega_3
    )\bigg].
\end{align}
%
If the $xy$ crystallographic reference frame is rotated with respect to the $PS$ reference frame in the experiment, it is more convenient to express the fields with their $P$ and $S$ components, as
\begin{align}\label{refchangepr}
    \text{E}_\text{pr}^x&=\text{E}^S_\text{pr}\sin\theta=\text{E}_\text{pr}\sin\theta,\nonumber\\
    \text{E}_\text{pr}^y&=\text{E}^S_\text{pr}\cos\theta=\text{E}_\text{pr}\cos\theta,
\end{align}
and
\begin{align}\label{refchangepu}
    \text{E}_\text{pu}^x&=\text{E}_\text{pu}^S\sin(\theta+\Delta\theta)+\text{E}_\text{pu}^P\cos(\theta+\Delta\theta),\nonumber\\
    \text{E}_\text{pu}^y&=\text{E}_\text{pu}^S\cos(\theta+\Delta\theta)-\text{E}_\text{pu}^P\sin(\theta+\Delta\theta).
\end{align}
Here, $\Delta\theta$ is the angle between the pump polarization at $t_{pp}=0$ and the $PS$ reference frame (see Fig.\ \ref{FigSframe}). We note that this angle is fixed in the experiment.
\begin{figure}[t]
    \centering
    \includegraphics[width=0.3\textwidth,keepaspectratio]{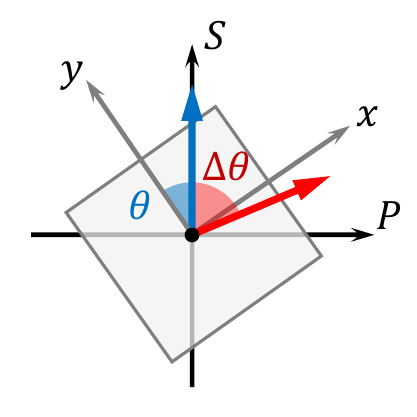}
    \caption{Polarization geometry. The probe (blue) defines a fixed $PS$ reference frame, while the sample is rotated by an angle $\theta$. The pump (red) is described in a reference frame rotated by a fixed angle $\Delta\theta$ with respect to $PS$.}
    \label{FigSframe}
\end{figure}\\
Finally, using Eq.\ \eqref{PchiE3pp2} with the rotations Eqs.\ \eqref{refchangepr} and \eqref{refchangepu}, we can rewrite the differential signal Eq.\ \eqref{DGtoP} as
\begin{align}\label{diffsign}
    \Delta\Gamma(t_{pp},\theta)\propto\int d\omega_1d\omega_2d\omega_3 e^{-i(\omega_2+\omega_3)t_{pp}}\text{E}_\text{pr}(\omega_1)\bigg[\big(F_{23}(\theta)+F_{32}(\theta)\big)\mathcal{S}_\parallel+
    \big(G_{23}(\theta)+G_{32}(\theta)\big)\mathcal{S}_\perp+\big(H_{23}-H_{32}\big)\mathcal{A}\bigg],
\end{align}
where we performed an integration over $\omega$, as dictated by the detection procedure of pump-probe experiments \cite{udina_prb19}, and $\int d\omega\,\delta(\omega-\omega_1-\omega_2-\omega_3)=1$.
As defined in Eq.\ (5) of the main text, $\mathcal{S}_{\parallel,\perp}$ and $\mathcal{A}$ denote, respectively, symmetric and antisymmetric combinations of the pump field components under exchange $\omega_2\leftrightarrow \omega_3$,
\begin{align}\label{fieldcomb}
    \mathcal{S}_\parallel&=\text{E}^S_\text{pu}(\omega_2)\text{E}^S_\text{pu}(\omega_3)-\text{E}^P_\text{pu}(\omega_2)\text{E}^P_\text{pu}(\omega_3),\nonumber\\
    \mathcal{S}_\perp&=\text{E}^P_\text{pu}(\omega_2)\text{E}^S_\text{pu}(\omega_3)+\text{E}^S_\text{pu}(\omega_2)\text{E}^P_\text{pu}(\omega_3),\nonumber\\    
    \mathcal{A}&=\text{E}^P_\text{pu}(\omega_2)\text{E}^S_\text{pu}(\omega_3)-\text{E}^S_\text{pu}(\omega_2)\text{E}^P_\text{pu}(\omega_3),
\end{align}
where the subscripts $\parallel$ and $\perp$ indicate that each product contain parallel or perpendicular components of the pump field. 
The coefficients of the symmetric combinations read explicitly
\begin{align}\label{Fkl}
    F_{kl}(\theta)=\frac{1}{2}\bigg[&\frac{\sin(4\theta)\cos(2\Delta\theta)}{2}\big(\chi_{ii;jj}(\omega;\omega_1,\omega_k,\omega_l)+\chi_{ij;ij}(\omega;\omega_1,\omega_k,\omega_l)+\chi_{ij;ji}(\omega;\omega_1,\omega_l,\omega_k)-\chi_{ii;ii}(\omega;\omega_1,\omega_k,\omega_l)\big)\nonumber\\
    +&\sin^2(2\theta)\sin(2\Delta\theta)\big(\chi_{ii;ii}(\omega;\omega_1,\omega_k,\omega_l)-\chi_{ii;jj}(\omega;\omega_1,\omega_k,\omega_l)-\chi_{ij;ij}(\omega;\omega_1,\omega_k,\omega_l)-\chi_{ij;ji}(\omega;\omega_1,\omega_l,\omega_k)\big)\nonumber\\
    +&\sin(2\Delta\theta)\big(\chi_{ij;ij}(\omega;\omega_1,\omega_k,\omega_l)+\chi_{ij;ji}(\omega;\omega_1,\omega_l,\omega_k)\big)\Big)\bigg]\equiv F(\omega_1,\omega_k,\omega_l,\theta)\equiv F_{\omega_k\omega_l}(\theta),
\end{align}
and 
\begin{align}\label{Gkl}
    G_{kl}(\theta)=\frac{1}{2}\bigg[&\frac{\sin(4\theta)\sin(2\Delta\theta)}{2}\big(\chi_{ii;jj}(\omega;\omega_1,\omega_k,\omega_l)+\chi_{ij;ij}(\omega;\omega_1,\omega_k,\omega_l)+\chi_{ij;ji}(\omega;\omega_1,\omega_l,\omega_k)-\chi_{ii;ii}(\omega;\omega_1,\omega_k,\omega_l)\big)\nonumber\\
    +&\sin^2(2\theta)\cos(2\Delta\theta)\big(\chi_{ii;ii}(\omega;\omega_1,\omega_k,\omega_l)-\chi_{ii;jj}(\omega;\omega_1,\omega_k,\omega_l)-\chi_{ij;ij}(\omega;\omega_1,\omega_k,\omega_l)-\chi_{ij;ji}(\omega;\omega_1,\omega_l,\omega_k)\big)\nonumber\\
    +&\cos(2\Delta\theta)\big(\chi_{ij;ij}(\omega;\omega_1,\omega_k,\omega_l)+\chi_{ij;ji}(\omega;\omega_1,\omega_l,\omega_k)\big)\Big)\bigg]\equiv G(\omega_1,\omega_k,\omega_l,\theta)\equiv G_{\omega_k\omega_l}(\theta).
\end{align}
Notice that when the combinations $F_{23}(\theta)+F_{32}(\theta)$ and $G_{23}(\theta)+G_{32}(\theta)$ are taken in Eq.\ \eqref{diffsign}, $\chi_{ij;ij}$ and $\chi_{ij;ji}$ only appear through their symmetric combination,
\begin{align}\label{chiS}
    \chi^{\mathcal{S}}_{ij}(\omega;\omega_1,\omega_2,\omega_3)=\chi_{ij;ij}(\omega;\omega_1,\omega_2,\omega_3)+\chi_{ij;ji}(\omega;\omega_1,\omega_2,\omega_3).
\end{align}
Moreover, for $\theta=0$ or $\theta=\pi/2$, i.e., with a probe aligned with one of the crystallographic axes, terms scaling with $\chi_{ii;ii}$ and $\chi_{ii;jj}$ vanish, leaving only the response mediated by $\chi^{\mathcal{S}}_{ij}$, as it is considered in the main text.\\
The coefficient of the antisymmetric combination, on the other hand, reads explicitly
\begin{align}\label{Hkl}
    H_{kl}=H(\omega_1,\omega_k,\omega_l)=\frac{1}{2}\big(\chi_{ij;ij}(\omega;\omega_1,\omega_k,\omega_l)+\chi_{ij;ji}(\omega;\omega_1,\omega_l,\omega_k)\big)=\chi_{ij;ij}(\omega;\omega_1,\omega_k,\omega_l),
\end{align}
with no dependence on the angle $\theta$. The combination appearing in Eq.\ \eqref{diffsign} can then be rewritten immediately as
\begin{align}\label{Hkldiff}
H_{23}-H_{32}=\chi_{ij;ij}(\omega;\omega_1,\omega_2,\omega_3)-\chi_{ij;ij}(\omega;\omega_1,\omega_3,\omega_2),
\end{align}
which is explicitly antisymmetric under the exchange $\omega_2\leftrightarrow\omega_3$. Using $\chi_{ij;ij}(\omega;\omega_1,\omega_3,\omega_2)=\chi_{ij;ji}(\omega;\omega_1,\omega_2,\omega_3)$, such frequency antisymmetry can be equivalently recast as an antisymmetry under the exchange of the Cartesian pump indices, leading to
\begin{align}\label{chiAS}
    H_{23}-H_{32}=\chi^{\mathcal{A}}_{ij}(\omega;\omega_1,\omega_2,\omega_3)=\chi_{ij;ij}(\omega;\omega_1,\omega_2,\omega_3)-\chi_{ij;ji}(\omega;\omega_1,\omega_2,\omega_3).
\end{align}
Both Eqs.\ \eqref{Hkldiff} and \eqref{chiAS} make it evident that the antisymmetric response is lost when Kleinman symmetry is enforced, corresponding to taking the static limit of the susceptibility in which $\chi_{ij;ij}(0;0,0,0)=\chi_{ij;ji}(0;0,0,0)$. \\
Eq.\ \eqref{diffsign} represents the differential intensity $\Delta\Gamma(t_{pp},\theta)$ measured for generic spectral content of the pump and probe fields, and for a generic polarization of the pump. We note that this expression assumes that all possible generated sidebands of the probe are collected by the photodetectors. A realistic detection setup has a limited spectral range, which can be taken into account with an appropriate function $R(\omega)$ when performing the integral over $\omega$.
\clearpage
\section*{Pump-probe signal under monochromatic pump pulses}
To further analyze the spectral response of the pump-probe experiment, we consider monochromatic fields at frequency $\Omega$,
\begin{align}\label{monfields}
    \text{E}_\text{pu}^P(\omega_s)&=\mathcal{E}_P\,\delta(\omega_s-\Omega)+\mathcal{E}_P^*\,\delta(\omega_s+\Omega),\nonumber\\
    \text{E}_\text{pu}^S(\omega_s)&=\mathcal{E}_S\,\delta(\omega_s-\Omega)+\mathcal{E}_S^*\,\delta(\omega_s+\Omega),
\end{align}
where $\mathcal{E}_{P,S}$ are the complex field amplitudes. By using Eq.\ \eqref{monfields} in Eq.\ \eqref{fieldcomb}, we then find
\begin{align}\label{fieldcomblin1}
    \mathcal{S}_\parallel&=\big(\mathcal{E}_S^2-\mathcal{E}_P^2\big)\delta(\omega_2-\Omega)\delta(\omega_3-\Omega)+\big(\mathcal{E}_S^{*2}-\mathcal{E}_P^{*2}\big)\delta(\omega_2+\Omega)\delta(\omega_3+\Omega)\nonumber\\
    &+\big(|\mathcal{E}_S|^2-|\mathcal{E}_P|^2\big)\big(\delta(\omega_2-\Omega)\delta(\omega_3+\Omega)+\delta(\omega_2+\Omega)\delta(\omega_3-\Omega)\big),
\end{align}
and
\begin{align}\label{fieldcomblin2}
    \mathcal{S}_\perp&=2\mathcal{E}_P\mathcal{E}_S\,\delta(\omega_2-\Omega)\delta(\omega_3-\Omega)+2\mathcal{E}_P^*\mathcal{E}_S^*\,\delta(\omega_2+\Omega)\delta(\omega_3+\Omega)\nonumber\\
    &+\big(\mathcal{E}_P\mathcal{E}_S^*+\mathcal{E}_P^*\mathcal{E}_S\big)\big(\delta(\omega_2-\Omega)\delta(\omega_3+\Omega)+\delta(\omega_2+\Omega)\delta(\omega_3-\Omega)\big),
\end{align}
for the symmetric combinations of the fields, and
\begin{align}\label{fieldcomblin3}
    \mathcal{A}&=\big(\boldsymbol{\mathcal{E}\times\boldsymbol{\mathcal{E}}^*}\big)_z\big(\delta(\omega_2-\Omega)\delta(\omega_3+\Omega)-\delta(\omega_2+\Omega)\delta(\omega_3-\Omega)\big),
\end{align}
for the antisymmetric combination. Here, $\boldsymbol{\mathcal{E}}=(\mathcal{E}_P,\mathcal{E}_S)$, and $(\boldsymbol{\mathcal{E}\times\boldsymbol{\mathcal{E}}^*})_z=\mathcal{E}_P\mathcal{E}_S^*-\mathcal{E}_P^*\mathcal{E}_S$.
\subsection*{Linearly polarized light}
For linearly polarized light, the $P$ and $S$ components of the electric field are in phase, such that $\mathcal{E}_P\mathcal{E}_S^*=\mathcal{E}_P^*\mathcal{E}_S$. As a consequence, the cross product $\boldsymbol{\mathcal{E}\times\boldsymbol{\mathcal{E}}}^*$ vanishes, and the antisymmetric contribution is identically zero.
Using Eqs.\ \eqref{fieldcomblin1} and \eqref{fieldcomblin2} into the differential signal $\Delta\Gamma(t_{pp},\theta)$ in Eq.\ \eqref{diffsign} one obtains the response to linearly polarized pump pulses, which exhibits oscillations at $\omega_2+\omega_3=\pm2\Omega$ and $\omega_2+\omega_3=0$. In Fourier space, $\Delta\Gamma(t_{pp},\theta)\to\Delta\bar\Gamma(\omega_{pp},\theta)$, we find for the three spectral components
\begin{align}\label{lin2Om}
    \Delta\bar\Gamma(2\Omega,\theta)\propto2\int d\omega_1\text{E}_\text{pr}(\omega_1)\bigg[F_{\Omega\Omega}(\theta)\big(\mathcal{E}_S^2-\mathcal{E}_P^2\big)+2G_{\Omega\Omega}(\theta)\mathcal{E}_P\mathcal{E}_S\bigg],
\end{align}
\begin{align}\label{lin-2Om}
    \Delta\bar\Gamma(-2\Omega,\theta)\propto2\int d\omega_1\text{E}_\text{pr}(\omega_1)\bigg[F_{-\Omega-\Omega}(\theta)\big(\mathcal{E}_S^{*2}-\mathcal{E}_P^{*2}\big)+2G_{-\Omega-\Omega}(\theta)\mathcal{E}_P^*\mathcal{E}_S^*\bigg],
\end{align}
\begin{align}\label{lin0}
    \Delta\bar\Gamma(0,\theta)\propto2\int d\omega_1\text{E}_\text{pr}(\omega_1)\bigg[&\big(F_{\Omega-\Omega}(\theta)+F_{-\Omega\Omega}(\theta)\big)\big(|\mathcal{E}_S|^2-|\mathcal{E}_P|^2\big)+\big(G_{\Omega-\Omega}(\theta)+G_{-\Omega\Omega}(\theta)\big)\big(\mathcal{E}_P\mathcal{E}_S^*+\mathcal{E}_P^*\mathcal{E}_S\big)\bigg].
\end{align}
Because only the symmetric combinations contribute to the pump-probe signal, the same result could be obtained enforcing Kleinman symmetry, as it was done in Ref.\ \cite{basini_prb24}. 
These three contributions contain all the classical third-order nonlinear effects, in the second-harmonic signal described by Eqs.\ \eqref{lin2Om} and \eqref{lin-2Om}, as well as in the optical rectification channel described by Eq.\ \eqref{lin0}. 
Also we note that all spectral components of the signal have fourfold angular dependence, coming directly from $F_{kl}(\theta)$ and $G_{kl}(\theta)$ in Eqs.\ \eqref{Fkl} and \eqref{Gkl} \cite{basini_prb24}. For the particular case of $\theta=0$, one has
\begin{align}\label{lin2Om2}
    \Delta\bar\Gamma(2\Omega,0)\propto\int d\omega_1 \chi^\mathcal{S}_{xy}(\omega_1+2\Omega;\omega_1,\Omega,\Omega)\text{E}_\text{pr}(\omega_1)\bigg[\sin(2\Delta\theta)\big(\mathcal{E}_S^2-\mathcal{E}_P^2\big)+2\cos(2\Delta\theta)\mathcal{E}_P\mathcal{E}_S\bigg],
\end{align}
\begin{align}\label{lin-2Om2}
    \Delta\bar\Gamma(-2\Omega,0)\propto\int d\omega_1\chi^\mathcal{S}_{xy}(\omega_1-2\Omega;\omega_1,-\Omega,-\Omega)\text{E}_\text{pr}(\omega_1)\bigg[\sin(2\Delta\theta)\big(\mathcal{E}_S^{*2}-\mathcal{E}_P^{*2}\big)+2\cos(2\Delta\theta)\mathcal{E}_P^*\mathcal{E}_S^*\bigg],
\end{align}
\begin{align}\label{lin02}
    \Delta\bar\Gamma(0,0)\propto2\int d\omega_1\chi^\mathcal{S}_{xy}(\omega_1;\omega_1,\Omega,-\Omega)\text{E}_\text{pr}(\omega_1)\bigg[&\sin(2\Delta\theta)\big(|\mathcal{E}_S|^2-|\mathcal{E}_P|^2\big)+\cos(2\Delta\theta)\big(\mathcal{E}_P\mathcal{E}_S^*+\mathcal{E}_P^*\mathcal{E}_S\big)\bigg].
\end{align}
\subsection*{Circularly polarized light}
For circularly polarized light it is convenient to recast the complex field amplitudes $\mathcal{E}_{P,S}$ in the circular basis $\mathcal{E}_{L,R}$ as
\begin{align}\label{lintocirc}
    \mathcal{E}_L&=\frac{1}{\sqrt{2}}(\mathcal{E}_P+i\mathcal{E}_S),\nonumber\\
    \mathcal{E}_R&=\frac{1}{\sqrt{2}}(\mathcal{E}_P-i\mathcal{E}_S).
\end{align}
With such a change of variables, we can rewrite Eqs.\ \eqref{fieldcomblin1}, \eqref{fieldcomblin2} and \eqref{fieldcomblin3} as
\begin{align}\label{fieldcombcirc1}
    \mathcal{S}_\parallel=&-\big(\mathcal{E}_L^2+\mathcal{E}_R^2\big)\delta(\omega_2-\Omega)\delta(\omega_3-\Omega)-\big(\mathcal{E}_L^{*2}+\mathcal{E}_R^{*2}\big)\delta(\omega_2+\Omega)\delta(\omega_3+\Omega)\nonumber\\
    &-\big(\mathcal{E}_L\mathcal{E}_R^*+\mathcal{E}_L^*\mathcal{E}_R\big)\big(\delta(\omega_2-\Omega)\delta(\omega_3+\Omega)+\delta(\omega_2+\Omega)\delta(\omega_3-\Omega)\big),
\end{align}
\begin{align}\label{fieldcombcirc2}
    \mathcal{S}_\perp=&-i(\mathcal{E}_L^2-\mathcal{E}_R^2)\,\delta(\omega_2-\Omega)\delta(\omega_3-\Omega)+i(\mathcal{E}_L^{*2}-\mathcal{E}_R^{*2})\,\delta(\omega_2+\Omega)\delta(\omega_3+\Omega)\nonumber\\
    &-i\big(\mathcal{E}_L\mathcal{E}_R^*-\mathcal{E}_L^*\mathcal{E}_R\big)\big(\delta(\omega_2-\Omega)\delta(\omega_3+\Omega)+\delta(\omega_2+\Omega)\delta(\omega_3-\Omega)\big),
\end{align}
and
\begin{align}\label{fieldcombcirc3}
    \mathcal{A}&=i\big(|\mathcal{E}_L|^2-|\mathcal{E}_R|^2\big)\big(\delta(\omega_2-\Omega)\delta(\omega_3+\Omega)-\delta(\omega_2+\Omega)\delta(\omega_3-\Omega)\big).
\end{align}
In the following and in the main text we neglect terms in $\mathcal{E}_L\mathcal{E}_R^*$ or $\mathcal{E}_L^*\mathcal{E}_R$, considering an experiment performed either with purely left-handed or right-handed circular light. As a consequence, one can immediately see that the symmetric terms only contribute to the response in $\pm 2\Omega$, while the antisymmetric one only contributes to the static response. Using Eqs.\ \eqref{fieldcombcirc1}, \eqref{fieldcombcirc2} and \eqref{fieldcombcirc3} in Eq.\ \eqref{diffsign}, one explicitly finds for the three spectral components
\begin{align}\label{circ2Om}
\Delta\bar\Gamma(2\Omega,\theta)\propto -2 \int d\omega_1 \text{E}_\text{pr}(\omega_1)\bigg[F_{\Omega\Omega}(\theta)(\mathcal{E}_L^2+\mathcal{E}_R^2)+iG_{\Omega\Omega}(\theta)(\mathcal{E}_L^2-\mathcal{E}_R^2)\bigg],
\end{align}
which reduces to Eq.\ (7) of the main text for $\theta=0$ and monochromatic probe field $\text{E}_\text{pr}(\omega_1)\propto\delta(\omega_1-\omega)+\delta(\omega_1+\omega)$, 
\begin{align}\label{circ-2Om}
\Delta\bar\Gamma(-2\Omega,\theta)\propto -2 \int d\omega_1 \text{E}_\text{pr}(\omega_1)\bigg[F_{-\Omega-\Omega}(\theta)(\mathcal{E}_L^{*2}+\mathcal{E}_R^{*2})-iG_{-\Omega-\Omega}(\theta)(\mathcal{E}_L^{*2}-\mathcal{E}_R^{*2})\bigg],
\end{align}
and
\begin{align}\label{circ0}
\Delta\bar\Gamma(0)\propto 2\int d\omega_1 \text{E}_\text{pr}(\omega_1)\bigg[i\chi_{ij}^\mathcal{A}(\omega_1;\omega_1,\Omega,-\Omega)\big(|\mathcal{E}_L|^{2}-|\mathcal{E}_R|^{2}\big)\bigg],
\end{align}
which corresponds to Eq.\ (8) of the main text for monochromatic probe. We note that the whole static response does not display any angular dependence, as it is given only by the antisymmetric term.
\\
Contributions scaling with $\mathcal{E}_{L}^2+\mathcal{E}_R^2$ and $\mathcal{E}_{L}^{*2}+\mathcal{E}_R^{*2}$ are canceled in the dichroic signal $\Delta\bar\Gamma_{R-L}=\frac{1}{2}(\Delta\bar\Gamma_R-\Delta\bar\Gamma_L)$, defined as the difference between pump-probe response to right and left polarized light. In particular, with equivalent pump field amplitude $\mathcal{E}_\text{pu}$ for both polarizations, we find 
\begin{align}\label{circ2Om2}
&\Delta\bar\Gamma_{R-L}(2\Omega,\theta)\propto 2i \int d\omega_1 \text{E}_\text{pr}(\omega_1)G_{\Omega\Omega}(\theta)\mathcal{E}_\text{pu}^2,\\
&\Delta\bar\Gamma_{R-L}(-2\Omega,\theta)\propto -2i \int d\omega_1 \text{E}_\text{pr}(\omega_1)G_{-\Omega-\Omega}(\theta)\mathcal{E}_\text{pu}^{*2},\\
&\Delta\bar\Gamma_{R-L}(0)\propto -2i\int d\omega_1 \text{E}_\text{pr}(\omega_1)\chi_{ij}^\mathcal{A}(\omega;\omega_1,\Omega,-\Omega)\big|\mathcal{E}_\text{pu}|^{2}.
\end{align}

We report in Table \ref{tab:monochr} the weights of the different spectral components, as obtained throughout this section, for $\theta=0$ and monochromatic probe.
\begin{table}[h]
\centering
\begin{tabular}{|l|c|c|}
\hline
    & Linearly polarized pump & Circularly polarized pump \\
    \hline
    $\omega_2+\omega_3=2\Omega$ 
    & $\chi^\mathcal{S}_{xy}[\sin(2\Delta\theta)(\mathcal{E}_S^2-\mathcal{E}_P^2)+2\cos(2\Delta\theta)\mathcal{E}_P\mathcal{E}_S]$
    & $\chi^\mathcal{S}_{xy}[\sin(2\Delta\theta)(\mathcal{E}_L^2+\mathcal{E}_R^2)+i\cos(2\Delta\theta)(\mathcal{E}_L^2-\mathcal{E}_R^2)]$ \\[2mm]
    $\omega_2+\omega_3=-2\Omega$
    & $\chi^\mathcal{S}_{xy}[\sin(2\Delta\theta)(\mathcal{E}_S^{*2}-\mathcal{E}_P^{*2})+2\cos(2\Delta\theta)\mathcal{E}_P^*\mathcal{E}_S^*]$
    & $\chi^\mathcal{S}_{xy}[\sin(2\Delta\theta)(\mathcal{E}_L^{*2}+\mathcal{E}_R^{*2})-i\cos(2\Delta\theta)(\mathcal{E}_L^{*2}-\mathcal{E}_R^{*2})]$\\[2mm]
    $\omega_2+\omega_3=0$
    & $\chi^\mathcal{S}_{xy}[\sin(2\Delta\theta)(|\mathcal{E}_S|^2-|\mathcal{E}_P|^2)+\cos(2\Delta\theta)(\mathcal{E}_P\mathcal{E}_S^*+\mathcal{E}_P^*\mathcal{E}_S)]$
    & $i\chi^\mathcal{A}_{xy}(|\mathcal{E}_L|^2-|\mathcal{E}_R|^2)$\\
    \hline
\end{tabular}
\caption{Schematic recap of the field combinations that contribute to the pump-probe response components for $\theta=0$.}
\label{tab:monochr}
\end{table}

\clearpage
\section*{Response to monochromatic pump pulses with small frequency mismatch}
As reported in Ref.\ \cite{basini_prb24}, when a frequency mismatch $\Delta\Omega$ between the two pump components $\text{E}_\text{pu}^P(\omega_s)$ and $\text{E}_\text{pu}^S(\omega_s)$ is present, a quasi-static response at $\Delta\Omega$ arises in the symmetric channel through difference-frequency generation of the circularly polarized pumps. Here we reproduce this result, and further show that the antisymmetric channel contributes exclusively to the quasi-static response for small mismatch $\Delta\Omega\ll\Omega$.\\
We again consider monochromatic fields, where now the $P$ component is shifted by $-\Delta\Omega$,
\begin{align}\label{monfieldsmism}
    \text{E}_\text{pu}^P(\omega_s)&=\mathcal{E}_P\,\delta(\omega_s-\Omega+\Delta\Omega)+\mathcal{E}_P^*\,\delta(\omega_s+\Omega-\Delta\Omega),\nonumber\\
    \text{E}_\text{pu}^S(\omega_s)&=\mathcal{E}_S\,\delta(\omega_s-\Omega)+\mathcal{E}_S^*\,\delta(\omega_s+\Omega),
\end{align}
and then change basis with Eqs.\ \eqref{lintocirc}.
As above, in the following we neglect terms that mix left and right pump components. With the frequency mismatch, Eqs.\ \eqref{fieldcombcirc1}, \eqref{fieldcombcirc2} and \eqref{fieldcombcirc3} are rewritten as 
\begin{align}\label{fieldcombmism1}
    \mathcal{S}_\parallel=-\frac{1}{2}&\big(\mathcal{E}_L^2+\mathcal{E}_R^2\big)\big[\delta(\omega_2-\Omega)\delta(\omega_3-\Omega)+\delta(\omega_2-\Omega+\Delta\Omega)\delta(\omega_3-\Omega+\Delta\Omega)\big]\nonumber\\
    -\frac{1}{2}&\big(\mathcal{E}_L^{*2}+\mathcal{E}_R^{*2}\big)\big[\delta(\omega_2+\Omega)\delta(\omega_3+\Omega)+\delta(\omega_2+\Omega-\Delta\Omega)\delta(\omega_3+\Omega-\Delta\Omega)\big]\nonumber\\
    -\frac{1}{2}&\big(|\mathcal{E}_L|^2+|\mathcal{E}_R|^2\big)\big[\delta(\omega_2-\Omega)\delta(\omega_3+\Omega)+\delta(\omega_2+\Omega)\delta(\omega_3-\Omega)\nonumber\\
    &-\delta(\omega_2-\Omega+\Delta\Omega)\delta(\omega_3+\Omega-\Delta\Omega)-\delta(\omega_2+\Omega-\Delta\Omega)\delta(\omega_3-\Omega+\Delta\Omega)\big],
\end{align}
\begin{align}\label{fieldcombmism2}
    \mathcal{S}_\perp=-\frac{i}{2}&(\mathcal{E}_L^2-\mathcal{E}_R^2)\big[\delta(\omega_2-\Omega+\Delta\Omega)\delta(\omega_3-\Omega)+\delta(\omega_2-\Omega)\delta(\omega_3-\Omega+\Delta\Omega)\big]\nonumber\\
    +\frac{i}{2}&(\mathcal{E}_L^{*2}-\mathcal{E}_R^{*2})\,\big[\delta(\omega_2+\Omega-\Delta\Omega)\delta(\omega_3+\Omega)+\delta(\omega_2+\Omega)\delta(\omega_3+\Omega-\Delta\Omega)\big]\nonumber\\
    +\frac{i}{2}&\big(|\mathcal{E}_L|^2-|\mathcal{E}_R|^2\big)\big[\delta(\omega_2-\Omega+\Delta\Omega)\delta(\omega_3+\Omega)+\delta(\omega_2+\Omega)\delta(\omega_3-\Omega+\Delta\Omega)\nonumber\\
    &-\delta(\omega_2-\Omega)\delta(\omega_3+\Omega-\Delta\Omega)-\delta(\omega_2+\Omega-\Delta\Omega)\delta(\omega_3-\Omega)\big],
\end{align}
and
\begin{align}\label{fieldcombmism3}
    \mathcal{A}=-\frac{i}{2}&\big(\mathcal{E}_L^2-\mathcal{E}_R^2\big)\big[\delta(\omega_2-\Omega+\Delta\Omega)\delta(\omega_3-\Omega)-\delta(\omega_2-\Omega)\delta(\omega_3-\Omega+\Delta\Omega)\big]\nonumber\\
    +\frac{i}{2}&\big(\mathcal{E}_L^{*2}-\mathcal{E}_R^{*2}\big)\big[\delta(\omega_2+\Omega-\Delta\Omega)\delta(\omega_3+\Omega)-\delta(\omega_2+\Omega)\delta(\omega_3+\Omega-\Delta\Omega)\big]\nonumber\\
    +\frac{i}{2}&\big(|\mathcal{E}_L|^2-|\mathcal{E}_R|^2\big)\big[\delta(\omega_2-\Omega+\Delta\Omega)\delta(\omega_3+\Omega)+\delta(\omega_2-\Omega)\delta(\omega_3+\Omega-\Delta\Omega)\nonumber\\
    &-\delta(\omega_2+\Omega)\delta(\omega_3-\Omega+\Delta\Omega)-\delta(\omega_2+\Omega-\Delta\Omega)\delta(\omega_3-\Omega)\big].
\end{align}
We now assume that the small frequency mismatch $\Delta\Omega\ll\Omega$ between the pump components does not significantly affect the third-order susceptibility, so that its dependence on $\Delta\Omega$ can be neglected, e.g., $\chi_{ij;kl}(\omega;\omega_1,\Omega,\Omega-\Delta\Omega)\simeq\chi_{ij;kl}(\omega;\omega_1,\Omega,\Omega)$ and analogously for all other combinations. Using Eqs.\ \eqref{fieldcombmism1}, \eqref{fieldcombmism2} and \eqref{fieldcombmism3} in Eq.\ \eqref{diffsign} we find
\begin{align}\label{diffsignmism}
    \Delta\Gamma(t_{pp},\theta)\propto\int d\omega_1&\text{E}_\text{pr}(\omega_1)\nonumber\\
    \times\bigg[&-F_{\Omega\Omega}(\theta)e^{-2i\Omega t_{pp}}\big(\mathcal{E}_L^2+\mathcal{E}_R^2\big)\big(1+e^{2i\Delta\Omega t_{pp}}\big)-F_{-\Omega-\Omega}(\theta)e^{2i\Omega t_{pp}}\big(\mathcal{E}_L^{*2}+\mathcal{E}_R^{*2}\big)\big(1+e^{-2i\Delta\Omega t_{pp}}\big)\nonumber\\
    &-2i G_{\Omega\Omega}(\theta)\big(\mathcal{E}_L^2-\mathcal{E}_R^2\big)e^{-i(2\Omega-\Delta\Omega)t_{pp}}+2i G_{-\Omega-\Omega}(\theta)\big(\mathcal{E}_L^{*2}-\mathcal{E}_R^{*2}\big)e^{i(2\Omega-\Delta\Omega)t_{pp}}\nonumber\\
    &-2\big(|\mathcal{E}_L|^2-|\mathcal{E}_R|^2\big)\big(G_{\Omega-\Omega}(\theta)+G_{-\Omega\Omega}(\theta)\big)\sin(\Delta\Omega\,t_{pp})\nonumber\\
    &+2i\big(|\mathcal{E}_L|^2-|\mathcal{E}_R|^2\big)\chi_{ij}^{\mathcal{A}}(\omega;\omega_1,\Omega,-\Omega)\cos(\Delta\Omega\,t_{pp})\bigg].
\end{align}
As discussed above, contributions scaling with $\mathcal{E}_{L}^2+\mathcal{E}_R^2$ and $\mathcal{E}_{L}^{*2}+\mathcal{E}_R^{*2}$ are canceled in the dichroic signal $\Delta\bar\Gamma_{R-L}$, so that only oscillations at $\pm(2\Omega-\Delta\Omega)$ and $\pm\Delta\Omega$ are present. In particular, in Fourier space,
\begin{align}
    &\Delta\bar\Gamma_{R-L}(2\Omega-\Delta\Omega,\theta)\propto 2i\int d\omega_1 \text{E}_\text{pr}(\omega_1)G_{\Omega\Omega}(\theta)\mathcal{E}_\text{pu}^2\\
    &\Delta\bar\Gamma_{R-L}(-2\Omega+\Delta\Omega,\theta)\propto-2i\int d\omega_1\text{E}_\text{pr}(\omega_1)G_{-\Omega-\Omega}(\theta)\mathcal{E}^{*2}_\text{pu}\\
    &\Delta\bar\Gamma_{R-L}(\Delta\Omega,\theta)\propto-i\int d\omega_1\text{E}_\text{pr}(\omega_1)\big[G_{\Omega-\Omega}(\theta)+G_{-\Omega\Omega}(\theta)+\chi^\mathcal{A}_{ij}(\omega;\omega_1,\Omega,-\Omega)\big]|\mathcal{E}_\text{pu}|^2\label{quasi-static}\\
    &\Delta\bar\Gamma_{R-L}(-\Delta\Omega,\theta)\propto i\int d\omega_1\text{E}_\text{pr}(\omega_1)\big[G_{\Omega-\Omega}(\theta)+G_{-\Omega\Omega}(\theta)-\chi^\mathcal{A}_{ij}(\omega;\omega_1,\Omega,-\Omega)\big]|\mathcal{E}_\text{pu}|^2.
\end{align}
One can immediately see that a symmetric contribution to the quasi-static response, scaling with the coefficients $G_{kl}(\theta)$ in Eq.\ \eqref{Gkl}, appears for finite $\Delta\Omega$. In particular, for $\theta=0$, $G_{\Omega-\Omega}(0)+G_{-\Omega\Omega}(0)=\cos(2\Delta\theta)\chi^\mathcal{S}_{ij}(\omega;\omega_1,\Omega,-\Omega)$, and one finds for the positive quasi-static component Eq.\ \eqref{quasi-static}
\begin{align}
    \Delta\bar\Gamma_{R-L}(\Delta\Omega,0)\propto-i\int d\omega_1\text{E}_\text{pr}(\omega_1)\big[\cos(2\Delta\theta)\,\chi^\mathcal{S}_{ij}(\omega;\omega_1,\Omega,-\Omega)+\chi_{ij}^\mathcal{A}(\omega;\omega,\Omega,-\Omega)\big]|\mathcal{E}_\text{pu}|^2.
\end{align}
On the other hand, the antisymmetric susceptibility does not contribute to the $\pm(2\Omega-\Delta\Omega)$ response. This remains valid as long as the susceptibility is approximately insensitive to the small frequency mismatch.
\clearpage
\section*{Details of the calculations on the $sp$-model}
\subsection*{Electronic Hamiltonian}
For an explicit calculation of the antisymmetric susceptibility, we consider a simple square lattice with two atoms per unit cell, labeled A and B in the following. To simulate a system with distinct orbital character of the valence and the conduction band, we assign $s$-type orbitals to site A, with on-site energy $-\Delta$, and $p_{x}$ and $p_y$ orbitals to site B, with on-site energy $+\Delta$. Finally, we include nearest-neighbor hopping with amplitude $t$. The system is shown in Fig. 2(a) of the main text.\\
We define $a(\textbf{R}_i)$ the annihilation operator on site A of the $i$-th unit cell, and $b_{x,y}(\textbf{R}_i+\hat x)$ the annihilation operator on site B of the same unit cell for orbitals $p_x$ and $p_y$. Here we set the nearest-neighbor distance as 1.
The Hamiltonian associated with this toy-model is, in real space,
\begin{align}
    H=&-\Delta\sum_i [a^\dagger(\textbf{R}_i) a(\textbf{R}_i)]+\Delta\sum_{i}[b_x^\dagger(\textbf{R}_i+\hat x) b_x(\textbf{R}_i+\hat x)+b_y^\dagger(\textbf{R}_i+\hat x) b_y(\textbf{R}_i+\hat x)]\nonumber\\
    &-t\sum_i[a^\dagger(\textbf{R}_i)b_x(\textbf{R}_i+\hat x)+b^\dagger_x(\textbf{R}_i+\hat x)a(\textbf{R}_i)]-t\sum_i[a^\dagger(\textbf{R}_i)b_y(\textbf{R}_i+\hat y)+b_y^\dagger(\textbf{R}_i+\hat y)a(\textbf{R}_i)]\nonumber\\
    &+t\sum_i[a^\dagger(\textbf{R}_i)b_x(\textbf{R}_i-\hat x)+b^\dagger_x(\textbf{R}_i-\hat x)a(\textbf{R}_i)]+t\sum_i[a^\dagger(\textbf{R}_i)b_y(\textbf{R}_i-\hat y)+b_y^\dagger(\textbf{R}_i-\hat y)a(\textbf{R}_i)].
\end{align}
Shifting to Fourier space, the Hamiltonian can be rewritten as 
\begin{align}\label{tbham2}
    H=\sum_{\textbf{k}}
    \begin{pmatrix}
        a_\textbf{k}^\dagger & b_{x,\textbf{k}}^\dagger & b_{y,\textbf{k}}^\dagger
    \end{pmatrix}\mathcal{H}_\textbf{k}\begin{pmatrix}
        a_\textbf{k} \\ b_{x,\textbf{k}} \\ b_{y,\textbf{k}}
    \end{pmatrix},
\end{align}
where
\begin{align}\label{tbham}
    \mathcal{H}_\textbf{k}=\begin{pmatrix}
        -\Delta & -2it\sin(\text{k}_x) & -2it\sin(\text{k}_y) \\
        2it\sin(\text{k}_x) & +\Delta & 0 \\
        2it\sin(\text{k}_y) & 0 & +\Delta
    \end{pmatrix}.
\end{align}
The tight-binding Hamiltonian describes three bands,
\begin{align}
    E_{1,2}(\textbf{k})&=\mp\sqrt{\Delta^2+4t^2\big(\sin^2(\text{k}_x)+\sin^2(\text{k}_y)\big)}\equiv\mp E_\textbf{k},\nonumber\\
    E_3(\textbf{k})&=\Delta.
\end{align}
Band 1 displays dominant $s$-like character, whereas bands 2 and 3 display dominant $p$-like character.\\ 
We can then define the electronic Green's function as $\tilde G_0^{-1}(i\nu_n)=i\nu_n\mathbb1-\mathcal{H}_\textbf{k}$, where $\mathbb1$ is the $3\times3$ identity matrix and $\nu_n=(2n+1)\pi T$ is the fermionic Matsubara frequency at temperature $T$. Explicitly,
\begin{align}\label{greenorb}
    \tilde G_0(i\nu_n,\textbf{k})=\frac{1}{(i\nu_n-E_1(\textbf{k}))(i\nu_n-E_2(\textbf{k}))(i\nu_n-E_3(\textbf{k}))}\begin{pmatrix}
        (i\nu_n-\Delta)^2 & (i\nu_n-\Delta)\epsilon_{\text{k}_x} & (i\nu_n-\Delta)\epsilon_{\text{k}_y} \\
        (i\nu_n-\Delta)\epsilon_{\text{k}_x}^* & (i\nu_n)^2-\Delta^2-|\epsilon_{\text{k}_y}|^2 & \epsilon_{\text{k}_x}^*\epsilon_{\text{k}_y} \\
        (i\nu_n-\Delta)\epsilon_{\text{k}_y}^* & \epsilon_{\text{k}_x}\epsilon_{\text{k}_y}^* & (i\nu_n)^2-\Delta^2-|\epsilon_{\text{k}_x}|^2
        \end{pmatrix},
\end{align}
where we defined $\epsilon_{\text{k}_x}=-2it\sin(\text{k}_x)$ and $\epsilon_{\text{k}_y}=-2it\sin(\text{k}_y)$.\\
To represent the Green's function in the band basis in which it is diagonal, we perform the Bogoliubov rotation $\tilde G_0(i\nu_n,\textbf{k})=U_\textbf{k}G_0(i\nu_n,\textbf{k})U_\textbf{k}^\dagger$, and obtain
\begin{align}
    G_0(i\nu_n,\textbf{k})=\begin{pmatrix}
        \frac{1}{i\nu_n-E_1(\textbf{k})} & 0 & 0 \\
        0 & \frac{1}{i\nu_n-E_2(\textbf{k})} & 0 \\
        0 & 0 & \frac{1}{i\nu_n-E_3(\textbf{k})}
    \end{pmatrix}.
\end{align}
The matrix that performs the rotation reads explicitly
\begin{align}
    U_\textbf{k}=\begin{pmatrix}
        u_\textbf{k} & v_\textbf{k} & 0 \\
        \frac{\epsilon_{\text{k}_x}}{2E_\textbf{k}u_\textbf{k}} & \frac{\epsilon_{\text{k}_x}^*}{2E_\textbf{k}v_\textbf{k}} & \frac{\epsilon_{\text{k}_y}^*}{2E_\textbf{k}u_\textbf{k}v_\textbf{k}} \\
        \frac{\epsilon_{\text{k}_y}}{2E_\textbf{k}u_\textbf{k}} & \frac{\epsilon_{\text{k}_y}^*}{2E_\textbf{k}v_\textbf{k}} & \frac{\epsilon_{\text{k}_x}}{2E_\textbf{k}u_\textbf{k}v_{\textbf{k}}}
    \end{pmatrix},
\end{align}
where $u_\textbf{k}$ and $v_\textbf{k}$ are the coherence factors with properties $u_\textbf{k}^2+v_\textbf{k}^2=1$, $u_\textbf{k}^2-v_\textbf{k}^2=\Delta/E_\textbf{k}$, and $2u_\textbf{k}v_\textbf{k}=\sqrt{|\epsilon_{\text{k}_x}|^2+|\epsilon_{\text{k}_y}|^2}/E_\textbf{k}$.
\subsection*{Light-matter interaction}
To introduce the electromagnetic field, represented by the vector potential $\textbf{A}$, we perform the Peierls substitution $\textbf{k}\to\textbf{k}+\frac{e}{c}\textbf{A}$ in the tight-binding Hamiltonian $\mathcal{H}_\textbf{k}$ in Eq.\ \eqref{tbham}, with $-e$ the electron charge and $c$ the light velocity. 
We then expand the resulting Hamiltonian in powers of $\textbf{A}$, to generate all orders of light-matter interactions:
\begin{align}\label{intham}
    \mathcal{H}_{\textbf{k}+\frac{e}{c}\textbf{A}}\simeq \mathcal{H}_\textbf{k}+\sum_n \frac{1}{n!}\frac{\partial^n\mathcal{H}_\textbf{k}}{\partial\textbf{k}^n}\cdot\left(\frac{e}{c}\textbf{A}\right)^n.
\end{align}
Since we are interested in a third-order nonlinear response, relevant vertices are generated up to $n=4$, which corresponds to taking 1-, 2-, 3- or 4-photons electronic transitions. For further simplification, we focus on the mixed components $\chi_{xy;xy}$ and $\chi_{xy;yx}$, relevant for the antisymmetric and symmetric susceptibilities, which means selecting processes involving two photons polarized along $x$ and two along $y$. Since mixed derivatives of $\mathcal{H}_\textbf{k}$ vanish in the $sp$-model by Eq.\ \eqref{tbham}, only one-photon (``paramagnetic''-like) terms and two-photons (``diamagnetic''-like) terms with photons of same polarization survive in the expansion. We note that to compute the $\chi_{xx;xx}$ component, relevant for the symmetric response at $\theta\neq0$, 3- and 4-photon vertices need to be included to preserve gauge invariance of the final result.
From Eq.\ \eqref{tbham} we can find, in the orbital basis, the velocity vertices
\begin{align}\label{veloorb}
    \tilde{\text{v}}_x=\frac{\partial\mathcal{H}_\textbf{k}}{\partial\text{k}_x}=\begin{pmatrix}
        0 & -2it\cos(\text{k}_x) & 0 \\
        2it\cos(\text{k}_x) & 0 & 0 \\
        0 & 0 & 0
    \end{pmatrix}\qquad \tilde{\text{v}}_y=\frac{\partial\mathcal{H}_\textbf{k}}{\partial\text{k}_y}=\begin{pmatrix}
        0 & 0 & -2it\cos(\text{k}_y) \\
        0 & 0 & 0 \\
        2it\cos(\text{k}_y) & 0 & 0
    \end{pmatrix}
\end{align}
that mediate paramagnetic interactions, and the density-like vertices
\begin{align}\label{densiorb}
    \tilde\rho_x=\frac{\partial^2\mathcal{H}_\textbf{k}}{\partial\text{k}_x^2}=\begin{pmatrix}
        0 & 2it\sin(\text{k}_x) & 0 \\
        -2it\sin(\text{k}_x) & 0 & 0 \\
        0 & 0 & 0
    \end{pmatrix}\qquad \tilde\rho_y=\frac{\partial^2\mathcal{H}_\textbf{k}}{\partial\text{k}_y^2}=\begin{pmatrix}
        0 & 0 & 2it\sin(\text{k}_y) \\
        0 & 0 & 0 \\
        -2it\sin(\text{k}_y) & 0 & 0
    \end{pmatrix}
\end{align}
that mediate diamagnetic interactions. We can then represent these vertices in the band basis as $\text{v}_i=U_\textbf{k}^\dagger\tilde{\text{v}}_iU_\textbf{k}$ and $\rho_i=U_\textbf{k}^\dagger\tilde{\rho}_iU_\textbf{k}$ respectively. More generally, since 
eqs.\ \eqref{veloorb} and \eqref{densiorb} have the same matrix structure, one can recast the $n$-th order derivatives of the tight-binding Hamiltonian as
\begin{align}
    U_\textbf{k}^\dagger \frac{\partial^n\mathcal{H}_\textbf{k}}{\partial \text{k}_x^n}U_\textbf{k}=\begin{pmatrix}
        \frac{\epsilon_{\text{k}_x}}{E_\textbf{k}} & -\frac{\epsilon_{\text{k}_x}}{E_\textbf{k}}\frac{\Delta}{\sqrt{|\epsilon_{\text{k}_x}|^2+|\epsilon_{\text{k}_y}|^2}} & -\frac{\epsilon_{\text{k}_y}}{2E_\textbf{k}v_\textbf{k}}\\
        -\frac{\epsilon_{\text{k}_x}}{E_\textbf{k}}\frac{\Delta}{\sqrt{|\epsilon_{\text{k}_x}|^2+|\epsilon_{\text{k}_y}|^2}} & -\frac{\epsilon_{\text{k}_x}}{E_\textbf{k}} & -\frac{\epsilon_{\text{k}_y}}{2E_\textbf{k}u_\textbf{k}}\\
        -\frac{\epsilon_{\text{k}_y}}{2E_\textbf{k}v_\textbf{k}} & -\frac{\epsilon_{\text{k}_y}}{2E_\textbf{k}u_\textbf{k}} & 0 \\
    \end{pmatrix}\frac{\partial^n\epsilon_{\text{k}_x}}{\partial\text{k}_x^n},
\end{align}
and
\begin{align}
    U_\textbf{k}^\dagger \frac{\partial^n\mathcal{H}_\textbf{k}}{\partial \text{k}_y^n}U_\textbf{k}=\begin{pmatrix}
        \frac{\epsilon_{\text{k}_y}}{E_\textbf{k}} & -\frac{\epsilon_{\text{k}_y}}{E_\textbf{k}}\frac{\Delta}{\sqrt{|\epsilon_{\text{k}_x}|^2+|\epsilon_{\text{k}_y}|^2}} & -\frac{\epsilon_{\text{k}_x}}{2E_\textbf{k}v_\textbf{k}}\\
        -\frac{\epsilon_{\text{k}_y}}{E_\textbf{k}}\frac{\Delta}{\sqrt{|\epsilon_{\text{k}_x}|^2+|\epsilon_{\text{k}_y}|^2}} & -\frac{\epsilon_{\text{k}_y}}{E_\textbf{k}} & -\frac{\epsilon_{\text{k}_x}}{2E_\textbf{k}u_\textbf{k}}\\
        -\frac{\epsilon_{\text{k}_x}}{2E_\textbf{k}v_\textbf{k}} & -\frac{\epsilon_{\text{k}_x}}{2E_\textbf{k}u_\textbf{k}} & 0 \\
    \end{pmatrix}\frac{\partial^n\epsilon_{\text{k}_y}}{\partial\text{k}_y^n}.
\end{align}

Having selected the relevant terms from the interaction Hamiltonian Eq.\ \eqref{intham}, we can employ a path-integral formalism in which the system is described by the partition function $\mathcal{Z}=\int \mathcal{D}[\boldsymbol{\psi}]\mathcal{D}[\textbf{A}]e^{-S[\boldsymbol{\psi},\textbf{A}]}$, where $\boldsymbol\psi$ is the electron spinor field defined by Eq.\ \eqref{tbham2}. The total light-matter action reads
\begin{align}\label{lmact}
    S[\boldsymbol{\psi},\textbf{A}]=-\sum_{i\nu_n,\textbf{k}}\boldsymbol{\psi}^T_\textbf{k}(-i\nu_n)\tilde G_0^{-1}(i\nu_n,\textbf{k})\boldsymbol{\psi}_\textbf{k}(i\nu_n)+\sum_{i\nu_n,\textbf{k}}\sum_{i\nu_n^\prime,\textbf{k}^\prime}\boldsymbol{\psi}_\textbf{k}^T(-i\nu_n)\Sigma_{\textbf{k}\textbf{k}^\prime}(i\nu_n-i\nu_n^\prime)\boldsymbol{\psi}_{\textbf{k}^\prime}(i\nu_n^\prime),
\end{align}
with 
\begin{align}\label{selfenergy}
    \Sigma_{\textbf{kk}^\prime}(i\nu_n-i\nu_n^\prime)=\delta_{\textbf{k}\textbf{k}^\prime}\bigg[\sqrt{\frac{T}{N}}\frac{e}{c}&\sum_{i\Omega_m}[\tilde{\text{v}}_x\text{A}_x(i\Omega_m)+\tilde{\text{v}}_y\text{A}_y(i\Omega_m)]\delta(i\Omega_m-i\nu_n+i\nu_n^\prime)\nonumber\\
    +\frac{T}{N}\frac{e^2}{2c^2}&\sum_{i\Omega_m}\sum_{i\Omega_n}[\tilde\rho_x\text{A}_x(i\Omega_m)\text{A}_x(i\Omega_n)+\tilde\rho_y\text{A}_y(i\Omega_m)\text{A}_y(i\Omega_n)]\delta(i\Omega_m+i\Omega_n-i\nu_n+i\nu_n^\prime)\bigg],
\end{align}
where $\Omega_m=2\pi m T$ are the bosonic Matsubara frequencies, and $N$ the number of unit cells. Following the standard procedure established in Refs.\ \cite{benfatto_prb04,udina_prb19},
we then integrate the electron field out of the partition function, to find the matter-mediated contribution to the effective action of the electromagnetic field:
\begin{align}\label{seff}
    S_\text{eff}[\textbf{A}]=\sum_{m\geq1}\frac{\text{Tr}[\tilde G_0\Sigma]^m}{m},
\end{align}
where $\text{Tr}$ is the trace, and retain only terms scaling with $\text{A}_x^2\text{A}_y^2$, that describe mixed third-order processes mediated by electronic transitions.
\subsection*{Diamagnetic-like antisymmetric kernel}
The diamagnetic-like contribution is contained in the $m=2$ term of $S_\text{eff} [\textbf{A}]$ in Eq.\ \eqref{seff}, when multiplying twice the second row of Eq.\ \eqref{selfenergy}. In particular, we are interested in
\begin{align}\label{sdia1}
    S_\text{dia}[\textbf{A}]=\frac{e^4}{4c^4}\sum_{i\Omega_m}\sum_{i\Omega_n}\sum_{i\Omega_l}\sum_{i\Omega_s}\frac{T}{N}\sum_{i\nu_n,\textbf{k}}\,&\text{Tr}\big[ \tilde G_0(i\nu_n,\textbf{k})\tilde \rho_y \tilde G_0(i\nu_n+i\Omega_m+i\Omega_n,\textbf{k})\tilde \rho_x\big]\nonumber\\
    &\times\text{A}_y(i\Omega_m)\text{A}_y(i\Omega_n)\text{A}_x(i\Omega_l)\text{A}_x(i\Omega_s)\delta_{-i\Omega_s,i\Omega_m+i\Omega_n+i\Omega_l},
\end{align}
where $i\Omega_m+i\Omega_n+i\Omega_l+i\Omega_s=0$ by energy conservation.
We then rotate both the Green's function and the density-like vertex to the band basis, and explicit the trace:
\begin{align}
    \text{Tr}&\big[ \tilde G_0(i\nu_n,\textbf{k})\tilde \rho_y \tilde G_0(i\nu_n+i\Omega_m+i\Omega_n,\textbf{k})\tilde \rho_x\big]\nonumber\\
    &=\text{Tr}\big[ G_0(i\nu_n,\textbf{k}) \rho_y  G_0(i\nu_n+i\Omega_m+i\Omega_n,\textbf{k}) \rho_x\big]\nonumber\\
    &=\sum_{ab} [G_0(i\nu_n,\textbf{k})]_{aa}[\rho_y]_{ab}[G_0(i\nu_n+i\Omega_m+i\Omega_n,\textbf{k})]_{bb}[\rho_x]_{ba},
\end{align}
where $a$ and $b$ run over band indices. We can thus rewrite the diamagnetic action as
\begin{align}\label{diaact}
    S_\text{dia}[\textbf{A}]=\frac{e^4}{4c^4}\sum_{i\Omega_m}\sum_{i\Omega_n}\sum_{i\Omega_l}\sum_{i\Omega_s}\frac{1}{N}\sum_\textbf{k}\sum_{ab}&\Big[[\rho_y]_{ab}[\rho_x]_{ba}\,d_{ab}(i\Omega_m+i\Omega_n)\Big]\nonumber\\
    \times&\text{A}_y(i\Omega_m)\text{A}_y(i\Omega_n)\text{A}_x(i\Omega_l)\text{A}_x(i\Omega_s)\delta_{-i\Omega_s,i\Omega_m+i\Omega_n+i\Omega_l},
\end{align}
where
\begin{align}\label{dab}
    d_{ab}(i\Omega_m+i\Omega_n)&=T\sum_{i\nu_n}[G_0(i\nu_n,\textbf{k})]_{aa}[G_0(i\nu_n+i\Omega_m+i\Omega_n,\textbf{k})]_{bb}\nonumber\\
    &=\frac{f(E_a(\textbf{k}))-f(E_b(\textbf{k}))}{i\Omega_m+i\Omega_n-(E_b(\textbf{k})-E_a(\textbf{k}))}
\end{align}
contains the frequency-dependent poles associated with the electronic transitions, with $f(E_a(\textbf{k}))$ the Fermi-Dirac distribution. 
We now compute the third-order nonlinear current, that describes the response of the system, as the functional derivative of Eq.\ \eqref{diaact} with respect to $\text{A}_x/c$:
\begin{align}\label{nlcurrdia}
    \text{J}_x(\omega)\propto -\frac{\delta S_\text{dia}[\textbf{A}]}{\delta\text{A}_x(-\omega)/c}=-\frac{e^4}{2c^3}\int d\omega_1 d\omega_2& d\omega_3\frac{1}{N}\sum_\textbf{k}\sum_{ab}\Big[[\rho_y]_{ab}[\rho_x]_{ba}\,d_{ab}(\omega_1+\omega_2)\Big]\nonumber\\
    \times&\text{A}_y(\omega_1)\text{A}_y(\omega_2)\text{A}_x(\omega_3)\delta(\omega_1+\omega_2+\omega_3-\omega),
\end{align}
where we performed the analytic continuation for all bosonic Matsubara frequencies, and $\omega_1+\omega_2+\omega_3-\omega=0$ by energy conservation.
So far, we did not distinguish explicitly between pump and probe fields when writing the vector potentials $\text A_i$. According to the convention established in the main text, we are interested in a nonlinear current along $x$, generated in response to a probe along $y$ and a circular pump having both $x$ and $y$ components. Thus, the functional derivative in Eq.\ \eqref{nlcurrdia} is performed with respect to the generated signal along $x$, and the remaining three fields are associated with a probe photon $\text A_y$ and two pump photons $\text{A}_x$ and $\text{A}_y$.
The kernel component $\text{K}_{xy;yx}^\text{dia}(\omega;\omega_1,\omega_2,\omega_3)$ is obtained as the coefficient of $(\text{A}/c)^3$ in the nonlinear current.

We now consider monochromatic pump and probe fields. For the antisymmetric kernel, we are interested in computing $\text{K}_{xy;xy}(\omega;\omega,\Omega,-\Omega)=\text{K}_{xy;yx}(\omega;\omega,-\Omega,\Omega)$, and $\text{K}_{xy;yx}(\omega;\omega,\Omega,-\Omega)$. When $\omega_2=\Omega$ and $\omega_3=-\Omega$, we obtain from the nonlinear current
\begin{align}
    \text{K}_{xy;yx}^\text{dia}(\omega;\omega,\Omega,-\Omega)= -\frac{e^4}{2}\frac{1}{N}\sum_\textbf{k}\sum_{ab}\,[\rho_y]_{ab}[\rho_x]_{ba}\,d_{ab}(\omega+\Omega),
\end{align}
while, for $\omega_2=-\Omega$ and $\omega_3=\Omega$,
\begin{align}
    \text{K}_{xy;yx}^\text{dia}(\omega;\omega,-\Omega,\Omega)=\text{K}_{xy;xy}^\text{dia}(\omega;\omega,\Omega,-\Omega)=-\frac{e^4}{2}\frac{1}{N}\sum_\textbf{k}\sum_{ab}\,[\rho_y]_{ab}[\rho_x]_{ba}\,d_{ab}(\omega-\Omega).
\end{align}
Finally, using the symmetry properties of the kernel, we can write the antisymmetric diamagnetic component as
\begin{align}\label{kdia}
    \text{K}_\text{dia}^\mathcal{A}(\omega;\omega,\Omega,-\Omega)&=\text{K}^\text{dia}_{xy;xy}(\omega;\omega,\Omega,-\Omega)-\text{K}^\text{dia}_{xy;yx}(\omega;\omega,\Omega,-\Omega)\nonumber\\
    &=\frac{1}{2}[\text{K}^\text{dia}_{xy;xy}(\omega;\omega,\Omega,-\Omega)-\text{K}^\text{dia}_{xy;yx}(\omega;\omega,\Omega,-\Omega)+\text{K}^\text{dia}_{yx;yx}(\omega;\omega,\Omega,-\Omega)-\text{K}^\text{dia}_{yx;xy}(\omega;\omega,\Omega,-\Omega)]\nonumber\\
    &=-\frac{e^4}{4}\frac{1}{N}\sum_\textbf{k}\sum_{ab}\big[[\rho_x]_{ab}[\rho_y]_{ba}+[\rho_y]_{ab}[\rho_x]_{ba}\big]\big[d_{ab}(\omega-\Omega)-d_{ab}(\omega+\Omega)\big],
\end{align}
which corresponds to Eq.\ (9) of the main text. 

For simulations of the antisymmetric susceptibility we set $T=0$, so that the Fermi-Dirac distributions in \eqref{dab} for the three bands are $f(E_1(\textbf{k}))=1$, $f(E_{2,3}(\textbf{k}))=0$. The sum over $\textbf{k}$ is, instead, performed numerically.
\subsection*{Paramagnetic-like antisymmetric kernel}
The paramagnetic-like contribution is contained in the $m=4$ term of $S_\text{eff}[\textbf{A}]$ in Eq.\ \eqref{seff}, when multiplying four times the first row of Eq.\ \eqref{selfenergy}. In particular, we are interested in
\begin{align}\label{spara1}
    S_\text{para}[\textbf{A}]=\frac{e^4}{4c^4}\sum_{i\Omega_m}&\sum_{i\Omega_n}\sum_{i\Omega_l}\sum_{i\Omega_s}\nonumber\\
    \frac{T}{N}\sum_{i\nu_n,\textbf{k}}&\text{Tr}\big[\tilde G_0(i\nu_n,\textbf{k})\tilde{\text{v}}_y\tilde G_0(i\nu_n+i\Omega_m,\textbf{k})\tilde{\text{v}}_y\tilde G_0(i\nu_n+i\Omega_m+i\Omega_n,\textbf{k})\tilde{\text{v}}_x \tilde G_0(i\nu_n+i\Omega_m+i\Omega_n+i\Omega_l, \textbf{k})\tilde{\text{v}}_x\big]\nonumber\\
    \times&\text{A}_y(i\Omega_m)\text{A}_y(i\Omega_n)\text{A}_x(i\Omega_l)\text{A}_x(i\Omega_s)\delta_{-i\Omega_s,i\Omega_m+i\Omega_n+i\Omega_l}+\text{perm.},
\end{align}
where ``perm.'' denotes all distinct permutations of the external field insertions, i.e., all possible ways of attaching the photon legs to the fermionic loop. We note that every permutation also changes the position of the velocity vertices in the trace, accordingly. Following the same steps of the diamagnetic-like case, we rewrite the trace as
\begin{align}
    \text{Tr}\big[\tilde G_0(i\nu_n&,\textbf{k})\tilde{\text{v}}_y\tilde G_0(i\nu_n+i\Omega_m,\textbf{k})\tilde{\text{v}}_y\tilde G_0(i\nu_n+i\Omega_m+i\Omega_n,\textbf{k})\tilde{\text{v}}_x \tilde G_0(i\nu_n+i\Omega_m+i\Omega_n+i\Omega_l, \textbf{k})\tilde{\text{v}}_x\big]\nonumber\\
    =\text{Tr}\big[&G_0(i\nu_n,\textbf{k}){\text{v}}_y G_0(i\nu_n+i\Omega_m,\textbf{k}){\text{v}}_y G_0(i\nu_n+i\Omega_m+i\Omega_n,\textbf{k}){\text{v}}_x  G_0(i\nu_n+i\Omega_m+i\Omega_n+i\Omega_l, \textbf{k}){\text{v}}_x\big]\nonumber\\
    =\sum_{abcd}[&G_0(i\nu_n,\textbf{k})]_{aa}[\text{v}_y]_{ab}[G_0(i\nu_n+i\Omega_m,\textbf{k})]_{bb}[\text{v}_y]_{bc}\nonumber\\
    \times[&G_0(i\nu_n+i\Omega_m+i\Omega_n,\textbf{k})]_{cc}[\text{v}_x]_{cd}[G_0(i\nu_n+i\Omega_m+i\Omega_n+i\Omega_l,\textbf{k})]_{dd}[\text{v}_x]_{da},
\end{align}
where $a$, $b$, $c$, and $d$ run over band indices. The paramagnetic contribution to the action can then be rewritten as
\begin{align}\label{spara}
    S_\text{para}[\textbf{A}]=\frac{e^4}{4c^4}\sum_{i\Omega_m}\sum_{i\Omega_n}\sum_{i\Omega_l}\sum_{i\Omega_s}\frac{1}{N}\sum_\textbf{k}\sum_{abcd}&\Big[[\text{v}_y]_{ab}[\text{v}_y]_{bc}[\text{v}_x]_{cd}[\text{v}_x]_{da}\,p_{abcd}(i\Omega_m,i\Omega_n,i\Omega_l)\Big]\nonumber\\
    &\times\text{A}_y(i\Omega_m)\text{A}_y(i\Omega_n)\text{A}_x(i\Omega_l)\text{A}_x(i\Omega_s)\delta_{-i\Omega_s,i\Omega_m+i\Omega_n+i\Omega_l}+\text{perm.},
\end{align}
where
\begin{align}\label{pabcd}
    p_{abcd}&(i\Omega_m,i\Omega_n,i\Omega_l)=T\sum_{i\nu_n}[G_0(i\nu_n,\textbf{k})]_{aa}[G_0(i\nu_n+i\Omega_m,\textbf{k})]_{bb}\nonumber\\
    &\times[G_0(i\nu_n+i\Omega_m+i\Omega_n,\textbf{k})]_{cc}[G_0(i\nu_n+i\Omega_m+i\Omega_n+i\Omega_l,\textbf{k})]_{dd}\nonumber\\
    =&-\Big[\frac{f(E_a(\textbf{k}))}{\big(i\Omega_m+E_a(\textbf{k})-E_b(\textbf{k})\big)\big(i\Omega_m+i\Omega_n+E_a(\textbf{k})-E_c(\textbf{k})\big)\big(i\Omega_m+i\Omega_n+i\Omega_l+E_a(\textbf{k})-E_d(\textbf{k})\big)}\nonumber\\
    &+\frac{f(E_b(\textbf{k}))}{\big(-i\Omega_m+E_b(\textbf{k})-E_a(\textbf{k})\big)\big(i\Omega_n+E_b(\textbf{k})-E_c(\textbf{k})\big)\big(i\Omega_n+i\Omega_l+E_b(\textbf{k})-E_d(\textbf{k})\big)}\nonumber\\
    &+\frac{f(E_c(\textbf{k}))}{\big(-i\Omega_m-i\Omega_n+E_c(\textbf{k})-E_a(\textbf{k})\big)\big(-i\Omega_n+E_c(\textbf{k})-E_b(\textbf{k})\big)\big(i\Omega_l+E_c(\textbf{k})-E_d(\textbf{k})\big)}\nonumber\\
    &+\frac{f(E_d(\textbf{k}))}{\big(-i\Omega_m-i\Omega_n-i\Omega_l+E_d(\textbf{k})-E_a(\textbf{k})\big)\big(-i\Omega_n-i\Omega_l+E_d(\textbf{k})-E_b(\textbf{k})\big)\big(-i\Omega_l+E_d(\textbf{k})-E_c(\textbf{k})\big)}\Big]
\end{align}
contains the frequency-dependent poles, analogously to $d_{ab}$ for the diamagnetic process. From Eq.\ \eqref{spara} we can compute the nonlinear current:
\begin{align}
    \text{J}_x(\omega)\propto - \frac{\delta S_\text{para}[\textbf{A}]}{\delta\text{A}_x(-\omega)/c}=-\frac{e^4}{2c^3}\int d\omega_1 d\omega_2 d\omega_3\frac{1}{N}\sum_\textbf{k}\sum_{abcd} &\Big[[\text{v}_y]_{ab}[\text{v}_{y}]_{bc}[\text{v}_x]_{cd}[\text{v}_x]_{da}\, p_{abcd}(\omega_1,\omega_2,\omega_3)\Big]\nonumber\\
    &\times\text{A}_y(\omega_1)\text{A}_y(\omega_2)\text{A}_x(\omega_3)\delta(\omega_1+\omega_2+\omega_3-\omega)+\text{perm.},
\end{align}
where we performed the analytic continuation for all bosonic Matsubara frequencies and $\omega_1+\omega_2+\omega_3-\omega=0$. The kernel component $\text{K}^\text{para}_{xy;yx}(\omega;\omega_1,\omega_2,\omega_3)$ is obtained as the coefficient of $(\text{A}/c)^3$ in the nonlinear current.

We now consider monochromatic fields. With the ultimate goal of obtaining compact expressions for the antisymmetric kernel, it is convenient to write all six permutations in the kernel components explicitly. In particular,
\begin{align}
    \text{K}^\text{para}_{xy;yx}(\omega;\omega,\Omega,-\Omega)=-\frac{e^4}{2}\frac{1}{N}\sum_\textbf{k}\sum_{abcd}\Big[&[\text{v}_y]_{ab}[\text{v}_{y}]_{bc}[\text{v}_x]_{cd}[\text{v}_x]_{da}\, p_{abcd}(\omega,\Omega,-\Omega)\nonumber\\
    +&[\text{v}_y]_{ab}[\text{v}_y]_{bc}[\text{v}_x]_{cd}[\text{v}_x]_{da}\, p_{abcd}(\Omega,\omega,-\Omega)\nonumber\\
    +&[\text{v}_x]_{ab}[\text{v}_y]_{bc}[\text{v}_y]_{cd}[\text{v}_x]_{da}\, p_{abcd}(-\Omega,\Omega,\omega)\nonumber\\
    +&[\text{v}_x]_{ab}[\text{v}_y]_{bc}[\text{v}_y]_{cd}[\text{v}_x]_{da}\, p_{abcd}(-\Omega,\omega,\Omega)\nonumber\\
    +&[\text{v}_y]_{ab}[\text{v}_x]_{bc}[\text{v}_y]_{cd}[\text{v}_x]_{da}\, p_{abcd}(\Omega,-\Omega,\omega)\nonumber\\
    +&[\text{v}_y]_{ab}[\text{v}_x]_{bc}[\text{v}_y]_{cd}[\text{v}_x]_{da}\, p_{abcd}(\omega,-\Omega,\Omega)\Big]
\end{align}
for $\omega_2=\Omega$ and $\omega_3=-\Omega$, and
\begin{align}
    \text{K}_{xy;yx}^\text{para}(\omega;\omega,-\Omega,\Omega)=\text{K}^\text{para}_{xy;xy}(\omega;\omega,\Omega,-\Omega)=-\frac{e^4}{2}\frac{1}{N}\sum_\textbf{k}\sum_{abcd}\Big[&[\text{v}_y]_{ab}[\text{v}_x]_{bc}[\text{v}_y]_{cd}[\text{v}_x]_{da}\, p_{abcd}(\omega,\Omega,-\Omega)\nonumber\\
    +&[\text{v}_x]_{ab}[\text{v}_y]_{bc}[\text{v}_y]_{cd}[\text{v}_x]_{da}\, p_{abcd}(\Omega,\omega,-\Omega)\nonumber\\
    +&[\text{v}_y]_{ab}[\text{v}_x]_{bc}[\text{v}_y]_{cd}[\text{v}_x]_{da}\, p_{abcd}(-\Omega,\Omega,\omega)\nonumber\\
    +&[\text{v}_y]_{ab}[\text{v}_y]_{bc}[\text{v}_x]_{cd}[\text{v}_x]_{da}\, p_{abcd}(-\Omega,\omega,\Omega)\nonumber\\
    +&[\text{v}_x]_{ab}[\text{v}_y]_{bc}[\text{v}_y]_{cd}[\text{v}_x]_{da}\, p_{abcd}(\Omega,-\Omega,\omega)\nonumber\\
    +&[\text{v}_y]_{ab}[\text{v}_y]_{bc}[\text{v}_x]_{cd}[\text{v}_x]_{da}\, p_{abcd}(\omega,-\Omega,\Omega)\Big]
\end{align}
for $\omega_2=-\Omega$ and $\omega_3=\Omega$. Notice that in every term the last matrix element $[\text{v}_x]_{da}$ is fixed on the $x$ component by our choice of performing the detection along $x$. Finally, we define the vectors $\textbf{v}=(\text{v}_x,\text{v}_y)$ and make use of the symmetry properties of the kernel to write the antisymmetric paramagnetic component as
\begin{align}\label{kpara}
    \text{K}_\text{para}^\mathcal{A}(\omega;\omega,\Omega,-\Omega)=\text{K}^\text{para}_{xy;xy}(\omega&;\omega,\Omega,-\Omega)-\text{K}^\text{para}_{xy;yx}(\omega;\omega,\Omega,-\Omega)\nonumber\\
    =\frac{1}{2}[\text{K}^\text{para}_{xy;xy}&(\omega;\omega,\Omega,-\Omega)-\text{K}^\text{para}_{xy;yx}(\omega;\omega,\Omega,-\Omega)+\text{K}^\text{para}_{yx;yx}(\omega;\omega,\Omega,-\Omega)-\text{K}^\text{para}_{yx;xy}(\omega;\omega,\Omega,-\Omega)]\nonumber\\
    =-\frac{e^4}{4}\frac{1}{N}&\sum_\textbf{k}\sum_{abcd}\big[([\textbf{v}]_{ab}\times[\textbf{v}]_{bc})\cdot([\textbf{v}]_{cd}\times[\textbf{v}]_{da})\big]\big[p_{abcd}(-\Omega,\Omega,\omega)-p_{abcd}(\Omega,-\Omega,\omega)\big]\nonumber\\
    +&\big[([\textbf{v}]_{ab}\times[\textbf{v}]_{cd})\cdot([\textbf{v}]_{bc}\times[\textbf{v}]_{da})\big]\big[p_{abcd}(-\Omega,\omega,\Omega)-p_{abcd}(\Omega,\omega,-\Omega)\big]\nonumber\\
    +&\big[([\textbf{v}]_{bc}\times[\textbf{v}]_{cd})\cdot([\textbf{v}]_{ab}\times[\textbf{v}]_{da})\big]\big[p_{abcd}(\omega,-\Omega,\Omega)-p_{abcd}(\omega,\Omega,-\Omega)\big],
\end{align}
which corresponds to Eq.\ (10) of the main text.

As it is the case for the diamagnetic kernel, the explicit calculation of Eq.\ \eqref{pabcd} is carried out at $T=0$, while the sum over $\textbf{k}$ is performed numerically.
\subsection*{Mixed diamagnetic-paramagnetic antisymmetric kernel}
The mixed contributions with one diamagnetic and two paramagnetic insertions are contained in the $m=3$ term of $S_\text{eff}[\textbf{A}]$ in Eq.\ \eqref{seff}, when multiplying twice the first row and once the second row of Eq.\ \eqref{selfenergy}. In particular, we are interested in
\begin{align}\label{smix1}
    S_\text{mix}^{(1)}[\textbf{A}]=\frac{e^4}{6c^4}\sum_{i\Omega_m}\sum_{i\Omega_n}\sum_{i\Omega_l}\sum_{i\Omega_s}
    \frac{T}{N}\sum_{i\nu_n,\textbf{k}}\text{Tr}&\big[\tilde{G_0}(i\nu_n,\textbf{k})\tilde{\text{v}}_y\tilde G_0(i\nu_n+i\Omega_m,\textbf{k})\tilde{\text{v}}_y\tilde G_0(i\nu_n+i\Omega_m+i\Omega_n,\textbf{k})\tilde\rho_x\big]\nonumber\\
    \times&\text{A}_y(i\Omega_m)\text{A}_y(i\Omega_n)\text{A}_x(i\Omega_l)\text{A}_x(i\Omega_s)\delta_{-i\Omega_s,i\Omega_m+i\Omega_n+i\Omega_l}
\end{align}
and
\begin{align}\label{smix2}
    S_\text{mix}^{(2)}[\textbf{A}]=\frac{e^4}{6c^4}\sum_{i\Omega_m}\sum_{i\Omega_n}\sum_{i\Omega_l}\sum_{i\Omega_s}
    \frac{T}{N}\sum_{i\nu_n,\textbf{k}}\text{Tr}&\big[\tilde{G_0}(i\nu_n,\textbf{k})\tilde{\rho}_y\tilde G_0(i\nu_n+i\Omega_m+i\Omega_n,\textbf{k})\tilde{\text{v}}_x\tilde G_0(i\nu_n+i\Omega_m+i\Omega_n+i\Omega_l,\textbf{k})\tilde{\text{v}}_x\big]\nonumber\\\times&\text{A}_y(i\Omega_m)\text{A}_y(i\Omega_n)\text{A}_x(i\Omega_l)\text{A}_x(i\Omega_s)\delta_{-i\Omega_s,i\Omega_m+i\Omega_n+i\Omega_l}.
\end{align}
The two actions have been explicitly separated having in mind that the detection along $x$ can be performed either on the diamagnetic insertion (process $S^{(1)}_\text{mix}$) or on a paramagnetic one (process $S^{(2)}_\text{mix}$). In the following we provide details for the calculation of the first contribution. 
\\
Following the same strategy detailed above, we rewrite the trace as
\begin{align}
    \text{Tr}&\big[\tilde{G_0}(i\nu_n,\textbf{k})\tilde{\text{v}}_y\tilde G_0(i\nu_n+i\Omega_m,\textbf{k})\tilde{\text{v}}_y\tilde G_0(i\nu_n+i\Omega_m+i\Omega_n,\textbf{k})\tilde\rho_x\big]\nonumber\\
    &=\text{Tr}\big[G_0(i\nu_n,\textbf{k})\text{v}_yG_0(i\nu_n+i\Omega_m,\textbf{k})\text{v}_yG_0(i\nu_n+i\Omega_m+i\Omega_n,\textbf{k})\rho_x\big]\nonumber\\
    &=\sum_{abc}\big[[G_0(i\nu_n,\textbf{k})]_{aa}[\text{v}_y]_{ab}[G_0(i\nu_n+i\Omega_m,\textbf{k})]_{bb}[\text{v}_y]_{bc}[G_0(i\nu_n+i\Omega_m+i\Omega_n,\textbf{k})]_{cc}[\rho_x]_{ca}\big],
\end{align}
where $a$, $b$, and $c$ run over band indices. The mixed action can thus be rewritten as
\begin{align}\label{smix}
    S_\text{mix}^{(1)}[\textbf{A}]=\frac{e^4}{6c^4}\sum_{i\Omega_m}\sum_{i\Omega_n}\sum_{i\Omega_l}\frac{1}{N}\sum_\textbf{k}\sum_{abc}\Big[[\text{v}_y]_{ab}[\text{v}_y]_{bc}[\rho_x]_{ca}\,m_{abc}(i\Omega_m,i\Omega_m+i\Omega_n)\Big]\text{A}_y(i\Omega_m)\text{A}_y(i\Omega_n)\text{A}_x(i\Omega_l)\text{A}_x(i\Omega_s),
\end{align}
where
\begin{align}\label{mabc}
    m_{abc}(i\Omega_m,i\Omega_m+i\Omega_n)=&T\sum_{i\nu_n}[G_0(i\nu_n,\textbf{k})]_{aa}[G_0(i\nu_n+i\Omega_m,\textbf{k})]_{bb}[G_0(i\nu_n+i\Omega_m+i\Omega_n,\textbf{k})]_{cc}\nonumber\\
    =&-\Big[\frac{f(E_a(\textbf{k}))}{\big(i\Omega_m+E_a(\textbf{k})-E_b(\textbf{k})\big)\big(i\Omega_m+i\Omega_n+E_a(\textbf{k})-E_c(\textbf{k})\big)}\nonumber\\
    &+\frac{f(E_b(\textbf{k}))}{\big(-i\Omega_m+E_b(\textbf{k})-E_a(\textbf{k})\big)\big(-i\Omega_m+(i\Omega_m+i\Omega_n)+E_b(\textbf{k})-E_c(\textbf{k})\big)}\nonumber\\
    &+\frac{f(E_c(\textbf{k}))}{\big(-i\Omega_m-i\Omega_n+E_c(\textbf{k})-E_a(\textbf{k})\big)\big(i\Omega_m-(i\Omega_m+i\Omega_n)+E_c(\textbf{k})-E_b(\textbf{k})\big)}\Big]
\end{align}
contains the frequency dependent poles, analogously to $d_{ab}$ and $p_{abcd}$. From Eq.\ \eqref{smix} we can compute the nonlinear current:
\begin{align}
    \text{J}^{(1)}_x(\omega)\propto-\frac{\delta S^{(1)}_\text{mix}[\textbf{A}]}{\delta\text{A}_x(-\omega)/c}=-\frac{e^4}{3c^3}\int d\omega_1d\omega_2 d\omega_3 \frac{1}{N}\sum_\textbf{k}\sum_{abc}\Big[&[\text{v}_y]_{ab}[\text{v}_y]_{bc}[\rho_x]_{ca}\,m_{abc}(\omega_1,\omega_1+\omega_2)\Big]\nonumber\\
    \times&\text{A}_y(\omega_1)\text{A}_y(\omega_2)\text{A}_x(\omega_3)\delta(\omega_1+\omega_2+\omega_3-\omega),
\end{align}
where we performed the analytic continuation for all bosonic Matsubara frequencies and $\omega_1+\omega_2+\omega_3-\omega=0$. The coefficient of $(\text{A}/c)^3$ yields the kernel component $\text{K}_{xy;yx}^{\text{mix}(1)}(\omega;\omega_1,\omega_2,\omega_3)$.\\
With monochromatic fields, we obtain
\begin{align}
    \text{K}^{\text{mix}(1)}_{xy;yx}(\omega;\omega,\Omega,-\Omega)=-\frac{e^4}{3}\frac{1}{N}\sum_\textbf{k}\sum_{abc}[\text{v}_y]_{ab}[\text{v}_y]_{bc}[\rho_x]_{ca}\,m_{abc}(\omega,\omega+\Omega),
\end{align}
for $\omega_2=\Omega$ and $\omega_3=-\Omega$, and
\begin{align}
    \text{K}^{\text{mix}(1)}_{xy;yx}(\omega;\omega,-\Omega,\Omega)=\text{K}^{\text{mix}(1)}_{xy;xy}(\omega;\omega,\Omega,-\Omega)=-\frac{e^4}{3}\frac{1}{N}\sum_\textbf{k}\sum_{abc}[\text{v}_y]_{ab}[\text{v}_y]_{bc}[\rho_x]_{ca}\,m_{abc}(\omega,\omega-\Omega),
\end{align}
for $\omega_2=-\Omega$ and $\omega_3=\Omega$. We can finally write the first antisymmetric mixed component as
\begin{align}
    \text{K}_{\text{mix}}^{\mathcal{A}(1)}(\omega;\omega,\Omega,-\Omega)&=\text{K}^{\text{mix}(1)}_{xy;xy}(\omega;\omega,\Omega,-\Omega)-\text{K}^{\text{mix}(1)}_{xy;yx}(\omega;\omega,\Omega,-\Omega)\nonumber\\
    &=\frac{1}{2}[\text{K}^{\text{mix}(1)}_{xy;xy}(\omega;\omega,\Omega,-\Omega)-\text{K}^{\text{mix}(1)}_{xy;yx}(\omega;\omega,\Omega,-\Omega)+\text{K}^{\text{mix}(1)}_{yx;yx}(\omega;\omega,\Omega,-\Omega)-\text{K}^{\text{mix}(1)}_{yx;xy}(\omega;\omega,\Omega,-\Omega)]\nonumber\\
    &=-\frac{e^4}{6}\frac{1}{N}\sum_\textbf{k}\sum_{abc}\big[[\text{v}_x]_{ab}[\text{v}_x]_{bc}[\rho_y]_{ca}+[\text{v}_y]_{ab}[\text{v}_y]_{bc}[\rho_x]_{ca}\big]\big[m_{abc}(\omega,\omega-\Omega)-m_{abc}(\omega,\omega+\Omega)\big].
\end{align}
Completely analogous calculations on the second contribution, from Eq.\ \eqref{smix2}, give
\begin{align}
    \text{K}_\text{mix}^{\mathcal{A}(2)}(\omega;\omega,\Omega,-\Omega)\!=\!-\frac{e^4}{6}\frac{1}{N}\sum_\textbf{k}\sum_{abc}\big[[\rho_x]_{ab}[\text{v}_y]_{bc}[\text{v}_y]_{ca}\!+\![\rho_y]_{ab}[\text{v}_x]_{bc}[\text{v}_x]_{ca}\big]\big[m_{abc}(\omega-\Omega,\omega)\!-\!m_{abc}(\omega+\Omega,\omega)\big].
\end{align}
The total mixed diamagnetic-paramagnetic antisymmetric kernel is then found as 
\begin{align}\label{kmix}
    \text{K}_{\text{mix}}^{\mathcal{A}}(\omega;\omega,\Omega,-\Omega)=\text{K}_{\text{mix}}^{\mathcal{A}(1)}(\omega;\omega,\Omega,-\Omega)+\text{K}_{\text{mix}}^{\mathcal{A}(2)}(\omega;\omega,\Omega,-\Omega).
\end{align}
As in the previous cases, Eq.\ \eqref{mabc} is evaluated at $T=0$, and the sum over $\textbf{k}$ is performed numerically.
\subsection*{Antisymmetric susceptibility}
The antisymmetric susceptibility, that mediates the light-induced Faraday effect, is found from the interaction kernels as 
\begin{align}\label{chiel}
    \chi^\mathcal{A}_{xy}(\omega;\omega,\Omega,-\Omega)=\frac{1}{\omega^2\Omega^2}\big[\text{K}_{\text{dia}}^{\mathcal{A}}(\omega;\omega,\Omega,-\Omega)+\text{K}_{\text{para}}^{\mathcal{A}}(\omega;\omega,\Omega,-\Omega)+\text{K}_{\text{mix}}^{\mathcal{A}}(\omega;\omega,\Omega,-\Omega)\big].
\end{align}
The symmetric susceptibility $\chi^\mathcal{S}_{xy}$ can be readily obtained from the kernel components derived above.

In the low-frequency limit, one finds that $\chi^\mathcal{A}_{xy}(\omega;\omega,\Omega,-\Omega)\sim\omega\Omega$. This behavior demonstrates that the light-induced Faraday effect is intrinsically dynamical, vanishing in the limit of either a static pump or a static probe. On the other hand, one finds that $\chi_{xy}^{\mathcal{S}}(\omega;\omega,\Omega,-\Omega)$ goes to a constant value for $\omega=0$ and $\Omega=0$. In Fig.\ \ref{FigS1} we show the ratio $\chi^\mathcal{A}_{xy}/\chi^\mathcal{S}_{xy}$ for a fixed value of the probe frequency $\omega/\Delta$.

We emphasize that the derivation of the antisymmetric kernels in the diamagnetic, paramagnetic, and mixed cases does not rely on model-specific details.
However, additional contributions to Eqs.\ \eqref{kdia}, \eqref{kpara}, and \eqref{kmix} may arise in systems admitting nonvanishing mixed $xy$ second-order vertices (or higher-order terms), whose presence depends on the specific structure of the tight-binding Hamiltonian $\mathcal{H}_\mathbf{k}$.
\vskip 7mm
\begin{figure}[h]
    \centering
    \includegraphics[width=0.5\textwidth,keepaspectratio]{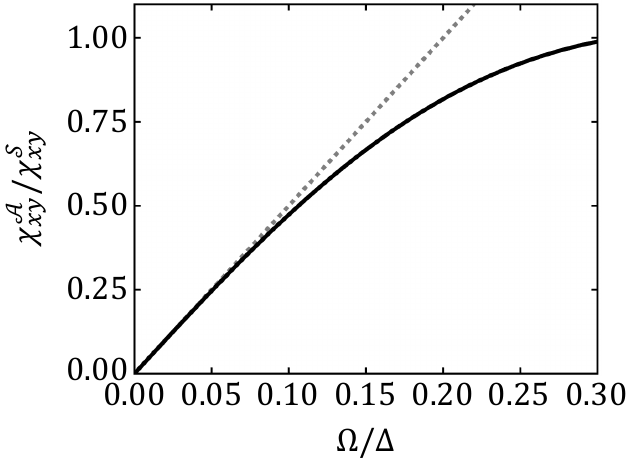}
    \caption{Ratio $\chi^\mathcal{A}_{xy}/\chi_{xy}^\mathcal{S}$ in the $sp$-model as a function of the pump frequency $\Omega/\Delta$, for fixed probe frequency $\omega/\Delta=1.5$ and for $t/\Delta=0.1$. Gray dashed line follows the small-frequency linear behavior of the antisymmetric susceptibility.}
    \label{FigS1}
\end{figure}

\clearpage

\section*{Details of the simulations of the pump-probe response with Gaussian fields}
Here we detail the calculations and approximations considered to reproduce the pump-probe response spectrum shown in Fig.\ 3(d) of the main text.\\
To simulate a realistic response we must consider pump fields with finite bandwidths. In particular, we consider Gaussian pulses with a slight frequency mismatch:
\begin{align}\label{gausspulse}
    \text{E}_\text{pu}^P(\omega_s)&=\mathcal{E}_P\,\mathcal{G}(\omega_s-\Omega+\Delta\Omega)+\mathcal{E}_P^*\,\mathcal{G}(\omega_s+\Omega-\Delta\Omega),\nonumber\\
    \text{E}_\text{pu}^S(\omega_s)&=\mathcal{E}_S\,\mathcal{G}(\omega_s-\Omega)+\mathcal{E}_S^*\,\mathcal{G}(\omega_s+\Omega),
\end{align}
with
\begin{align}
    \mathcal{G}(\omega)=e^{-\frac{\omega^2\tau^2}{2}},
\end{align}
and $\tau$ represents the time duration of the pulse, which controls the bandwidth of the Gaussian pulse. We then closely follow the derivation below Eq.\ \eqref{monfieldsmism}, and change basis to $\mathcal{E}_L$ and $\mathcal{E}_R$ with Eqs.\ \eqref{lintocirc}. As we discussed above, only terms coming from the field combinations $\mathcal{S}_\perp$ and $\mathcal{A}$ of Eq.\ \eqref{fieldcomb} contribute to the differential dichroism $\Delta\Gamma_{R-L}$, which read for Gaussian pulses
\begin{align}
    \mathcal{S}_\perp=-\frac{i}{2}&(\mathcal{E}_L^2-\mathcal{E}_R^2)\big[\mathcal{G}(\omega_2-\Omega+\Delta\Omega)\mathcal{G}(\omega_3-\Omega)+\mathcal{G}(\omega_2-\Omega)\mathcal{G}(\omega_3-\Omega+\Delta\Omega)\big]\nonumber\\
    +\frac{i}{2}&(\mathcal{E}_L^{*2}-\mathcal{E}_R^{*2})\,\big[\mathcal{G}(\omega_2+\Omega-\Delta\Omega)\mathcal{G}(\omega_3+\Omega)+\mathcal{G}(\omega_2+\Omega)\mathcal{G}(\omega_3+\Omega-\Delta\Omega)\big]\nonumber\\
    +\frac{i}{2}&\big(|\mathcal{E}_L|^2-|\mathcal{E}_R|^2\big)\big[\mathcal{G}(\omega_2-\Omega+\Delta\Omega)\mathcal{G}(\omega_3+\Omega)+\mathcal{G}(\omega_2+\Omega)\mathcal{G}(\omega_3-\Omega+\Delta\Omega)\nonumber\\
    &-\mathcal{G}(\omega_2-\Omega)\mathcal{G}(\omega_3+\Omega-\Delta\Omega)-\mathcal{G}(\omega_2+\Omega-\Delta\Omega)\mathcal{G}(\omega_3-\Omega)\big],
\end{align}
and
\begin{align}
    \mathcal{A}=-\frac{i}{2}&\big(\mathcal{E}_L^2-\mathcal{E}_R^2\big)\big[\mathcal{G}(\omega_2-\Omega+\Delta\Omega)\mathcal{G}(\omega_3-\Omega)-\mathcal{G}(\omega_2-\Omega)\mathcal{G}(\omega_3-\Omega+\Delta\Omega)\big]\nonumber\\
    +\frac{i}{2}&\big(\mathcal{E}_L^{*2}-\mathcal{E}_R^{*2}\big)\big[\mathcal{G}(\omega_2+\Omega-\Delta\Omega)\mathcal{G}(\omega_3+\Omega)-\mathcal{G}(\omega_2+\Omega)\mathcal{G}(\omega_3+\Omega-\Delta\Omega)\big]\nonumber\\
    +\frac{i}{2}&\big(|\mathcal{E}_L|^2-|\mathcal{E}_R|^2\big)\big[\mathcal{G}(\omega_2-\Omega+\Delta\Omega)\mathcal{G}(\omega_3+\Omega)+\mathcal{G}(\omega_2-\Omega)\mathcal{G}(\omega_3+\Omega-\Delta\Omega)\nonumber\\
    &-\mathcal{G}(\omega_2+\Omega)\mathcal{G}(\omega_3-\Omega+\Delta\Omega)-\mathcal{G}(\omega_2+\Omega-\Delta\Omega)\mathcal{G}(\omega_3-\Omega)\big],
\end{align}
where we have neglected terms in $\mathcal E_L\mathcal{E}_R^*$ and $\mathcal E_L^*\mathcal{E}_R$. Fixing $\theta=0$ and $\Delta\theta=0$, the relevant part of the differential intensity Eq.\ \eqref{diffsign} can be written as
\begin{align}\label{diffintgauss}
    \Delta\Gamma(t_{pp},0)\propto&\int d\omega_1d\omega_2d\omega_3\,e^{-i(\omega_2+\omega_3)t_{pp}}\text{E}_\text{pr}(\omega_1)\big[\chi^\mathcal{S}_{xy}(\omega;\omega_1,\omega_2,\omega_3)\mathcal{S}_\perp+\chi^\mathcal{A}_{xy}(\omega;\omega_1,\omega_2,\omega_3)\mathcal{A}\big]\nonumber\\
    \simeq \frac{i}{2}&\int d\omega_1d\omega_2d\omega_3\,e^{-i(\omega_2+\omega_3)t_{pp}}\text{E}_\text{pr}(\omega_1)\nonumber\\
    \times\bigg[&-(\mathcal{E}_L^2-\mathcal{E}_R^2)\chi^\mathcal{S}_{xy}(\omega+2\Omega;\omega,\Omega,\Omega)\big[\mathcal{G}(\omega_2-\Omega+\Delta\Omega)\mathcal{G}(\omega_3-\Omega)+\mathcal{G}(\omega_2-\Omega)\mathcal{G}(\omega_3-\Omega+\Delta\Omega)\big]\nonumber\\
    &+(\mathcal{E}_L^{*2}-\mathcal{E}_R^{*2})\chi^\mathcal{S}_{xy}(\omega-2\Omega;\omega,-\Omega,-\Omega)\big[\mathcal{G}(\omega_2+\Omega-\Delta\Omega)\mathcal{G}(\omega_3+\Omega)+\mathcal{G}(\omega_2+\Omega)\mathcal{G}(\omega_3+\Omega-\Delta\Omega)\big]\nonumber\\
    &+\big(|\mathcal{E}_L|^2-|\mathcal{E}_R|^2\big)\chi^\mathcal{S}_{xy}(\omega;\omega,\Omega,-\Omega)\big[\mathcal{G}(\omega_2-\Omega+\Delta\Omega)\mathcal{G}(\omega_3+\Omega)-\mathcal{G}(\omega_2-\Omega)\mathcal{G}(\omega_3+\Omega-\Delta\Omega)\nonumber\\
    &+\mathcal{G}(\omega_2+\Omega)\mathcal{G}(\omega_3-\Omega+\Delta\Omega)-\mathcal{G}(\omega_2+\Omega-\Delta\Omega)\mathcal{G}(\omega_3-\Omega)\big]\nonumber\\
    &+\big(|\mathcal{E}_L|^2-|\mathcal{E}_R|^2\big)\chi^\mathcal{A}_{xy}(\omega;\omega,\Omega,-\Omega)\big[\mathcal{G}(\omega_2-\Omega+\Delta\Omega)\mathcal{G}(\omega_3+\Omega)+\mathcal{G}(\omega_2-\Omega)\mathcal{G}(\omega_3+\Omega-\Delta\Omega)\nonumber\\
    &+\mathcal{G}(\omega_2+\Omega)\mathcal{G}(\omega_3-\Omega+\Delta\Omega)+\mathcal{G}(\omega_2+\Omega-\Delta\Omega)\mathcal{G}(\omega_3-\Omega)\big]\bigg].
\end{align}
In writing this expression, we evaluated the components of the susceptibility at the central frequencies of the pump ($\Omega$ and $\Omega-\Delta\Omega$) and probe ($\omega$). This approximation is appropriate for narrowband pump pulses, $\Omega\tau\gg1$, and is adopted to enable an efficient numerical evaluation of the response, as the integrals over $\omega_2$ and $\omega_3$ reduce to Gaussian form.
We then assume that $\chi^\mathcal{S}_{xy}(\omega;\omega_1,\omega_2,\omega_3)$ is weakly dependent on the frequency arguments for small variations $\pm\Omega$ around the central frequency of the probe $\omega$, and in particular approximately equivalent to $\chi^\mathcal{S}_{xy}(\omega;\omega,\Omega,-\Omega)$.
We can thus rewrite Eq.\ \eqref{diffintgauss} as
\begin{align}\label{diffintgauss2}
    \Delta\Gamma(t_{pp},0)
    \simeq& \frac{i}{2}\chi^\mathcal{S}_{xy}(\omega;\omega,\Omega,-\Omega)\int d\omega_1d\omega_2d\omega_3\,e^{-i(\omega_2+\omega_3)t_{pp}}\text{E}_\text{pr}(\omega_1)\nonumber\\
    \times\bigg[&-(\mathcal{E}_L^2-\mathcal{E}_R^2)\big(\mathcal{G}(\omega_2-\Omega+\Delta\Omega)\mathcal{G}(\omega_3-\Omega)+\mathcal{G}(\omega_2-\Omega)\mathcal{G}(\omega_3-\Omega+\Delta\Omega)\big)\nonumber\\
    &+(\mathcal{E}_L^{*2}-\mathcal{E}_R^{*2})\big(\mathcal{G}(\omega_2+\Omega-\Delta\Omega)\mathcal{G}(\omega_3+\Omega)+\mathcal{G}(\omega_2+\Omega)\mathcal{G}(\omega_3+\Omega-\Delta\Omega)\big)\nonumber\\
    &+\big(|\mathcal{E}_L|^2-|\mathcal{E}_R|^2\big)\big(\mathcal{G}(\omega_2-\Omega+\Delta\Omega)\mathcal{G}(\omega_3+\Omega)-\mathcal{G}(\omega_2-\Omega)\mathcal{G}(\omega_3+\Omega-\Delta\Omega)\nonumber\\
    &+\mathcal{G}(\omega_2+\Omega)\mathcal{G}(\omega_3-\Omega+\Delta\Omega)-\mathcal{G}(\omega_2+\Omega-\Delta\Omega)\mathcal{G}(\omega_3-\Omega)\big)\nonumber\\
    &+\big(|\mathcal{E}_L|^2-|\mathcal{E}_R|^2\big)\frac{\chi^\mathcal{A}_{xy}(\omega;\omega,\Omega,-\Omega)}{\chi^\mathcal{S}_{xy}(\omega;\omega,\Omega,-\Omega)}\big(\mathcal{G}(\omega_2-\Omega+\Delta\Omega)\mathcal{G}(\omega_3+\Omega)+\mathcal{G}(\omega_2-\Omega)\mathcal{G}(\omega_3+\Omega-\Delta\Omega)\nonumber\\
    &+\mathcal{G}(\omega_2+\Omega)\mathcal{G}(\omega_3-\Omega+\Delta\Omega)+\mathcal{G}(\omega_2+\Omega-\Delta\Omega)\mathcal{G}(\omega_3-\Omega)\big)\bigg].
\end{align}
As it is the case for monochromatic pumps, the terms in $\mathcal{E}_L^2-\mathcal{E}_R^2$ and $\mathcal{E}_L^{*2}-\mathcal{E}_R^{*2}$ contribute to the response centered at $\pm(2\Omega-\Delta\Omega)$, while terms in $|\mathcal{E}_L|^2-|\mathcal{E}_R|^2$ contribute to the quasi-static response centered at $\pm\Delta\Omega$.\\
To explicitly calculate the response as in Fig.\ 3(d) of the main text, we fix the ratio $\chi^\mathcal{A}_{xy}(\omega;\omega,\Omega,-\Omega)/\chi^\mathcal{S}_{xy}(\omega;\omega,\Omega,-\Omega)$ to a reasonable value obtained from calculations in the $sp$-model, and we take a narrowband Gaussian probe field with central frequency $\omega\gg\Omega$. The Fourier-transformed dichroic signal $\Delta\bar\Gamma_{R-L}(\omega_{pp},0)$ obtained from Eq.\ \eqref{diffintgauss2} provides the full pump-probe response, while the symmetric response is obtained neglecting the antisymmetric contribution. 
\clearpage
\section*{Phonon-mediated antisymmetric response}
\subsection*{Electron-phonon interaction and linear phonon excitation}
We here compute the antisymmetric susceptibility that mediates the light-induced Faraday effect in presence of IR-active phonons driven by the pump field. By definition, in the dipole approximation an IR-active phonon $\textbf{Q}$ transforms as a polar vector, and thus can be treated on the same footing as the vector potential $\textbf{A}$.
In the path-integral formalism, the system is described by $\mathcal{Z}=\int \mathcal{D}[\boldsymbol\psi]\mathcal{D}[\textbf{A}]\mathcal{D}[\textbf{Q}]e^{-S[\boldsymbol{\psi},\textbf{A},\textbf{Q}]}$, where the light-matter action reads
\begin{align}\label{lmactphon}
    S[\boldsymbol{\psi},\textbf{A},\textbf{Q}]=&-\sum_{i\nu_n,\textbf{k}}\boldsymbol{\psi}^T_\textbf{k}(-i\nu_n)\tilde G_0^{-1}(i\nu_n,\textbf{k})\boldsymbol{\psi}_\textbf{k}(i\nu_n)+\sum_{i\nu_n,\textbf{k}}\sum_{i\nu_n^\prime,\textbf{k}^\prime}\boldsymbol{\psi}_\textbf{k}^T(-i\nu_n)\Sigma_{\textbf{k}\textbf{k}^\prime}(i\nu_n-i\nu_n^\prime)\boldsymbol{\psi}_{\textbf{k}^\prime}(i\nu_n^\prime)\nonumber\\
    &-\sum_{i\Omega_m}\frac{\Omega_m^2+\omega_0^2(T)}{2\omega_0(T)}|\textbf{Q}(i\Omega_m)|^2,
\end{align}
where
\begin{align}\label{selfenergyphon}
    \Sigma_{\textbf{kk}^\prime}(i\nu_n-i\nu_n^\prime)\simeq\delta_{\textbf{k}\textbf{k}^\prime}\bigg[\sqrt{\frac{T}{N}}\frac{e}{c}&\sum_{i\Omega_m}[\tilde{\text{v}}_x\text{A}_x(i\Omega_m)+\tilde{\text{v}}_y\text{A}_y(i\Omega_m)]\delta(i\Omega_m-i\nu_n+i\nu_n^\prime)\nonumber\\
    +\frac{T}{N}\frac{e^2}{2c^2}&\sum_{i\Omega_m}\sum_{i\Omega_n}[\tilde\rho_x\text{A}_x(i\Omega_m)\text{A}_x(i\Omega_n)+\tilde\rho_y\text{A}_y(i\Omega_m)\text{A}_y(i\Omega_n)]\delta(i\Omega_m+i\Omega_n-i\nu_n+i\nu_n^\prime)\nonumber\\
    +\sqrt{\frac{T}{N}}&\sum_{i\Omega_m}[\tilde g_x\text{Q}_x(i\Omega_m)+\tilde g_y \text{Q}_y(i\Omega_m)]\delta(i\Omega_m-i\nu_n+i\nu_n^\prime)
    \nonumber\\
    +\frac{T}{N}\frac{e}{c}&\sum_{i\Omega_m}\sum_{i\Omega_n}[\tilde h_x\text{Q}_x(i\Omega_m)\text{A}_x(i\Omega_n)+\tilde h_y\text{Q}_y(i\Omega_m)\text{A}_y(i\Omega_n)]\delta(i\Omega_m+i\Omega_n-i\nu_n+i\nu_n^\prime)\bigg].
\end{align}
In particular, the $|\textbf Q|^2$ term in the second row of Eq.\ \eqref{lmactphon} describes the bare phonon, which we take dispersionless at a temperature-dependent frequency $\omega_0(T)$. The coefficients $\tilde g_i$, written in the orbital basis, mediate the electron-phonon interactions, while $\tilde h_i$ represent mixed phonon-photon vertices.
The transverse-optical IR-active phonon of the $sp$-model results in electron-phonon vertices $\tilde g_i$ and $\tilde h_i$ with a similar matrix structure to the paramagnetic ($\tilde{\text{v}}_i$) and diamagnetic ($\tilde\rho_i$) light-matter couplings of the purely electronic model, respectively.
In writing Eq.\ \eqref{selfenergyphon} we have immediately neglected terms that do not contribute to the two-phonon susceptibility $\chi^{\mathcal{A},\text{ph}}_{xy}$ shown in Fig.\ 4(a) of the main text. For example, terms corresponding to two-phonon-mediated electronic transitions $\sim\boldsymbol{\psi}\text Q_x\text{Q}_x\boldsymbol{\psi}$ and $\sim\boldsymbol{\psi}\text Q_y\text Q_y\boldsymbol{\psi}$ are only relevant when the fields can directly excite two phonons with the same polarization. However, the probe is assumed at a much higher frequency with respect to the characteristic phonon frequency, and we are interested in components of the susceptibility with mixed pump indices, $\chi_{xy;xy}$ and $\chi_{xy;yx}$, so that these vertices do not enter in the calculations. On the other hand, we neglect terms mixing $x$ and $y$ components at the same vertex, $\sim \boldsymbol{\psi} \text{Q}_x \text{Q}_y \boldsymbol{\psi}$, consistently with the structure of the electron–photon interaction obtained within the $sp$-model.
\\
We then integrate the electron field out of the partition function, and find the effective action that couples the electromagnetic field to the phonon field:
\begin{align}\label{seffphon}
    S_\text{eff}[\textbf{A},\textbf{Q}]=\sum_{m\geq1}\frac{\text{Tr}[\tilde G_0\Sigma]^m}{m}-\sum_{i\Omega_m}\frac{\Omega_m^2+\omega_0^2(T)}{2\omega_0(T)}|\textbf{Q}(i\Omega_m)|^2.
\end{align}
The relevant processes needed to calculate the phonon-mediated diagram are shown in Fig.\ \ref{FigS2}. We first focus on the linear couplings $\text Q_x\text A_x$ and $\text Q_y\text A_y$, that encode the direct excitation of the phonons via the pump fields. Subsequently, we evaluate the nonlinear electronic 4-point loop $\Pi^\mathcal{A}_{xy}$, composed by the terms scaling with $\text{Q}_x\text{Q}_y\text{A}_x\text{A}_y$, that describe mixed third-order processes with two phonons and two photons. As in the electronic case, we can distinguish between diamagnetic, paramagnetic, and mixed contributions.
\begin{figure}[t]
    \centering
    \includegraphics[width=0.98\textwidth,keepaspectratio]{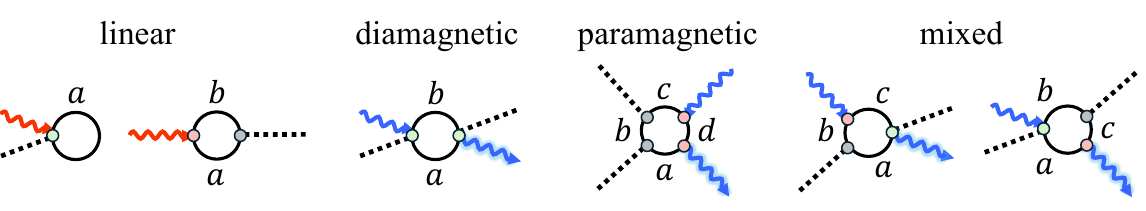}
    \caption{Diagrammatic representation of the processes contributing to the phonon-mediated diagram depicted in Fig.\ 4(a) of the main text. Red wavy lines denote the pump fields, Blue wavy lines denote the probe field, black dashed lines denote the phonon. Black solid lines represent electronic propagators. Red dots denote velocity vertices, gray dots represent the electron-phonon vertices, and green dots denote  the mixed phonon-photon vertices. The probe photon on which the detection is performed is highlighted with a blue shadow.}
    \label{FigS2}
\end{figure}
\subsection*{Linear coupling}
The linear couplings $\text Q_x\text A_x$ and $\text Q_y\text A_y$ are contained in the $m=1$ and $m=2$ terms of the expansion, and read 
\begin{align}
    S_{c}[\textbf{A},\textbf{Q}]&=\frac{e}{c}\sum_{i\Omega_m}\sum_{i=x,y}\frac{T}{N}\sum_{i\nu_n,\textbf{k}}\,\text{Tr}[\tilde G_0(i\nu_n,\textbf{k})\tilde h_i]\,\text{A}_i(i\Omega_m)\text{Q}_i(-i\Omega_m)\nonumber\\
    &+\frac{e}{c}\sum_{i\Omega_m}\sum_{i,j=x,y}\frac{T}{N}\sum_{i\nu_n,\textbf{k}}\text{Tr}[\tilde G_0(i\nu_n,\textbf{k})\tilde{\text{v}}_i\tilde G_0(i\nu_n+i\Omega_m,\textbf{k})\tilde{g}_j]\,\text{A}_i(i\Omega_m)\text{Q}_j(-i\Omega_m)\nonumber\\
    &=\frac{e}{c}\sum_{i\Omega_m}\sum_{i=x,y}\frac{T}{N}\sum_{i\nu_n,\textbf{k}}\text{Tr}[G_0(i\nu_n,\textbf{k})h_i]\text{A}_i(i\Omega_m)\text{Q}_i(-i\Omega_m)\nonumber\\
    &+\frac{e}{c}\sum_{i\Omega_m}\sum_{i,j=x,y}\frac{T}{N}\sum_{i\nu_n,\textbf{k}}\text{Tr}[G_0(i\nu_n,\textbf{k}){\text{v}}_i G_0(i\nu_n+i\Omega_m,\textbf{k}){g}_j]\,\text{A}_i(i\Omega_m)\text{Q}_j(-i\Omega_m),
\end{align}
where, following the notation defined above, $g_i=U_\textbf{k}^\dagger\tilde g_iU_\textbf{k}$ and $h_i=U_\textbf{k}^\dagger\tilde h_iU_\textbf{k}$ represent the electron-phonon vertices in the band basis. We can thus define 
\begin{align}
    Z_{ij}(i\Omega_m)=e\bigg[\frac{T}{N}\sum_{i\nu_n,\textbf{k}}\text{Tr}[G_0(i\nu_n,\textbf{k})h_i]+\frac{T}{N}\sum_{i\nu_n,\textbf{k}}\text{Tr}[G_0(i\nu_n,\textbf{k})\text{v}_iG_0(i\nu_n+i\Omega_m,\textbf{k})g_j]\bigg],
\end{align}
so that the action of the linear coupling becomes
\begin{align}\label{slincoup}
    S_c[\textbf{A},\textbf{Q}]=\frac{1}{c}\sum_{i,j=x,y}\sum_{i\Omega_m}Z_{ij}(i\Omega_m)\text{A}_i(i\Omega_m)\text{Q}_{j}(-i\Omega_m).
\end{align}
%
In the following we consider an isotropic and frequency-independent coupling, $Z_{ij}(i\Omega_m)=Z\,\delta_{ij}$. We note that the quantity $Z$ here defined is linked to the usual Born effective charge $Z_B$ as $Z_B=Z\sqrt{M/\omega_0(T)}$, with $M$ the ionic mass \cite{bistoni_2D19}.
\subsection*{Diamagnetic-like phonon-mediated antisymmetric kernel}
The diamagnetic-like contribution is contained in the $m=2$ term of $S_\text{eff}[\textbf{A},\textbf{Q}]$ in Eq.\ \eqref{seffphon}, when multiplying twice the fourth row of Eq.\ \eqref{selfenergyphon}. In particular,
\begin{align}
    S_\text{dia}[\textbf{A},\textbf{Q}]=\frac{e^2}{c^2}\sum_{i\Omega_m}\sum_{i\Omega_n}\sum_{i\Omega_l}\sum_{i\Omega_s}\frac{T}{N}\sum_{i\nu_n,\textbf{k}}&\text{Tr}[\tilde G_0(i\nu_n,\textbf{k})\tilde h_y\tilde G_0(i\nu_n+i\Omega_m+i\Omega_n,\textbf{k})\tilde h_x]\nonumber\\
    &\times\text{Q}_y(i\Omega_m)\text{A}_y(i\Omega_m)\text{Q}_x(i\Omega_l)\text{A}_x(i\Omega_s)\delta_{-i\Omega_s,i\Omega_m+i\Omega_n+i\Omega_l},
\end{align}
where $i\Omega_m+i\Omega_n+i\Omega_l+i\Omega_s=0$ by energy conservation. Following the same derivation detailed below Eq.\ \eqref{sdia1}, the two components of the diamagnetic kernel read
\begin{align}
    \text{K}^{\text{dia},\text{ph}}_{xy;yx}(\omega;\omega,\Omega,-\Omega)=-\frac{e^2}{2}\frac{1}{N}\sum_\textbf{k}\sum_{ab}[h_y]_{ab}[h_x]_{ba}\,d_{ab}(\omega+\Omega),
\end{align}
and
\begin{align}
    \text{K}^{\text{dia},\text{ph}}_{xy;xy}(\omega;\omega,\Omega,-\Omega)=-\frac{e^2}{2}\frac{1}{N}\sum_\textbf{k}\sum_{ab}[h_y]_{ab}[h_x]_{ba}\,d_{ab}(\omega-\Omega),
\end{align}
where $d_{ab}$ is defined as in Eq.\ \eqref{dab}.
In this way, the antisymmetric diamagnetic component reads
\begin{align}
    \text{K}^{\mathcal{A},\text{ph}}_\text{dia}(\omega;\omega,\Omega,-\Omega)=-\frac{e^2}{4}\frac{1}{N}\sum_\textbf{k}\sum_{ab}\big[[h_x]_{ab}[h_y]_{ba}+[h_y]_{ab}[h_x]_{ba}\big]\big[d_{ab}(\omega-\Omega)-d_{ab}(\omega+\Omega)\big].
\end{align}
\subsection*{Paramagnetic-like phonon-mediated antisymmetric kernel}
The paramagnetic-like contribution is contained in the $m=4$ term of $S_\text{eff}[\textbf{A},\textbf{Q}]$ in Eq.\ \eqref{seffphon}, when multiplying twice the first row with twice the third row of Eq.\ \eqref{selfenergyphon}. In particular,
\begin{align}
    S_\text{para}[\textbf{A},\textbf{Q}]=\frac{e^2}{4c^2}\sum_{i\Omega_m}&\sum_{i\Omega_n}\sum_{i\Omega_l}\sum_{i\Omega_s}\nonumber\\
    \frac{T}{N}\sum_{i\nu_n,\textbf{k}}&\text{Tr}\big[\tilde G_0(i\nu_n,\textbf{k})\tilde g_y\tilde G_0(i\nu_n+i\Omega_m,\textbf{k})\tilde{\text{v}}_y\tilde G_0(i\nu_n+i\Omega_m+i\Omega_n,\textbf{k})\tilde g_x \tilde G_0(i\nu_n+i\Omega_m+i\Omega_n+i\Omega_l, \textbf{k})\tilde{\text{v}}_x\big]\nonumber\\
    \times&\text{Q}_y(i\Omega_m)\text{A}_y(i\Omega_n)\text{Q}_x(i\Omega_l)\text{A}_x(i\Omega_s)\delta_{-i\Omega_s,i\Omega_m+i\Omega_n+i\Omega_l}+\text{perm.},
\end{align}
where ``perm.'' denotes all distinct permutations of the external field insertions, i.e., all possible ways of attaching the photon and phonon legs to the fermionic loop.
Following the same derivation detailed below Eq.\ \eqref{spara1}, the two components of the paramagnetic kernel read
\begin{align}
    \text{K}^\text{para,ph}_{xy;yx}(\omega;\omega,\Omega,-\Omega)=-\frac{e^2}{4}\frac{1}{N}\sum_\textbf{k}\sum_{abcd}\Big[&[\text{v}_y]_{ab}[g_{y}]_{bc}[g_x]_{cd}[\text{v}_x]_{da}\, p_{abcd}(\omega,\Omega,-\Omega)\nonumber\\
    +&[g_y]_{ab}[\text{v}_y]_{bc}[g_x]_{cd}[\text{v}_x]_{da}\, p_{abcd}(\Omega,\omega,-\Omega)\nonumber\\
    +&[g_x]_{ab}[g_y]_{bc}[\text{v}_y]_{cd}[\text{v}_x]_{da}\, p_{abcd}(-\Omega,\Omega,\omega)\nonumber\\
    +&[g_x]_{ab}[\text{v}_y]_{bc}[g_y]_{cd}[\text{v}_x]_{da}\, p_{abcd}(-\Omega,\omega,\Omega)\nonumber\\
    +&[g_y]_{ab}[g_x]_{bc}[\text{v}_y]_{cd}[\text{v}_x]_{da}\, p_{abcd}(\Omega,-\Omega,\omega)\nonumber\\
    +&[\text{v}_y]_{ab}[g_x]_{bc}[g_y]_{cd}[\text{v}_x]_{da}\, p_{abcd}(\omega,-\Omega,\Omega)\Big],
\end{align}
and
\begin{align}
    \text{K}^\text{para,ph}_{xy;xy}(\omega;\omega,\Omega,-\Omega)=-\frac{e^2}{4}\frac{1}{N}\sum_\textbf{k}\sum_{abcd}\Big[&[\text{v}_y]_{ab}[g_x]_{bc}[g_y]_{cd}[\text{v}_x]_{da}\, p_{abcd}(\omega,\Omega,-\Omega)\nonumber\\
    +&[g_x]_{ab}[\text{v}_y]_{bc}[g_y]_{cd}[\text{v}_x]_{da}\, p_{abcd}(\Omega,\omega,-\Omega)\nonumber\\
    +&[g_y]_{ab}[g_x]_{bc}[\text{v}_y]_{cd}[\text{v}_x]_{da}\, p_{abcd}(-\Omega,\Omega,\omega)\nonumber\\
    +&[g_y]_{ab}[\text{v}_y]_{bc}[g_x]_{cd}[\text{v}_x]_{da}\, p_{abcd}(-\Omega,\omega,\Omega)\nonumber\\
    +&[g_x]_{ab}[g_y]_{bc}[\text{v}_y]_{cd}[\text{v}_x]_{da}\, p_{abcd}(\Omega,-\Omega,\omega)\nonumber\\
    +&[\text{v}_y]_{ab}[g_y]_{bc}[g_x]_{cd}[\text{v}_x]_{da}\, p_{abcd}(\omega,-\Omega,\Omega)\Big],
\end{align}
where $p_{abcd}$ is defined as in Eq.\ \eqref{pabcd}.
Notice that we have fixed the detection on an $x$ photon field, which fixes the last matrix element $[\text{v}_x]_{da}$. By defining the vectors $\textbf{v}=(\text v_x,\text v_y)$ and $\boldsymbol{g}=(g_x,g_y)$, the antisymmetric paramagnetic component can be written in a compact way as
\begin{align}
    \text{K}^{\mathcal{A},\text{ph}}_\text{para}(\omega;\omega,\Omega,-\Omega)=-\frac{e^2}{8}\frac{1}{N}\sum_\textbf{k}\sum_{abcd}&\big[([\boldsymbol{g}]_{ab}\times[\boldsymbol{g}]_{bc})\cdot([\textbf{v}]_{cd}\times[\textbf{v}]_{da})\big]\big[p_{abcd}(-\Omega,\Omega,\omega)-p_{abcd}(\Omega,-\Omega,\omega)\big]\nonumber\\
    +&\big[([\boldsymbol{g}]_{ab}\times[\boldsymbol{g}]_{cd})\cdot([\textbf{v}]_{bc}\times[\textbf{v}]_{da})\big]\big[p_{abcd}(-\Omega,\omega,\Omega)-p_{abcd}(\Omega,\omega,-\Omega)\big]\nonumber\\
    +&\big[([\boldsymbol{g}]_{bc}\times[\boldsymbol{g}]_{cd})\cdot([\textbf{v}]_{ab}\times[\textbf{v}]_{da})\big]\big[p_{abcd}(\omega,-\Omega,\Omega)-p_{abcd}(\omega,\Omega,-\Omega)\big].
\end{align}
\subsection*{Mixed diamagnetic-paramagnetic phonon-mediated antisymmetric kernel}
The mixed diamagnetic-paramagnetic contributions are contained in the $m=3$ term of $S_\text{eff}[\textbf{A},\textbf{Q}]$ in Eq.\ \eqref{seffphon}, when multiplying the first, the third, and the fourth row of Eq.\ \eqref{selfenergyphon}. In particular, we can distinguish between two contributions,
\begin{align}
    S_\text{mix}^{(1)}[\textbf{A},\textbf{Q}]=\frac{e^2}{3c^2}\sum_{i\Omega_m}\sum_{i\Omega_n}\sum_{i\Omega_l}\sum_{i\Omega_s}\frac{T}{N}\sum_{i\nu_n,\textbf{k}}\text{Tr}&\big[\tilde G_0(i\nu_n,\textbf{k})\tilde{g}_y\tilde G_0(i\nu_n+i\Omega_m,\textbf{k})\tilde{\text{v}}_y\tilde G_0(i\nu_n+i\Omega_m+i\Omega_n,\textbf{k})\tilde h_x\big]\nonumber\\
    \times&\text{Q}_y(i\Omega_m)\text{A}_y(i\Omega_n)\text{Q}_x(i\Omega_l)\text{A}_x(i\Omega_s)\delta_{-i\Omega_s,i\Omega_m+i\Omega_n+i\Omega_l},
\end{align}
corresponding to a detection performed on a diamagnetic insertion, and
\begin{align}
    S_\text{mix}^{(2)}[\textbf{A},\textbf{Q}]=\frac{e^2}{3c^2}\sum_{i\Omega_m}\sum_{i\Omega_n}\sum_{i\Omega_l}\sum_{i\Omega_s}
    \frac{T}{N}\sum_{i\nu_n,\textbf{k}}\text{Tr}&\big[\tilde{G_0}(i\nu_n,\textbf{k})\tilde{h}_y\tilde G_0(i\nu_n+i\Omega_m+i\Omega_n,\textbf{k})\tilde{g}_x\tilde G_0(i\nu_n+i\Omega_m+i\Omega_n+i\Omega_l,\textbf{k})\tilde{\text{v}}_x\big]\nonumber\\\times&\text{Q}_y(i\Omega_m)\text{A}_y(i\Omega_n)\text{Q}_x(i\Omega_l)\text{A}_x(i\Omega_s)\delta_{-i\Omega_s,i\Omega_m+i\Omega_n+i\Omega_l},
\end{align}
corresponding to a detection performed on a paramagnetic insertion.
Following the same derivation detailed below Eqs.\ \eqref{smix1} and \eqref{smix2}, one finds the antisymmetric contributions coming from these two actions, that read
\begin{align}
    \text{K}^{\mathcal{A}(1),\text{ph}}_\text{mix}(\omega;\omega,\Omega,-\Omega)=-\frac{e^2}{6}\frac{1}{N}\sum_\textbf{k}\sum_{abc}\big[[\text{v}_x]_{ab}[g_x]_{bc}[h_y]_{ca}+[\text{v}_y]_{ab}[g_y]_{bc}[h_x]_{ca}\big]\big[m_{abc}(\omega,\omega-\Omega)-m_{abc}(\omega,\omega+\Omega)\big],
\end{align}
and 
\begin{align}
    \text{K}^{\mathcal{A}(2),\text{ph}}_\text{mix}(\omega;\omega,\Omega,-\Omega)=-\frac{e^2}{6}\frac{1}{N}\sum_\textbf{k}\sum_{abc}\big[[h_x]_{ab}[g_x]_{bc}[\text{v}_y]_{ca}+[h_y]_{ab}[g_y]_{bc}[\text{v}_x]_{ca}\big]\big[m_{abc}(\omega-\Omega,\omega)-m_{abc}(\omega+\Omega,\omega)\big],
\end{align}
where $m_{abc}$ is defined as in Eq.\ \eqref{mabc}. The total mixed diamagnetic-paramagnetic antisymmetric kernel is then found as 
\begin{align}
    \text{K}^{\mathcal{A},\text{ph}}_\text{mix}(\omega;\omega,\Omega,-\Omega)=\text{K}^{\mathcal{A}(1),\text{ph}}_\text{mix}(\omega;\omega,\Omega,-\Omega)+\text{K}^{\mathcal{A}(2),\text{ph}}_\text{mix}(\omega;\omega,\Omega,-\Omega).
\end{align}
\subsection*{Antisymmetric phonon-mediated susceptibility}
The total antisymmetric electronic loop reads
\begin{align}
    \Pi^\mathcal{A}_{xy}(\omega;\omega,\Omega,-\Omega)=\frac{1}{\omega^2\Omega^2}\big[\text{K}^{\mathcal{A},\text{ph}}_\text{dia}(\omega;\omega,\Omega,-\Omega)+\text{K}^{\mathcal{A},\text{ph}}_\text{para}(\omega;\omega,\Omega,-\Omega)+\text{K}^{\mathcal{A},\text{ph}}_\text{mix}(\omega;\omega,\Omega,-\Omega)\big].
\end{align}
This response function does not by itself represent a physical optical response, as it requires two phonon insertions. The corresponding response function for the process shown in Fig.\ 4(a) of the main text is obtained by attaching the phonon propagators to two pump photons via the linear coupling in Eq.\ \eqref{slincoup}, and integrating the phonon field out of the partition function. In this way, one finally finds
\begin{align}\label{chiph}
    \chi^{\mathcal{A},\text{ph}}_{xy}(\omega;\omega,\Omega,-\Omega)=Z^2 D(\Omega)D(-\Omega)\,\Pi^\mathcal{A}_{xy}(\omega;\omega,\Omega,-\Omega),
\end{align}
where 
\begin{align}
    D(\Omega)=\frac{2\omega_0(T)}{(\Omega+i0^+)^2-\omega_0^2(T)}
\end{align}
is the bare-phonon propagator. Eq.\ \eqref{chiph} corresponds to Eq.\ (13) of the main text, once a temperature-dependent phenomenological damping parameter $\gamma(T)$ is introduced in the phonon propagator to describe its inverse lifetime.

The symmetric electronic loop $\Pi^\mathcal{S}_{xy}$, and the corresponding optical response function $\chi^\mathcal{S}_{xy}$, can be readily obtained from the kernel components derived above with analogous calculations.
\subsection*{Relative magnitude between all-electronic and phonon-mediated processes}
The relative magnitude between the fully electronic process described by $\chi^{\mathcal{A}}_{xy}$ in Eq.\ \eqref{chiel} and the phonon-mediated contribution described by $\chi^{\mathcal{A},\text{ph}}_{xy}$ in Eq.\ \eqref{chiph} can be estimated from their ratio. We take the Fermi velocity $\text{v}_\text{F}$ as the characteristic magnitude scale of the velocity vertices, while the electron-phonon vertex scales as $\beta\,\text{v}_{F}\sqrt{\frac{1}{M\omega_0(T)}}$ \cite{bistoni_2D19}, where $\beta$ is a dimensionless parameterization of the electron-phonon coupling strength. One then finds
\begin{align}
    \frac{\chi^{\mathcal{A},\text{ph}}_{xy}(\omega;\omega,\Omega,-\Omega)}{\chi^{\mathcal{A}}_{xy}(\omega;\omega,\Omega,-\Omega)}\sim \frac{\alpha\text{v}_\text{F}^2\Big(\beta\,\text{v}_\text{F}\sqrt{\frac{1}{M\omega_0(T)}}\Big)^2Z^2}{\text{v}_\text{F}^4}D(\Omega)D(-\Omega)=\frac{\alpha\beta^2Z^2}{M\omega_0(T)}D(\Omega)D(-\Omega),
\end{align}
where $\alpha$ is a dimensionless constant of order 1.
Introducing the Born effective charge as $Z=Z_B\sqrt{\frac{\omega_0(T)}{M}}$, we obtain
\begin{align}
    \frac{\chi^{\mathcal{A},\text{ph}}_{xy}(\omega;\omega,\Omega,-\Omega)}{\chi^{\mathcal{A}}_{xy}(\omega;\omega,\Omega,-\Omega)}\sim \frac{\alpha\beta^2Z_B^2}{M^2}D(\Omega)D(-\Omega).
\end{align}
An \textit{ab-initio} estimate of the ionic mass $M$, the electron-phonon coupling $\beta$ and the Born effective charge $Z_B$ would therefore provide a quantitative estimate of the relative strength between the two processes. In the main text, we use $\sqrt{\alpha}\beta Z_B/M=0.06\,\Omega$, which is of the correct order of magnitude of what expected in SrTiO$_3$. Spectra reported in Fig.\ 4 of the main text are obtained with Gaussian pulses, following calculations analogous to those discussed below Eq.\ \eqref{gausspulse}.
\clearpage
\section*{Estimate of the relative strength between IFE-induced and light-induced Faraday response}
As discussed in Section V of the main text, the nonlinear polarization induced by an inverse Faraday effect (IFE) and a light-induced Faraday effect can be written, in cylindrical geometry and MKS units, as
\begin{align}
        \langle \text{P}^x \rangle\simeq i\varepsilon_0\bigg[\frac{1}{c^2}\big[\chi^{(\text{F})}\big]^2+\chi^{\mathcal{A}}_{xy}\bigg]\,\mathcal{E}_\text{pr}^y(|{\mathcal{E}_L}|^2-|{\mathcal{E}_R}|^2).
\end{align}
The relative strength between the two effects is thus given by the ratio $\frac{1}{c^2}\chi^{\text{(F)}}/\chi_{xy}^\mathcal{A}$. Assuming a reasonable estimate of $\chi^\mathcal{A}_{xy}/\chi^\mathcal{S}_{xy}=0.1$, we compute an order-of-magnitude estimate of the ratio $\frac{1}{c^2}\chi^{\text{(F)}}/\chi_{xy}^\mathcal{S}$ in static conditions, in which both the Faraday and the symmetric susceptibility are finite.

Working within the many-body diagrammatic theory presented above, the Faraday response is composed by 3-legged response functions where two $\textbf{A}$ insertions are associated with orthogonal electric fields and one is associated with a magnetic field, i.e., the curl of the vector potential, along the $z$ component. The resulting current reads \cite{fukuyama_rpf69}
\begin{align}
   \text{J}_x(\omega)\propto\chi_{xyz}^{(\text{F})}(\omega;\omega,0)\,\text{A}_y(\omega)(i\text{q}_x\text{A}_y(0)-i\text{q}_y\text{A}_x(0))_z,
\end{align}
where $\textbf{q}$ is the light momentum. Building on the results of Ref.\ \cite{fukuyama_rpf69}, the Faraday tensor $\chi_{xyz}^{(\text{F})}$ can be expressed as the small-$\textbf{q}$ expansion of the linear electric susceptibility $\chi^e(\textbf{q})$. This leads to
\begin{align}
    \chi^{(\text{F})}_{xyz}(i\Omega_m;i\Omega_m,0)&=\frac{1}{\varepsilon_0}\frac{e}{\hbar}\frac{T}{N}\sum_{i\nu_n,\textbf{k}}\sum_{ab} \bigg(e^2[\text{v}_x]_{ab}\frac{\partial [\text{v}_y]_{ba}}{\partial \text{k}_y}-e^2[\text{v}_y]_{ab}\frac{\partial[\text{v}_x]_{ba}}{\partial\text{k}_y}\bigg)\nonumber\\
    &\times\bigg[[G_0(i\nu_n,\textbf{k})]_{aa}\frac{\partial [G_0(i\nu_n\!+\!i\Omega_m,\textbf{k})]_{bb}}{\partial\text{k}_x}\!-\!\frac{\partial [G_0(i\nu_n,\textbf{k})]_{aa}}{\partial\text{k}_x}[G_0(i\nu_n\!+\!i\Omega_m,\textbf{k})]_{bb}\bigg].   
\end{align}
We can then evaluate $\partial_{\text{k}_i}[G_0(i\nu_n,\textbf{k})]_{aa}=-\frac{1}{(i\nu_n-E_a(\textbf{k}))^2}\big[\frac{\partial E_{a}(\textbf{k})}{\partial \text{k}_i}\big]_{aa}=-\hbar[G_0(i\nu_n,\textbf{k})]_{aa}^2[\text{v}_i]_{aa}$.

In isotropic systems the velocity is proportional to $\textbf{k}/m$, with $m$ the effective mass, and thus $[\text{v}_i]_{ab}=\zeta_{ab\,}\hbar\text{k}_i/m$. In the following we take $[\text{v}_i]_{ab}=\text{v}_F$, where $\text{v}_F$ is the Fermi velocity, and we assume possible transitions only at the direct gap $\Delta$.
Analogously, $\partial_{\text{k}_j}[\text{v}_i]_{ab}=\zeta_{ab}\delta_{ij}\hbar/m$, and we thus take $\partial_{\text{k}_j}[\text{v}_i]_{ab}=\delta_{ij}\hbar/m_e$, where $m_e$ is the electron mass. In the static approximation, the sum over the Matsubara frequency gives a $1/\Delta^2$ scaling, while the sum over the internal momentum $\textbf{k}$ gives a $1/a^3$ scaling, with $a$ the lattice spacing. In this case,
\begin{align}
    \chi^{(\text{F})}_{xyz}(0;0,0)\sim \frac{1}{\varepsilon_0}\frac{e\hbar}{m_e}\frac{e^2\text{v}_F^2}{a^3\Delta^2}\sim \frac{\mu_B}{\varepsilon_0} \frac{e^2\text{v}_F^2}{a^3\Delta^2}
\end{align}
where $\mu_B=e\hbar/2m_e$ is the Bohr magneton.

Under the same approximations applied to the diagrammatic calculations detailed in previous sections, the symmetric optical susceptibility scales as
\begin{align}
    \chi_{xy}^\mathcal{S}(0;0,0,0)\sim \frac{1}{\varepsilon_0}\frac{e^4\text{v}_F^4}{a^3\Delta^3}.
\end{align}
We can thus estimate the ratio as
\begin{align}
    \frac{[\chi^{(\text{F})}_{xyz}(0;0,0)]^2/c^2}{\chi_{xy}^\mathcal{S}(0;0,0,0)}\sim \frac{1}{\varepsilon_0c^2}\frac{\mu_B^2}{a^3}\frac{1}{\Delta}=\frac{4\pi E_\text{mag}}{\Delta},
\end{align}
where $E_\text{mag}=\frac{\mu_0}{4\pi}\frac{\mu_B^2}{a^3}$ is the characteristic magnetic energy associated with the dipole-dipole interaction between two Bohr magnetons separated by a distance $a$.

\end{document}